\documentclass[reprint, amsmath,amssymb, aps]{revtex4-2}
\usepackage{color}    
\usepackage{dcolumn}
\usepackage{bm}
\usepackage{mathrsfs}
\usepackage{amsmath}
\usepackage{amssymb}
\usepackage{mathtools}
\usepackage{graphicx}
\usepackage{longtable}
\usepackage{url}
\usepackage{physics}
\usepackage{float}
\usepackage{amsthm} 
\usepackage{stmaryrd, bbm, slashed, simplewick} 
\usepackage[hidelinks]{hyperref}
\usepackage{cleveref}
\usepackage{csquotes}
\usepackage{setspace}
\DeclareGraphicsExtensions{.eps}
\usepackage{tikz}

\makeatletter
\newcommand{\fmarki}{*}
\newcommand{\fmarkii}{\ensuremath{\dagger}}
\newcommand{\fmarkiii}{\ensuremath{\ddagger}}
\newcommand{\fmarkiiii}{\ensuremath{$\mathsection$}}

\def\@fnsymbol#1{{\ifcase#1\or \fmarki\or \fmarkii\or \fmarkiii\or \fmarkiiii \else\@ctrerr\fi}}
\makeatother

\renewcommand{\fmarki}{*}
\renewcommand{\fmarkii}{$\flat$}
\renewcommand{\fmarkiii}{$\mathsection$}
\renewcommand{\fmarkiiii}{$\natural$}

\begin{document}

\preprint{APS/123-QED}

\title{Entanglement Negativity on Random Spin Networks}

\author{Goffredo Chirco}
\email{goffredo.chirco@unina.it}
\author{Simone Cepollaro}
\email{simone.cepollaro-ssm@unina.it}
\altaffiliation[Also at]{ Scuola Superiore Meridionale, Largo S. Marcellino 10, 80138 Napoli, Italy.}
\author{Gianluca Cuffaro}
\email{g.cuffaro@studenti.unina.it}
\author{Vittorio D'Esposito}
\email{vittorio.desposito@unina.it}
\affiliation{Dipartimento di Fisica ``Ettore Pancini'', Universit\`a di Napoli Federico II, Napoli, Italy; INFN, Sezione di Napoli, Italy;}

\begin{abstract}
We investigate multipartite entanglement for quantum states of $3d$ space geometry, described via generalised \emph{random} spin networks with fixed areas, in the context of background independent approaches to quantum gravity. We focus on entanglement \emph{negativity} as a well defined witness of quantum correlations for \emph{mixed} states, in our setting describing generic subregions of the boundary of a quantum $3d$ region of space. 
In particular, we consider a generic tripartition of the boundary of an open spin network state and we compute the typical R\'enyi negativity of two boundary subregions $A$ and $B$ immersed in the environment $C$, explicitly for a set of simple open random spin network states. We use the random character of the spin network to exploit replica and random average techniques to derive the typical R\'enyi negativty 
via a classical generalised Ising model correspondence generally used for random tensor networks in the large bond regime.  For trivially correlated random spin network states, with only local entanglement between spins located on the network edges, we find that typical log negativity displays a holographic character, in agreement with the results for random tensor networks, in large spin limit. When non-local bulk entanglement between intertwiners at the vertices is considered the negativity increases, while at the same time the holographic scaling is generally perturbed by the bulk contribution.

\begin{description}
\item[Keywords]
quantum spin networks, quantum gravity, multipartite entanglement negativity,\\ entanglement/geometry correspondence \end{description}
\end{abstract}

\maketitle

\section{\label{sec:level1}Introduction}

In the last decade, exciting new insights towards the problem of the emergence of classical spacetime geometry in quantum gravity came from ideas and tools of quantum information theory, under the radical perspective that classical spacetime might emerge from
the entanglement of the degrees of freedom of a non-perturbative quantum level of description~\cite{VanRaamsdonk:2010pw, Swingle:2009bg, Bianchi_2014, articleRaa, Lashkari:2013koa, Faulkner_2014, Cao:2016mst, Jacobson_2016, PhysRevD.95.024031, swingle2014universality, Qi:2013caa}. 
The idea of entanglement as the \emph{fabric of spacetime} and its interplay with holography is advanced in the framework of the AdS/CFT correspondence ~\cite{Maldacena:1997re}, as a natural consequence of the holographic entropy formula of Ryu and Takayanagi (the RT formula hereafter) and Maldacena's extended black hole picture~\cite{PhysRevLett.96.181602, Engelhardt:2014gca, Juan_Maldacena_2003}, which together suggested the possibility of reconstructing spacetime geometry in terms of a hierarchy of correlations of the holographic dual CFT state. 

In this context, quantitative studies of spacetime reconstruction from entanglement have been carried on for discretized toy models of holography, initially identifying specific tensor network MERA \cite{Vidal:2007hda} decompositions of the CFT state with bulk AdS-like discrete geometry \cite{Swingle:2012wq,Bhattacharyya:2016hbx,Bao:2015uaa,Bao:2018pvs, Bao:2019fpq,Yang:2015uoa}. Diverse classes of tensor network states satisfying the RT formula, like the quantum error-correcting HaPPY code~\cite{Pastawski:2015qua}, perfect tensors and their generalisations in terms of random tensor networks~\cite{Yang:2015uoa, Hayden:2016cfa} have raised a lot of interest toward a generalisation of the holographic scheme beyond the AdS/CFT duality. However, in most cases, the very meaning of assigning a geometry to a tensor network remains unclear, while the very holographic reconstructing power of such models is often undermined by flat entanglement spectra (see e.g.~\cite{Chen:2021lnq}), eventually providing an over-simplified modelling of the holographic duality.

More recently, the idea of a holographic geometry/entanglement correspondence has drawn a lot of attention in background independent approaches to quantum gravity, including loop quantum gravity (LQG) and its covariant generalisations, spin foams, simplicial models, and higher rank random matrix models generalisations like group field theory (GFT) ~\cite{thiemann_2007, rovelli_vidotto_2014, Perez:2012wv, Oriti:2013aqa, Ambjorn:2015rva, Bonzom:2011zz}. In this framework, different approaches share a kinematic description of quantum (space) geometry in terms of \emph{spin network} states, namely symmetric tensor network states defined on graphs labelled by spin representations and intertwining operators~\cite{Penrose_71, penrose_rindler_1984, Rovelli:1995ac}. On the one hand, spin networks realise networks of frame transformations, which operationally encode the 3d space manifold description into purely combinatorial and algebraic variables. On the other hand, such networks can be described as quantum many-body-like collections of \emph{fundamental} quanta of space. Such quanta are glued by quantum correlations to constitute discrete spatial geometries and dynamically interacting to give rise to spacetime manifolds of arbitrary topology~\cite{Donnelly:2008vx, Donnelly:2011hn,Baytas:2018wjd,
Livine:2017fgq, Delcamp:2016eya, Bianchi:2016hmk,
Bianchi:2015fra}. 
Geometric and topological features of such spin tensor networks are inherently related to their entanglement structure and can be studied in terms of well defined quantum geometry operators. 
 
While it is the very role of holography to be still unclear and to some extent little explored in background-independent quantum gravity, also in this context one expects holographic entanglement to play a key role in characterising ``physical'' quantum geometry states~\cite{PhysRevD.90.044044,Chirco_2015, Hamma2015AreaLF, Bianchi2016LoopEA, Bianchi:2019pvv}, possibly reflecting signatures of Einstein's equations or invariance for diffeomorphisms at the quantum level, as well as  constraining coarse-graining and effective emerging dynamics in the continuum limit~\cite{Livine:2006xk, Feller:2017jqx, Anza:2016fix, Chirco:2018fns, Chirco:2017xjb}.

Recent results in this framework have focused on \emph{local} realizations of a generalised holographic duality~\cite{Raju:2019qjq,Dittrich:2017hnl,Dittrich:2018xuk}, also by looking at classes of \emph{open} spin network states modelling quantum regions of space with boundaries~\cite{Chen_2021}. In this sense, much work has investigated the validity of the RT formula as a signature of holography for bipartite entanglement in pure boundary spin network states. 
Examples of bulk-from-boundary reconstruction in the light of holography have been proposed in~\cite{Chen:2022rty}, supporting the possibility of direct mapping of the hierarchy of correlations of the spin networks to a hierarchy of geometrical observables on quantum geometry states.

Further, in GFT, a convenient dictionary between the spin networks and the tensor networks formalism has been proposed in~\cite{Chirco_2018} (see also~\cite{Colafranceschi:2020ern} for a recent review), providing a tentative bridge between the entanglement geometry correspondence in AdS/CFT and background independent quantum gravity.

In particular, building on the quantum many-body analogy, recent work in GFT along the lines of~\cite{Hayden:2016cfa} has investigated the notion of \emph{random} spin network state, with randomness intended as the result of some coarse-graining induced by quantum gravitational dynamics. The statistical description of random spin networks has been then exploited to investigate the holographic character of the entanglement entropy in relation to quantum \emph{typicality}, in the regime of large spins dimension~\cite{Chirco2018a, Chirco_2020,
Chirco:2021chk,
Colafranceschi:2021acz, Colafranceschi:2022dig}.

In this work, we extend previous results on bipartite entanglement entropy for random spin networks to the case of \emph{multipartite} entanglement, by focusing on \emph{mixed} boundary spin network states. For such states, we shift the focus from entanglement entropy to entanglement \emph{negativity}~\cite{Peres_1996, PhysRevA.58.883,PhysRevA.60.3496, Eisert_99, PhysRevA.65.032314, Plenio_05, PhysRevB.94.035152}, an entanglement witness well-defined for mixed states. Our motivation is twofold. On the one hand,  we are interested in investigating computable measures of the mutual entanglement between two subsystems in a mixed state. Characterising the correlations of mixed quantum geometry states is of fundamental importance for consistently dealing with entanglement under coarse-graining, with the hope to shed light on the role of entanglement in open issues like the dynamical process of emergence of classical spacetime geometry, as well as quantum black hole physics and quantum chaotic dynamics in quantum gravity. On the one hand, quantifying multipartite entanglement is necessary to better characterise \emph{nonlocal} correlations in quantum geometry, hence to start considering hierarchies of correlations and its geometric interpretation in the perspective of a spacetime reconstruction from entanglement.
 
In particular, inspired by the recent work~\cite{Kudler_21, Dong_21}, we study the mutual entanglement of a random boundary spin network state over a tripartite boundary Hilbert space corresponding to three boundary regions $A, B, C$. We quantify the entanglement of two subregions $(A, B)$ by measuring the typical logarithmic negativity of the reduced mixed state $\rho_{AB}$. The calculation is given in explicit details for a set of very simple states.

For trivially correlated states, we find that the entanglement negativity of the two boundary regions is proportional to the sum of the areas of the minimal surfaces homologous to $A$ and $B$, which extend in the spin network bulk,
$$ \log\overline{\mathcal{N}}_{AB}\propto (\abs{\gamma_A}+\abs{\gamma_B})-\abs{\gamma_C}\, .
$$ 
with a negative contribution given by the bulk surface area $\abs{\gamma_C}$ testifying the presence of the environment $C$.  This is a natural generalisation of the RT formula for the entanglement entropy of a pure bipartite state $\rho_{AB}$. 

The tripartite case, however, displays a richer entanglement phases structure~\cite{Shapourian_21}. For states with non-trival bulk correlations, the presence of an environment reflects in the appearance of new internal bulk domain, which we call \emph{transition region} $T$, corresponding to the so called quantum islands in recent literature. When the bulk is highly entangled, the minimal surfaces are prevented from entering the graph and they end up coinciding with the outer boundary surface. In this case, we have that $\gamma_C=\qty(\gamma_A\cup\gamma_B)^c=\gamma_{AB}$, and the log negativity 
$$ \log\overline{\mathcal{N}}_{AB}\propto (\abs{\gamma_A}+\abs{\gamma_B})-\abs{\gamma_{AB}}\, .
$$ 
can be interpreted as a holographic formula for the quantum mutual information $I_{A:B}$ of the two subsystems.

In the proposed derivation, we see how local entanglement between spin states located on the network’s edges and non-local, gauge-invariant entanglement between intertwiners at the vertices play two different roles. Respectively, spin entanglement has a structural role in defining the connectivity of the graph, which naturally affect the definition of the minimal bulk surfaces. On the other hand, intertwiner entanglement, which defines correlations between actual geometry excitations in the bulk, generally tends to raise the negativity of the boundary by favouring the formation of transitory regions in the bulk, while at the same time tends to break up the holographic behaviour.\\

The paper is organised as follows. Section~\ref{sec:level2} provides a set of preliminary notions. First, we introduce the notion of spin network vertex state and its dual interpretation as a quantum of $3d$ geometry. Within the GFT formalism, we then focus on generalised \emph{open} spin networks describing a $3d$ quantum geometry with boundary. Here, we briefly recall the dictionary between spin networks and tensor networks to introduce a class of tensor product spin network states analogue to symmetric PEPs. We use this class of states as \emph{locally entangled backgrounds} from which non-trivially correlated states are defined via bulk to boundary mappings. Finally, we characterise the notion of random spin network states and we define the measure of negativity we are going to compute. 

In Section~\ref{setting} we define our working setting. We are interested in quantifying the entanglement of two generic subregions of a boundary random spin network state. We consider a tripartition of the boundary in three subregions $\{A, B, C\}$, and define a reduced boundary mixed state $\rho_{AB}$ by tracing out the $C$ system, intended as a generic \emph{environment}.
Due to the random character of the states, $k$-th order R\'enyi log-negativity is computed in expected value. Expected momenta of the reduced density matrix get mapped, via averaging, to partition functions of a generalised Ising model~\cite{Dong_21,Shapourian_21}, where Ising spin variables are replaced by elements of the permutation group attached to each vertex. This effectively turns the random spin network ensemble into a Cayley graph, where elements of the permutation group at each vertex interact pairwise throughout the graph according to the tripartite boundary conditions. 
In the large spin (typical) regime, computing the log-negativity of the boundary state amounts to finding the minimal free energy configurations of such dual statistical model.

In Section~\ref{compute}, we define a specially simple class of random spin networks. We explicitly compute the negativity for two simple examples given by open spin network states defined on open tree graphs with two and three vertices. A brief discussion of the results follows in Section \ref{discussion}. We provide further auxiliary material in four appendices. In Appendix \ref{permutation appendix} we recall the notion of geodesic for a Cayley metric on the permutation group. Appendix \ref{mapping} is dedicated to the technical details of the statistical mapping to the generalised Ising model. In Appendix~\ref{spinn}, we show how the third order negativity can be mapped to an alternative Ising model by decomposing the symmetric group $\mathbb{S}_3$ in terms of swap operators. The result is the natural extension of the Ising model one obtains for the $2$-nd Rényi entropy via statistical mapping~\cite{Chirco:2021chk}. Finally, in Appendix~\ref{Hamiltonian appendix}, we give the details of the computation of the $k$-th order Hamiltonians discussed in Section~\ref{compute}.

\section{\label{sec:level2}Preliminaries}
\begin{figure}
	\includegraphics[width=0.45\textwidth]{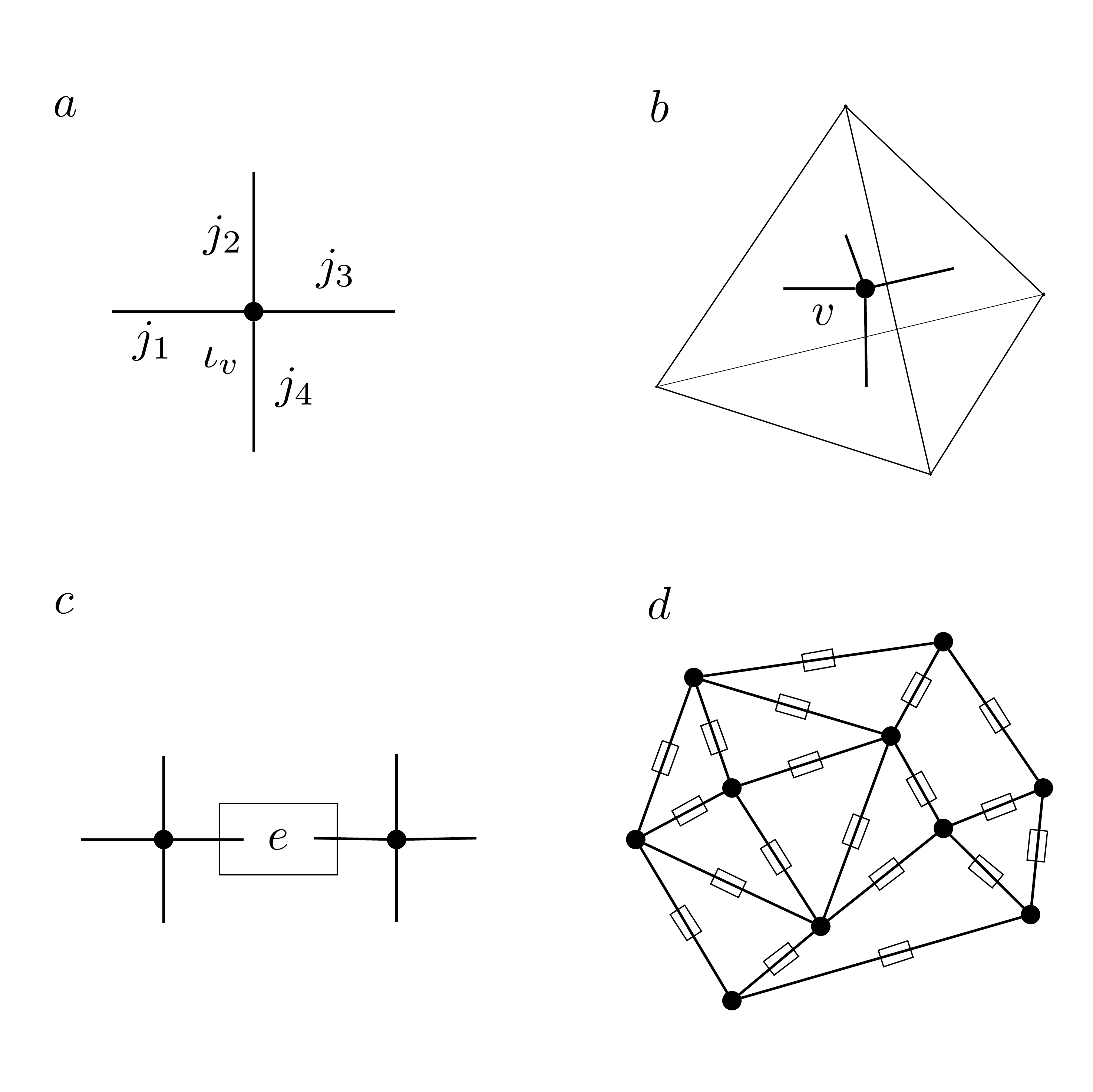}
	\caption{Construction of a spin network. Panel (a) shows a single spin network vertex and (b) is the dual description in terms of a tetrahedron. Panel (c) shows the gluing of two such vertices and panel (d) represents a spin network with multiple vertices.}
	\label{fig:sn}
\end{figure} 
\subsection{\label{sec:sub1} Quantum Spin Network States}
Let us review how spin network states are constructed as portions of $3d$ space quantised and glued together to form extended quantum geometry states.

Consider an elementary region of $3d$ space defined by a convex polyhedron geometry with $d$ faces. This is a $(d-1)$-simplex, comprised by $d$ vectors $\{\vec{N}_i\}_{i=1}^d$ in $\mathbb{R}^3$ with norms $\{j_i\}$, satisfying the closure constraint $\sum_i \vec{N}_i=\vec{0}$.  The space of such vectors modulo rotations gives the phase space (or space of shapes) of a $(d-1)$- simplex with fixed areas in three dimensions~\cite{Barrett_1998, baez1999quantum, PhysRevD.83.044035} 
\begin{equation}
S_p=\Big \{ \{|\vec{N}_i|= j_i\}_{i=1}^d \Big| \sum_i  \vec{N}_i=\vec{0} \Big\}\Big/SO(3)\, ,
\end{equation} 
which has the structure of a symplectic manifold. The quantization of $S_p$ consists in the quantum reduction of the product of irreducible representations spaces $V^{j_i}$ of $SU(2)$, which quantise the 2-sphere $S^2_{j_i}$ of radius $j_i$ associated to each element of $\{|\vec{N}_i|= j_i\}$. The result is the so called \emph{intertwiner space}
\begin{equation}
I_{\{j_i\}}\equiv \text{Inv}_{SU(2)}\Big[\bigotimes_{i=1}^d V^{j_i}\Big]\, ,
\end{equation}
which can be intended as the space of a quantum polyhedron with fixed areas values. The symplectic volume of the space tells us in how many ways we can recouple $d$ spins into a singlet state (of recoupled spin $J=0$) invariant under $SU(2)$. A basis state on $I_{\{j_i\}}$ is defined by $\ket{\{j_i\},\iota}$, labelled by the set of given recoupled spins $\{j_i\}$, $j_\in \frac{\mathbb{N}}{2}$, and the intertwiner number $\iota$. Hereafter, we use $\ket{\iota}$ to ease the notation. The Hilbert space of the quantum polyhedron (with all possible values of surface areas) is achieved by a direct sum over the spins, that is 
\begin{equation}
{H}_I\equiv \bigoplus_{\{j_i\}} I_{\{j_i\}} = \bigoplus_{\{j_i\}, \iota} \mathbb{C} \ket{\{j_i\},\iota}\, .
\end{equation}
Collections of quantum polyhedra can be attached (or glued) to form a network via \emph{edge} maps realised by \emph{bivalent} intertwiners, consisting in $SU(2)$-invariant singlet states of two spins 
\begin{equation}
\ket{e} \equiv \sum_m \frac{(-1)^{j-m}}{\sqrt{2j+1}} \ket{j,m}\otimes \ket{j,-m} \in \text{Inv}_{SU(2)} \big[V^j \otimes V^j\big]
\end{equation}
where $\ket{j,m}$ defines the usual spin basis in the representation space $V^j$, with magnetic momentum $m$ running by integer step from $-j$ to $+j$. It is convenient to represent a quantum polyhedron state dually as a star graph with a single vertex of valence $d$, where each face is dual to a spin. Given the tensor product of two intertwiner spaces $I_{AB}=I_{\{j_i,j\}}^A \otimes I_{\{k_i,j\}}^B$ and a basis state $\ket{\iota_A}\otimes \ket{\iota_B} \in I_{AB}$, the mapping
\begin{multline}
e: \ket{\iota_A} \otimes \ket{\iota_B} \in I_{AB} \mapsto  \sum_m \frac{(-1)^{j-m}}{\sqrt{2j+1}} \braket{j,m}{\iota_A}\otimes\\ \otimes \braket{j,-m}{\iota_B} \in \bigotimes_{\{i,\ell\}}V^{j_i} \otimes V^{k_\ell}
\end{multline}
realises the gluing of two quantum polyhedra along a face by projecting two spins from A and B into a singlet edge state. The gluing results in the creation of a simple \emph{open} graph given by two vertices connected by an edge. 
More generally, given $V$ vertices and $E$ edges, with $E\subset V$, a spin network state is a closed graph $\gamma=(V,E)$, with edges $e \in E$ labelled by irreducibile representations (irreps) of $SU(2)$, and vertices $v \in V$ labelled by $SU(2)$ intertwiner states $\iota_v$, i.e. $SU(2)$-invariant tensors recoupling the representations carried by the edges attached to the vertex (\cref{fig:sn}). 

In particular, a spin network state $\ket{\psi_\gamma}$ associated to a closed graph $\gamma$ can be obtained as a projection of a generic state $ \ket{\psi}$ in the tensor product space of $V$ intertwiners, 
\begin{equation}
I_V^{\{J\}}\equiv \bigotimes_{v=1}^V I^v_{\{j_i\}_v},
\end{equation} 
with $\{J\} \equiv (\{j_i\}_1, \dots, \{k_i\}_V)$, on a set of edge states according to the given connectivity of the graph $\gamma$. We write
\begin{equation}
\ket{\psi_\gamma}= \left(\bigotimes_{e \in E} \bra{e}\right)\, \ket{\psi}
\label{eq:proj}
\end{equation}
This notation will be particularly useful in the next sections.

Notice that one can further relate couples of spins shared by two vertices by a $SU(2)$ group element $g\in SU(2)$, by dressing the edge state as follows
\begin{equation}
\ket{e[g_e]} \equiv \sum_{a,b} {(-1)^{j_e-b}}\, D^{j_e}_{a,b}(g_e)\, \ket{j_e,a}\otimes \ket{j_e,b} 
\label{eq:edgehol}
\end{equation}
where $D^j(g)$ is a Wigner matrix representing the group element $g\in SU(2)$. As  $D^j(g)$ is a unitary matrix, this corresponds to a local unitary transformation on one of the two spins~\cite{PhysRevD.97.026009}. The resulting spin network becomes a function of the edges group elements,
\begin{multline}
\ket{\psi_\gamma[J,\{g_e\}]} = \sum_{\{\iota_v\}, a,b} \hat{\phi}_{\{J\}}^{\{\iota_v\}}\, \prod_{e} D^{j_e}_{a,b}(g_e)\, \ket{j_e,a}\otimes \\ \otimes \ket{j_e,b} \, \prod_{v}\ket{\iota_v}
\label{eq:glued}
\end{multline}

A further sum over the spin sectors leads to the full spin network Hilbert space
\begin{equation}
H_\gamma \equiv \bigoplus_{\{J\}} I_V^{\{J\}} = \bigoplus_{\{j_e, \iota_v\}} \mathbb{C}\,\ket{\{j_e, \iota_v\}_{\gamma}}\, .
\label{eq:fullintsp}
\end{equation} 
Remarkably, one can show that 
$$H_\gamma \cong L^2\left(SU(2)^{\times E}/SU(2)^{\times V}\right)\, ,$$
wherein 
\begin{equation}
\sum_{ a,b} \prod_{e} D^{j_e}_{a,b}(g_e)\, \ket{j_e,a}\otimes \ket{j_e,b} \, \prod_{v}\ket{\iota_v}
\end{equation}
defines a spin network basis state.

\begin{figure}
	\includegraphics[width=0.45\textwidth]{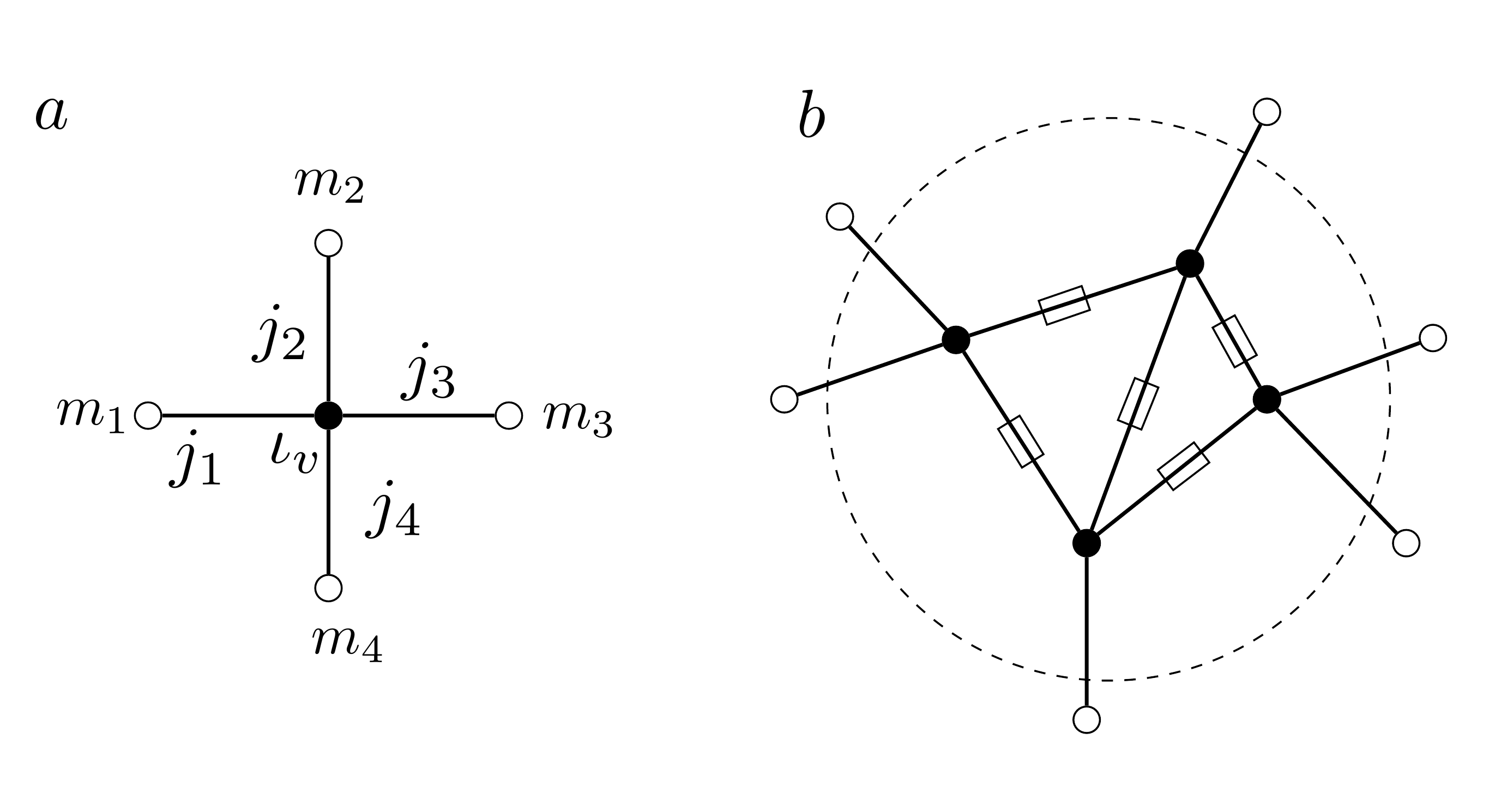}
	\caption{Construction of an open spin network by vertices gluing. The white dots represent the boundary degrees of freedom.}
	\label{fig:osn}
\end{figure} 

In loop quantum gravity (LQG), quantum states of geometry in $H_\gamma $  are given by wave-functions $\Psi_\gamma\equiv \sum_J\ket{\psi_\gamma[J,\{g_e\}]}$, by construction invariant under the $SU(2)$-action at each vertex $v$ of the graph,
\begin{eqnarray}
\Psi_\gamma: SU(2)^{\times E}  &\to& \mathbb{C} \\ \nonumber
\{g_e\}_{e \in \gamma} &\mapsto& \Psi(\{g_e\}_{e \in \gamma}) = \Psi(\{h_{t(e)}\,g_e\, h_{s(e)^{-1}}\}_{e \in \gamma}) \,, \
\end{eqnarray}
$\forall\; h_v\in SU(2) $, with $t(e)$ and $s(e)$ respectively referring to the target and source vertices of the edge $e$.
With the appropriate scalar product \cite{thiemann_2007, rovelli_vidotto_2014}, $H_\gamma$ gives the kinematical Hilbert space of quantum geometry in LQG, where the $SU(2)$ group elements $\{g_e\}$ represent $SU(2)$ holonomies of the Ashtekar-Barbero connection field along directed edges linking the two nodes. In this context, spin network (directed, embedded) graphs represent networks of frame transformations, which partially encode the operational content of the gravitational field reduced on $3d$ spacetime slices~\cite{thiemann_2007, Thiemann2003, rovelli_vidotto_2014}.

\subsection{Open Graph States}

Differently from LQG, in the next sections we will consider quantum states of geometry   with support on \emph{open} (undirected) graphs, namely graphs with sets of uncontracted edges, modelling quantum regions of space with \emph{boundaries}. In this sense, a generalised description of the spin network state requires enlarging the notion of Hilbert space given in \eqref{eq:fullintsp}.
Let us proceed as follows. Looking at the form of the spin network state in \eqref{eq:glued}, we see that by breaking each entangled pair of spins forming an edge we are left with a collection of \emph{single vertex open spin network states} (see \cref{fig:osn}), 
\begin{equation}
\ket{f_v}= \sum_{\iota_v,\{m_i\} }{f_{v}^{\iota_v, {\{j_i\}}_v}}_{\{m_i\}}\, \ket{\{j_i\},\iota}\otimes \bigotimes_i^d\ket{j_i,m_i}\, 
\label{vstate}
\end{equation}
defined in the single vertex product space  
\begin{equation}
B^v_{\{j_i\}_v}\equiv \text{Inv}_{SU(2)}\Big[\bigotimes_{i=1}^d V^{j_i}\Big]\otimes  \bigotimes_{i=1}^d V^{j_i}\, .
\label{key}
\end{equation}
One can think of such states as quantum polyhedron states dressed with extra boundary \emph{edge modes}, one per face. 

In this case, the sum over the spins, carried on at each vertex independently, leads to the full open vertex Hilbert space
\begin{equation}
H_v \equiv \bigoplus_{\{j_i\}}  B^v_{\{j_i\}} \cong L^2(SU(2)^d/SU(2))\, ,
\end{equation}
where wave functions $\phi_v(g_1, \dots, g_d)$ can be intended as \emph{group fields}~\cite{Oriti2018c, Oriti:2014uga}. 
A collection of (distinguishable) $V$ open vertices is associated to the separable Hilbert space 
\begin{equation}
{H}_V \equiv \bigotimes_{v=1}^V H_v\, .
\end{equation}
In particular, we have that the embedding $H_\gamma \subset H_V$ is  faithful, hence any closed spin network wavefunction in $H_\gamma$ is expressible in terms of functions in $H_V $, modulo gluing conditions realised, again, via the action of bivalent intertwiners on each fixed spin sector~\cite{Oriti:2013aqa}. 

Generally, in the case of open graphs the result of the gluing procedure in \eqref{eq:proj} is a mapping of the bulk degrees of freedom of $\ket{\Psi_\gamma}$ on the boundary space of unrecoupled spins. We have
\begin{eqnarray}
\Psi_\gamma: SU(2)^{\times E}  &\to& \mathcal{F}\left(SU(2)^{\{e \in \partial \gamma\}}\right) \\ \nonumber
\{g_e\}_{e \in \gamma} &\mapsto& \Psi(\{g_e\}_{e \in \gamma}) = \Psi(\{g_e\, h_{s(e)^{-1}}\}_{e \in \gamma}) \,, \
\end{eqnarray}
with $\ket{\Psi_\gamma}$ now defined in the boundary Hilbert space $\mathcal{H}_{\partial \gamma}=\bigotimes_{e \in \partial \gamma}\bigoplus_{j_e \in \frac{\mathbb{N}}{2}} V^{j_e}$.
Notice that in this case the gauge invariant property of the wavefunction reduces to $SU(2)$-covariance on the boundary.

The generalisation allows to describe the state of any generic subregion of a closed spin network, by cutting out the spin network along the subregion boundary edges. In the following sections, we will limit our analysis to spaces of fixed spins. Nevertheless, the open graph generalisation will be central in our analysis, as we will look at the entanglement structure of a generalised open spin network state through the correlations of its boundary edges modes.

\subsection{Analogy with symmetric Projected Entangled Pairs Tensor Networks} 

Consider an open graph $\gamma=(V,E, \partial \gamma)$, where we separate the set of edges $E$ into a set of internal edges $E$ and boundary edges $\partial \gamma$ for future convenience. In the following, we focus on a class of spin network states defined via \eqref{eq:proj} from product states of individual vertex states with fixed spins $\ket{\psi}=\bigotimes_v \ket{f_v}$, glued according to the connectivity pattern of $\gamma$. The resulting spin network states read 
\begin{equation}
\ket{\phi_\gamma} =\left(\bigotimes_{e \in E} \bra{e}\right)\bigotimes_v \ket{f_v}
 \in \bigotimes_{e \in \partial \gamma} V^{j_e}
 \label{eq:tn}
\end{equation}
(notice we are always setting the edge holonomies introduced in~\eqref{eq:edgehol} to the identity).
As first advanced in \cite{Colafranceschi:2021acz, Perez_2006}, states like $\ket{\phi_\gamma}$ in~\eqref{eq:tn} are analogue to peculiar symmetric tensor networks, where single vertex states are identified with tensor states: the $d$ vertex edge spin numbers $\{m_i\}$ correspond then to tensor indices (virtual indices), while the intertwiner number $\iota$ plays the role of the physical index of the tensor. Now, the  definition of edge states in $\ket{\phi_\gamma}$ as singlet states given by maximally entangled pairs of spins, characterises $\ket{\phi_\gamma}$ as \emph{symmetric} projected entangled pair states ({\bfseries sPEPS})~\cite{ORUS2014117}. 

The analogy suggests to think of spin networks as \emph{entanglement networks}, providing a direct relation between the connectivity (topology) of the quantum geometry state and its local entanglement structure.

\begin{figure}
	\includegraphics[width=0.45\textwidth]{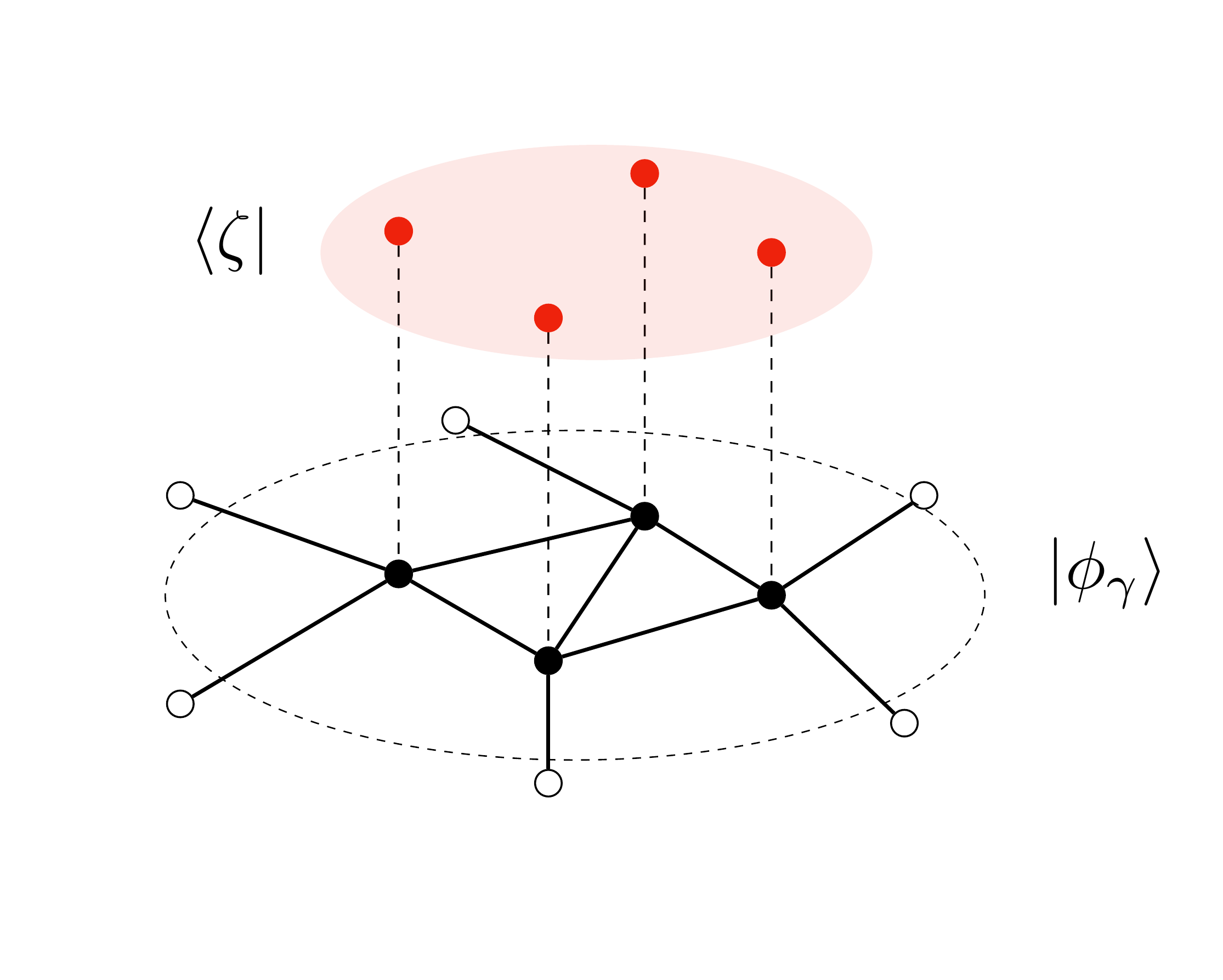}
	\caption{Bulk to boundary mapping of an open spin network state $\ket{\phi_\gamma}$ via a projection on a bulk state $\ket{\zeta}$.}
	\label{fig:bb}
\end{figure} 
 
States with richer, non-local, structure of correlations can be built from the tensor network analogue in \eqref{eq:tn} as follows. Given a generic state $\ket{\zeta}$ in the tensor product space of $V$ intertwiners $I_V^{\{J\}}$, that is  
\begin{equation}\label{input}
\ket{\zeta}=\sum_{\iota_1,...,\iota_V} \zeta_{\iota_1,...,\iota_V}\bigotimes_v \ket{\{j_i\}_v~\iota_v},
\end{equation}
we can construct an open spin network state  via the projection 
\begin{equation}\begin{split}\label{output}
\ket{\phi_{\partial \gamma}(\zeta)} &\equiv  \bra{\zeta}\phi_{\gamma}\rangle\\&=\left(\bra{\zeta} \otimes \bigotimes_{e \in E} \bra{e}\right)\bigotimes_v \ket{f_v}\\&=\sum_{\{m_e \in E_{\partial\gamma}\}}  \left(\phi_{\partial \gamma}(\zeta)\right)_{\{m_{e}\}} \bigotimes_{e\in \partial \gamma}\ket{j_e m_e}\, ,
\end{split}
\end{equation}
with coefficients
\begin{multline}
\left(\phi_{\partial \gamma}(\zeta)\right)_{\{m_{e\in\partial\gamma}\}} =\sum_{\{\iota\}}\sum_{\{n_{e\in E}\}} \sum_{\{p\}} \zeta^*_{\iota_1,...,\iota_V} \\ \prod_v(f_v)_{\{n\}_v~\iota_v}^{\{j\}_v} \prod_{e\in E} \delta_{n_v^i p^{vw}_i}\delta_{n_w^i p^{vw}_i}.
\end{multline}
In particular, one can think of  the boundary state in \eqref{output} as the \emph{output} of a map defined between the intertwiner space $I_V^{\{J\}}$ and the boundary (fixed) spins Hilbert space, i.e.
\begin{equation}\begin{split}
M[\phi_\gamma]~:&~I_V^{\{J\}} \equiv  \bigotimes_v I^{\{j\}_v}~\rightarrow~H_{\partial\gamma}^{\{J_\partial\}}\equiv \bigotimes_{e\in \partial \gamma}V^{j_e}\, ,
\end{split} 
\end{equation}
such that
\begin{equation}\begin{split}
M[\phi_\gamma]\ket{\zeta}=\langle \zeta |\phi_\gamma\rangle=\ket{\phi_{\partial \gamma}(\zeta)}.
\end{split}
\end{equation}
By feeding the bulk with an input state we set correlations among the intertwiners, which are eventually encoded in the boundary state \cite{Chirco:2021chk}. 

States constructed as $\ket{\phi_{\partial \gamma}(\zeta)}$ show that the entanglement  among spin magnetic indices is responsible for the connectivity of the graph and plays a role in the definition of the mapping. Differently, entanglement among intertwiners describes correlations between the geometry (volumes, see \cite{rovelli_vidotto_2014, PhysRevD.97.026009}) of the $3d$ space quantum geometries defined at each vertex.

In order to enhance the separation of roles between the two levels of entanglement, in the next sections, we consider spin network graph states $\ket{\phi_\gamma}$ comprised by random vertex states. On the one hand, building on the tensor network analogy, we know that such random tensor network states can be used to realise isometric bulk--to--boundary mappings which display holographic entanglement behaviour. On the other, we know that such states are characterised by a flat entanglement spectrum. Therefore, we can use random $\ket{\phi_\gamma}$ as convenient holographic entanglement background mappings on top of which non-trivial correlations will be induced by the choice of the input bulk state $\ket{\zeta}$.

\subsection{\label{sec:sub3random} Random Spin Networks as Entanglement Background}
Following~\cite{Hayden:2016cfa}, we assume that each single vertex state $\ket{f_v}$ in the boundary state  \eqref{output} is chosen \emph{independently} at \emph{random} from its single vertex space, with respect to the uniform probability measure. This is equivalent to take at each vertex $v$ of the graph $\gamma$ an arbitrary reference state $\ket{0} \in B_{\{j_i\}_v}^v $  and define $\ket{f_v} = U \ket{0}$ with $U$ a unitary operator. Accordingly, the random average of any function $f(\ket{f_v})$ is equivalent to an integration over the unitary $U$ with respect to the Haar measure $\mu=\mathrm{d}U$, with normalization $\int \mathrm{d}U=1$.

The \emph{random} density operator, 
\begin{multline}\label{ran}
\rho_{\partial \gamma}(\zeta)= \ketbra{\phi_{\gamma}(\zeta)}{\phi_{\gamma}(\zeta)} \\=\left(\ketbra{\zeta}{\zeta} \otimes \prod_{e \in E} \ketbra{e}{e} \right)\left(\prod_v \ketbra{f_v}{f_v}\right)\, ,
\end{multline}
can be read as a partial trace carried over the bulk intertwiner numbers and internal edges magnetic numbers, that is a sum over all but the boundary spins (dangling legs of the graph). In compact form, we write the boundary density matrix
\begin{equation}\label{tranb}
\rho_{\partial \gamma}(\zeta)= \text{Tr}_b \left( \rho_b(\zeta)\, \prod_v \ketbra{f_v}{f_v}\right)\,,
\end{equation}
with a bulk graph density matrix $\rho_B$ defined by
\begin{equation}
 \rho_b(\zeta)\equiv \left (\ketbra{\zeta}{\zeta}\otimes \prod_{e \in E} \ketbra{e}{e}\right)= \rho_\zeta \otimes \prod_{e \in E} \rho_{e}\,.
\end{equation}
The state $\rho_{\partial \gamma}$ is a linear function of the independent pure states of each intertwiner $\ketbra{f_v}{f_v}$. It defines an \emph{open} random spin network state describing a $3d$ quantum geometry with a boundary. 
 
We are interested in characterising the entanglement of generic subregions of the boundary. To this aim, we consider a multi-partition of the boundary in three subregions $\{A, B,C\}$, and focus on a reduced state obtained by tracing out $C$ as a generic \emph{environment}. Thereby, we look at the quantum correlation of the mixed state for $\{A,B\}$ via a measure of log-negativity.

\subsection{\label{sec:sub3} Entanglement Negativity}
Quantum  negativity is a good measure of mutual entanglement between two subsystems in a mixed state~\cite{Peres_1996, PhysRevA.58.883,PhysRevA.60.3496, Eisert_99, PhysRevA.65.032314, Plenio_05, PhysRevB.94.035152}. As such, it allows to generalise von Neumann entanglement entropy measures to the case of multipartite systems.

Let us then recall the definition of quantum negativity as follows. Consider a bipartite system, with total Hilbert space $\mathcal{H}=\mathcal{H}_A\otimes\mathcal{H}_B$, described by the generic density operator $\rho_{AB}$. Given an orthonormal basis $\ket{i}_A$ in $\mathcal{H}_A$ and similarly $\ket{j}_B$ in $\mathcal{H}_B$, the {\bfseries partial transpose} of $\rho_{AB}$ with respect to one of the subsystems, say B, is defined as 
\begin{equation}
(\rho_{i_Aj_B,k_Al_B})^{T_B}=\rho_{i_Al_B,k_Aj_B}\,.
\end{equation}
The eigenvalues of the partially transposed reduced density matrix $\rho_{AB}^{T_B}$ are real since the partial transposition is an Hermitian and trace-preserving map. Yet $\rho_{AB}^{T_B}$ is not completely positive, \emph{i.e.} it may have negative eigenvalues. If $\rho_{AB}$ is not entangled (separable), it is easy to see that $\rho_{AB}^{T_B}$ remains a positive semidefinite operator. Thereby, the presence of negative eigenvalues in the partial transpose is an indicator of quantum correlations in $\rho_{AB}$. The measure is designed to distinguish quantum correlated mixed
states from classically correlated ones~\cite{Shapourian_21}. 

In particular, one can quantify the amount of entanglement of a state according to the number of negative eigenvalues of the partial transpose in terms of the measures of \emph{negativity} and \emph{logarithmic negativity}  \cite{Plenio_05}, respectively defined as
\begin{equation}
    N(\rho_{AB})\equiv\frac{\lVert\rho_{AB}^{T_B}\rVert_1 -1}{2}=\sum_{i:\lambda_i<0}\abs{\lambda_i} \, ,\label{Neg}
\end{equation}
where $\lVert \cdot\rVert_1$ is the trace norm, and
\begin{equation}
    {E}_N(\rho_{AB})\equiv \log \lVert\rho_{AB}^{T_B}\rVert_1 \, .
\end{equation}
Both quantities are entanglement monotone under general positive partial transpose (PPT) preserving operations \cite{Peres_1996}. In this sense, we are considering the negativity as a faithful measure of quantum entanglement for our states. Nevertheless, we shall remark here that the PPT condition is not generally a necessary and sufficient separability criterion. There are generally entangled PPT states which belong to the class of \emph{bound entangled states} that have a non-negative partial transpose~\cite{HORODECKI19961, PhysRevLett.97.080501}.

In the next sections, we shall focus on logarithmic negativity (or log negativity) in particular.

Analogously to the R\'enyi generalization of von Neumann entropy, the latter can be computed in the limit $k \to 1$ of a $k$-the R\'enyi negativity measure~\cite{Dong_21}, defined by
\begin{equation}
N_k(\rho_{AB})=\Tr\qty[\qty(\rho_{AB}^{T_B})^k]\, .
\end{equation}

Notice that, since $\rho_{AB}^{T_B}$ has negative eigenvalues, even and odd moments should be treated separately
\begin{equation}
    \begin{split}
        N_k^{(\text{odd})}(\rho_{AB})&=\sum_i \text{sgn}(\lambda_i)\abs{\lambda_i}^k\\
        N_k^{(\text{even})}(\rho_{AB})&=\sum_i\abs{\lambda_i}^k
    \end{split}
\end{equation}
The logarithmic negativity only depends on the absolute values of eigenvalues, hence it can be recovered in the $k\to 1$ limit of the logarithm of the analytic continuation of the momenta for even $k$. If we set $k=2n$, we have then
\begin{equation}
    {E}_N(\rho_{AB})=\lim_{n\to\frac{1}{2}}\log N_{2n}(\rho_{AB})
\end{equation}
In the next sections, we focus on the even $k$-th R\'enyi negativity.
We compute the averaged logarithmic negativity of the random mixed boundary state $\rho_{AB}$, that is
\begin{equation}
 \overline{{E}_N(\rho_{AB})} \equiv \mathbb{E}_{\mu} \big[{E}_N(\rho_{AB}) \big] \, ,
\end{equation}
where we use the bar hereafter to denote the random average with respect to the Haar probability measure $\mu$.
In the following sections, we assume that quantum typicality is reached for our spin network system, such that the approximation
\begin{equation}
 \overline{{E}_N(\rho_{AB})} \simeq \lim_{n\to\frac{1}{2}}\log \overline{N_{2n}(\rho_{AB})}  \, ,
\end{equation}
holds in the large spin limit. Therefore, getting $\overline{E_N(\rho_{AB})}$ amounts to compute the expected value of the even momenta of the partial transposed matrix $\rho_{AB}^{T_B}$. As shown in~\cite{Dong_21}, such computation is mapped to the evaluation of partition functions of classical generalized Ising models, with a $\text{Sym}_n$ permutation group element at each vertex and cyclic, anti-cyclic and identity permutations as boundary pinning fields.

\section{Tripartite boundary spin network state}\label{setting}
Let us define the tripartition of the boundary into three regions $A$, $B$ and $C$ by dividing  the set of boundary edges $\partial \gamma$ into three subsets $E_A$, $E_B$ and $E_C$. Accordingly, within the fixed spins setting, the boundary Hilbert space factorises as follows
\begin{equation}
H_{\partial\gamma}^{\{J_\partial\}}= \bigotimes_{e\in E_A}V^{j_e} \otimes \bigotimes_{e\in E_B}V^{j_e}\otimes \bigotimes_{e\in E_C}V^{j_e}\,.
\end{equation}
Starting from the boundary density matrix in~\eqref{ran},
via a trace over the spin numbers in $C$ we get the reduced (mixed) boundary state
\begin{equation}
\rho_{AB}(\zeta)\equiv \text{Tr}_C [\rho_{\partial \gamma}(\zeta)]\,.
\label{eq:roab}
\end{equation}
We want to measure the quantum correlation between the subregions $A$ and $B$ via a measure of $k$-th R\'enyi negativity of the reduced state, that is 
\begin{equation}
N_k(\rho_{AB}(\zeta))=\Tr\qty[\qty(\rho_{AB}(\zeta)^{T_B})^k\Big/\qty(\text{Tr}\qty[\rho_{AB}(\zeta)])^k]\, , \label{negz}
\end{equation}
where the denominator in \eqref{negz} ensures the normalization of $\rho_{AB}(\zeta)$. Note that the traces above are over boundary indices.

For $\rho_{AB}(\zeta)$ is now a random density matrix, the negativity measure must be computed in expected value with respect to the uniform Haar measure $\mu$. In the large spin regime, as the trace concentrates, we can approximate the negativity by its \emph{typical value}, which we eventually assume being expressed as a ratio of expected values of the $k$-th moment and the $k$-th power of the partition function of $\rho_{AB}^{T_B}(\zeta)$. We have
\begin{figure}
    \includegraphics[width=0.3\textwidth]{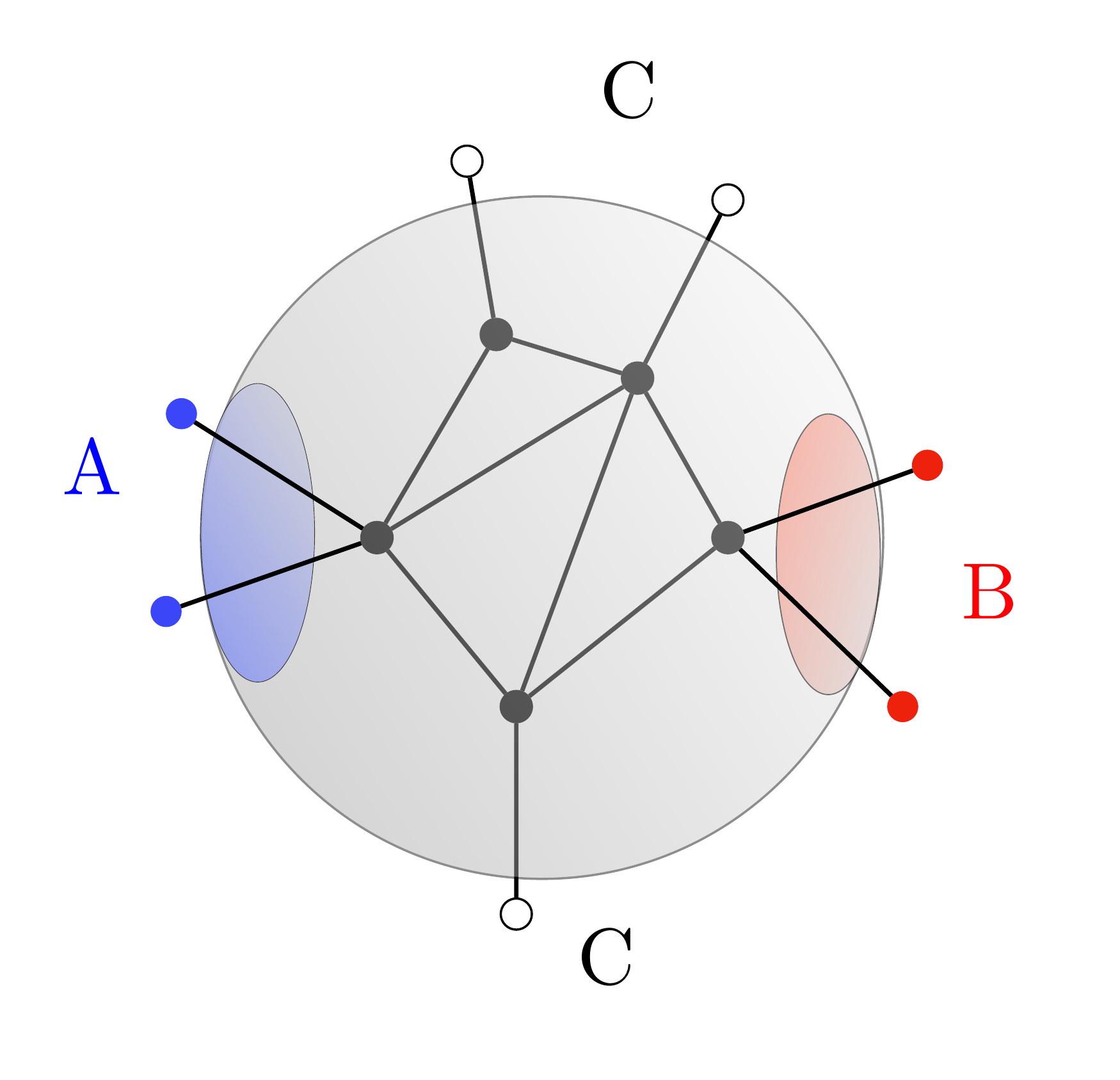}
	\caption{$\{A,B,C\}$ tripartition of an open spin network state. The boundary edges are dual to the surface of the quantum $3d$ region. The boundary spin are attached to virtual vertices, whose color indicates the different boundary set. Hereafter, we indicate as $C$ (grey) the environment over which we compute the partial trace; $B$ (red) the region over which we take the partial transpose; $A$ (blue) the remaining part of the boundary.}
	\label{fig:tripar}
\end{figure}
\begin{equation}
N_k(\rho_{AB}(\zeta)) \simeq \frac{\overline{\Tr\left[\qty(\rho_{AB}^{T_B}(\zeta))^k\right]}}{\overline{\left(\Tr\left[\qty(\rho_{AB}(\zeta))\right]\right)^k}} \equiv \frac{\overline{Z^{(k)}_1}}{\overline{Z^{(k)}_0}}\,.
\label{454}
\end{equation} 
Before dealing with the expectation value, via \emph{replica} trick we can linearize the partial transpose matrix as follows:
\begin{multline}
\text{Tr}\left[ \left(\rho_{AB}^{T_B}(\zeta)\right)^k\right]= \text{Tr}\left[\rho_{AB}(\zeta)^{\otimes k} \, P_A\qty(X)\otimes P_B\qty(X^{-1})\right]\\
\qquad = \text{Tr}\left[\rho_{\partial \gamma}(\zeta)^{\otimes k}\, P_A\qty(X) \otimes P_B\qty(X^{-1})\otimes P_C\qty(\mathbbm{1})\right]\, ,
\end{multline}
where $P_{I}(\sigma)$ denotes a unitary representation of the permutation $\sigma$, {$I=A,B,C$} and $X$, $X^{-1}$ and $\mathbbm{1}$ are the cyclic, anti-cyclic and identity permutations. Specifically, $P_C\qty(\mathbbm{1})$ and $P_B\qty(X^{-1})$ respectively implement the partial trace and the partial transpose, while $P_A\qty(X)$ realizes the replica trick, namely it allows us to linearize under the trace. We give in \cref{permutation appendix} a thorough discussion of the permutation group and all the notions that will be relevant in the following.

For the linearity of the trace, the average over the random tensors can be carried out before taking the partial trace. Then, by \eqref{tranb}, we get
\begin{equation}\begin{split}
\overline{Z_1^{(k)}}&=\Tr\Bigg[\rho_\zeta^{\otimes k}\otimes\rho_E^{\otimes k}\, \qty(\bigotimes_v\overline{\qty(\ket{f_v}\bra{f_v})^{\otimes k}}) \cdot \\
\quad & \cdot P_A\qty(X)\otimes P_B\qty(X^{-1})\otimes P_C\qty(\mathbbm{1})\Bigg]\,, \\
\overline{Z^{(k)}_0}&=\Tr\qty[\rho_\zeta^{\otimes k}\otimes\rho_E^{\otimes k}\, \qty(\bigotimes_v\overline{\qty(\ket{f_v}\bra{f_v})^{\otimes k}})]\, ,
\end{split}\label{parti}
\end{equation}
where the trace here runs over both bulk and boundary indices and $\rho_E = \bigotimes_{e \in E} \ket{e}\bra{e}$.

The average of the $k$ copies of the vertex state in \eqref{parti} by integration over the Haar measure, results (Schur’s lemma, see~\cite{38}) into the sum over unitary representations of the permutation group $g_v$ acting on the $k$ copies of the single vertex Hilbert space $B^v_{\{j_i\}_v}$,
\begin{equation}
\overline{\qty(\ket{f_v}\bra{f_v})^{\otimes k}}=\frac{\qty(D_v-1)!}{\qty(D_v+k-1)!}\sum_{g_v\in S_k}P_v(g_v)\, ,
\end{equation}
with dimension $\text{dim}(B^v_{\{j_i\}_v}) \equiv D_v = \prod_{i} d_{j^v_i} D_{\vec{j}_v}$.  

By performing the average individually on each independent random vertex state, eventually we obtain
\begin{multline}\label{z1 with permutations}
\overline{Z^{(k)}_1}=\mathcal{C}\Tr\Bigg[\rho_\zeta^{\otimes k}\otimes\rho_E^{\otimes k}\, \qty(\bigotimes_v \sum_{g_v\in S_k}P_v(g_v)) \cdot \\ \quad \cdot  P_A\qty(X) \otimes P_B\qty(X^{-1})\otimes P_C\qty(\mathbbm{1})\Bigg]\, , \end{multline}
where the trace factorizes over the Hilbert spaces of a) internal edges, b) boundary spins and c) bulk intertwiners, while $\mathcal{C} = \prod_v\qty[\frac{\qty(D_v-1)!}{\qty(D_v+k-1)!}]$. $\overline{Z^{(k)}_0}$ has the same form with $X$ and $X^{-1}$ replaced by $\mathbbm{1}$.

Now, to compute the action of the permutation operators on the different degrees of freedom at each vertex, we factorize $P_v(g_v)$ into three different sets of operators,
\begin{equation}
    P_v(g_v)= P_{v,0}(g_v)\otimes\bigotimes_{e^i_{vw} \in E} P_{v,i}(g_v) \otimes \bigotimes_{e^i_{v\bar{v}} \in \partial \gamma} P_{v,i}(g_v)\, ,\label{permvert}
\end{equation}
where $P_{v,0}(g_v)$ acts on $k$ copies of the intertwiner space, $ \bigotimes_{e^i_{vw} \in E}P_{v,i}(g_v)$ acts on $k$ copies of the internal links and $\bigotimes_{e^i_{v\bar{v}} \in \partial \gamma} P_{v,i}(g_v)$ acts on the boundary semi-edges, with $\bar{v}$ representing a virtual vertex which is connected to $v$ by the boundary edge $e^i_{v\bar{v}}$. We have indicated with $e^i_{vw} \in E$ an internal link which connects the vertex $v$ to the vertex $w$ along the $i^{\text{th}}$ edge.

As a consequence of this factorization, the whole computation of the trace can be decomposed in three contributions, respectively for the a) internal edges, b) boundary spins and c) bulk intertwiners. The three contributions are explicitly computed in \cref{mapping}.

The result is that $\overline{Z^{(k)}_{1/0}}$ can be written as

\begin{equation}
     \overline{Z^{(k)}_{1/0}} = \sum_{\{g_v\}} e^{-A^{(k)}_{1/0}\big[\{g_v\}\big]}\, , \label{Partition}
\end{equation}
where
\begin{multline}\label{A1 bulk generico}
    A^{(k)}_1\big[\{g_v\}\big] = \sum_{e^i_{vw} \in E}\Delta(g_v,g_w)\log d_{j^i_{vw}} + \\
    +\sum_{e^i_{v\bar{v}} \in A} \Delta(g_v, X)\log d_{j^i_v} +  \sum_{e^i_{v\bar{v}} \in B} \Delta(g_v, X^{-1})\log d_{j^i_v} +\\+ \sum_{e^i_{v\bar{v}} \in C} \Delta(g_v, \mathbbm{1})\log d_{j^i_v} + A(\zeta) + \xi 
   \end{multline}
   and 
   \begin{multline}\label{A0 bulk generico}
   A^{(k)}_0\big[\{g_v\}\big] = \sum_{e^i_{vw} \in E}\Delta(g_v,g_w)\log d_{j^i_{vw}} + \\   + \sum_{e^i_{v\bar{v}} \in \partial \gamma} \Delta(g_v, \mathbbm{1})\log d_{j^i_v} +  A(\zeta) + \xi \, ,
\end{multline}
$\xi$ being a constant term and

\begin{equation}
    A(\zeta) = -\log \qty[\Tr_{\dot{\gamma}} \Bigg\{\ketbra{\zeta}{\zeta}^{\otimes k}\qty(\bigotimes_v P_{v,0}(g_v))\Bigg\}]
\end{equation}
is the bulk state contribution. This is obtained from \eqref{bulk contribution appendix} by setting $\Omega \equiv \dot{\gamma}$. In \eqref{A1 bulk generico} and \eqref{A0 bulk generico}, $\Delta\qty(g,h)$ indicates the Cayley distance on the permutation group between the permutations $g$ and $h$, see \cref{permutation appendix}.

Remarkably, for the random character of the network, the computation of the typical $k$-th R\'enyi negativity is mapped to the evaluation of the partition functions of a \emph{generalized Ising Model} defined by the action $A^{(k)}_1$. The latter describes a two-body interaction between permutation elements, which therefore act as \emph{generalized spins}, attached to the spin network vertices. These interactions are described by the Cayley distance on the permutation group and the pinning fields $X$, $X^{-1}$ and $\mathbbm{1}$ are permutations attached to virtual vertices playing the role of boundary conditions. The strength of the interactions is given by $\log d$, $d$ being the dimension of the link, semi-link or intertwiner space according to the term we consider. This action prefers neighbouring ``spins" to be parallel. This means that in the large dimension limit, namely the strong coupling or ``low temperature" regime, the dominant configurations that minimize the action contain large domains separated by domain walls, which in turn give the energy cost of the configuration.

Differently from \cite{Dong_21}, the actions  \eqref{A1 bulk generico} and \eqref{A0 bulk generico} contain new terms due to the internal degrees of freedom, \emph{i.e.} the intertwiners, that characterize the spin network structure of the graph. Inserting a bulk state we end up with an additional contribution to the actions deriving from the bulk correlations. In turn, these give a relevant contribution to the analysis of the minimal surfaces.

\section{k-th order R\'enyi Negativity}
\label{compute}
We shall now explicitly calculate the $k$-th order Rényi negativity for boundary spin network states defined by the mapping
\begin{equation}\begin{split}
M[\phi_\gamma]\ket{\zeta}=\langle \zeta |\phi_\gamma\rangle=\ket{\phi_{\partial \gamma}(\zeta)}\, ,
\end{split}
\end{equation}
for a specially simple class of $\phi_\gamma$ and $\ket{\zeta}$.

Let us first define the bulk state. We divide the set of the bulk vertices in a region $\Omega\subseteq\dot{\gamma}$ and its complement $\bar{\Omega}$. We shall consider a state $\ket{\zeta_\Omega}$ where intertwiners are entangled, while considering a direct product state for the intertwiners in the region $\bar{\Omega}$. Accordingly, we define the bulk state as the product state over the two regions
\begin{equation}
     \ket{\zeta} = \ket{\zeta_\Omega} \otimes \qty(\bigotimes_{v\in \bar{\Omega}} \ket{\zeta_v}) \; \, \in \, \mathcal{H}_{\dot{\gamma}}\,.
\end{equation}

It is easy to see that non-correlated intertwiners give no contribution to the Ising partition function, as shown in \cref{mapping}. The partition functions read
\begin{widetext}
\begin{eqnarray}
   \overline{Z^{(k)}_1} &=& \Tr\Bigg\{\rho_{\zeta_\Omega}^{\otimes k}\otimes\rho_E^{\otimes k}\, \qty(\bigotimes_{v\in \Omega}\overline{\qty(\ket{f_v}\bra{f_v})^{\otimes k}}\otimes \bigotimes_{v\in \bar{\Omega}} \overline{\qty(\ket{f_v(\zeta)}\bra{f_v(\zeta)})^{\otimes k}}) \,  P_A\qty(X)\otimes P_B\qty(X^{-1})\otimes P_C\qty(\mathbbm{1})\Bigg\}\,,\\ 
    \overline{Z^{(k)}_0} &=& \Tr\Bigg\{\rho_{\zeta_{\Omega}}^{\otimes k}\otimes\rho_E^{\otimes k}\,\qty(\bigotimes_{v\in \Omega}\overline{\qty(\ket{f_v}\bra{f_v})^{\otimes k}}\otimes\bigotimes_{v\in \bar{\Omega}}\overline{\qty(\ket{f_v(\zeta)}\bra{f_v(\zeta)})^{\otimes k}})\Bigg\}\,,
\end{eqnarray}
\end{widetext}
where $\ket{f_v(\zeta)}$ denotes the contraction of the random single vertex state with an individual intertwiner state in $\bar{\Omega}$,
\begin{equation}
\ket{f_v(\zeta)}=\bra{\zeta_v}\ket{f_v}  = \sum_{\{m_i\}} f_v(\zeta)^{\{j_v\}}_{\{m_i\}}\, \bigotimes_{i}\ket{j_i, m_i}\,,
\end{equation}
where $f_v(\zeta)^{\{j_v\}}_{\{m\}} = \sum_\iota f_v\,^{\{j_v\}}_{\{m\}\iota} \qty(\zeta^{v}_{\iota})^*$ are the coefficients of the boundary state.

In the region $\Omega$, we construct $\ket{\zeta_{\Omega}}$ in terms of products of maximally entangled pairs of intertwiner states connecting disjoint couples of vertices in the bulk. We write
\begin{equation}
    \ket{\zeta_\Omega} = \bigotimes_{\langle v,w\rangle \in \Omega}\ket{e^\iota_{vw}} \,,
    \end{equation}
    with 
    \begin{equation}
   \ket{e^\iota_{vw}} \equiv \bigotimes_{\langle v,w\rangle \in \Omega} \frac{1}{\sqrt{D_{\{j\}_{vw}}}}\sum_\iota \ket{\{j\}_{vw},\iota_v}\otimes\ket{\{j\}_{vw},\iota_w}\,.\label{483}
\end{equation}
With the choice of Bell-like pairs of intertwiner (see~\cite{PhysRevD.99.086013} for a similar description), the bulk entropy contribution reduces to  
\begin{equation}
    A(\Omega) = \sum_{\langle vw\rangle \in \Omega}\Delta(g_v,g_w)\log D_{\{j\}_{vw}}\,,\label{484}
\end{equation}
namely the equivalent of an extra edge contribution to the action, with two main differences: the bond weight, which is now given by the logarithm of the minimal intertwiner space dimension for each pair $\langle vw\rangle \in \Omega$, and the \emph{non-local} nature of the bond, which can now connect non-adjacent vertices.

Concerning the spin network state $\ket{\phi_\gamma}$, we restrict our analysis to open \emph{regular} graphs, where all vertices have the same valence and \emph{homogeneous} spin colouring, with all spins fixed to the same value $j$. Homogeneity in spins makes all intertwiner spaces maximally symmetric, with maximal dimension $D_{\{j\}_{vw}}=d$. This implies that edges and intertwiner Hilbert spaces have equal dimension, that is
\begin{equation}
    d_{j^i_{vw}}=d_{j_v^i}=D_{\{j\}_{vw}}=d\,.
\end{equation}
In the generalised Ising model description, this allows us to introduce a unique \emph{temperature parameter} $\beta=\log d$ for the whole system and write the action $A^{(k)}_1$ as
\begin{equation}
    A^{(k)}_1\big[\{g_v\}\big]=\beta\mathcal{H}_k\big[\{g_v\}\big]\, ,
\end{equation}
where $\mathcal{H}_k\big[\{g_v\}\big]$ is the Ising-like Hamiltonian. We will denote with $\mathcal{H}_k\big[\{g_v\}\big]$ and $\mathcal{H}_k^c\big[\{g_v\}\big]$ the Hamiltonians corresponding respectively to a tensor product bulk state (no correlations) and to a bulk state exhibiting link-wise intertwiner correlations in the region $\Omega$. With this notation, we have respectively
\begin{multline}
    \mathcal{H}_k=\sum_{e^i_{vw}\in E}\Delta(g_v,g_w)+\sum_{\partial A}\Delta(g_v,X)+\\
    +\sum_{\partial B}\Delta(g_v,X^{-1})+\sum_{\partial C}\Delta(g_v,\mathbbm{1})
\end{multline}
and
\begin{multline}
    \mathcal{H}_k^c=\sum_{e^i_{vw}\in E}\Delta(g_v,g_w)+\sum_{\partial A}\Delta(g_v,X)+\\ +\sum_{\partial B}\Delta(g_v,X^{-1})+\sum_{\partial C}\Delta(g_v,\mathbbm{1})+\sum_{\langle vw\rangle\in\Omega} \Delta(g_v,g_w)\,.\label{488}
\end{multline}
We see that, for large $\beta$, the leading contribution in \eqref{Partition} are the ones associated to the configuration of permutations on the graph that minimizes the Hamiltonian. 

In the following sections we compute the minimal value of the Hamiltonians $\mathcal{H}_k$ and $\mathcal{H}_k^c$ for $k=3,4$ for two examples of open spin network states defined on open tree graphs with two and three vertices.

\subsection{Open tree-graph with 2 vertices}\label{two vertices graph}
Consider an open graph $\gamma$ composed by two $4$-valent vertices, $v$ and $w$, glued by an edge, and a generic tripartition of its boundary as follows 
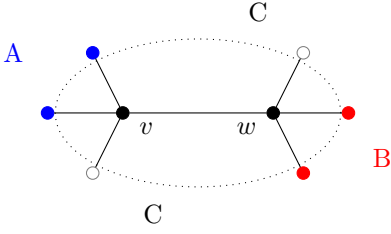
\begin{figure}[H]
    \centering
\begin{tikzpicture}
\tikzset{
a/.style={scale=.5, circle,minimum size=.1cm,fill=blue,draw=blue},
b/.style={scale=.5, circle,minimum size=.1cm,fill=red,draw=red},
c/.style={scale=.5, circle,minimum size=.1cm,fill=white!20,draw=gray,
                },
    v/.style={scale=.5, circle,minimum size=.1cm,fill=black,draw}}

\node[v] (0) at (1,0) {};
\node[v] (1) at (3,0) {};
\node[a] (2) at (0,0) {};
\node[a] (3) at (0.6,0.8) {};
\node[c] (4) at (0.6,-0.8) {};
\node[c] (5) at (3.4,0.8) {};
\node[b] (6) at (4,0) {};
\node[b] (7) at (3.4,-0.8) {};
 
\foreach \from/\to in {0/1, 2/0,3/0,4/0,5/1,6/1,7/1}
    \draw (\from) -- (\to);
 
\draw (1.1,0) node[anchor=north west]{$v$};
\draw (2.9,0) node[anchor=north east]{$w$};

\draw (2.8,1.1) node[above]{C} ; 
\draw (1.4,-1.1) node[below]{C} ; 
\draw[blue] (-0.2,0.8)node[anchor=east]{A} ;
\draw[red] (4.2,-0.6) node[anchor=west]{B} ; 
 \draw[dotted] (2,0) ellipse (54 pt and 28 pt); 
\end{tikzpicture}
\caption{Tripartion of a tree graph with two vertices.}
\label{treegraph2ver}
\end{figure}

The vertex $v$ has two boundary legs in A and one in C and similarly $w$ has two boundary legs in B and one in C; hence, the corresponding $k$-th Hamiltonian can be written as
\begin{multline}
\mathcal{H}_k=\Delta(g_v,g_w)+2\Delta(g_v,X)+\\ +2\Delta(g_w,X^{-1})+\Delta(g_v,\mathbbm{1})+\Delta(g_w,\mathbbm{1})\label{489}
\end{multline}
If we consider a bulk with entangled vertices in this simple case, the contribution of $A(\Omega)$ to the Hamiltonian consists in an additional $\Delta(g_v,g_w)$ local internal edge contribution. The $k$-th Hamiltonian for the correlated case reads
\begin{multline}
\mathcal{H}_k^c=2\Delta(g_v,g_w)+2\Delta(g_v,X)+2\Delta(g_w,X^{-1})+\\ +\Delta(g_v,\mathbbm{1})+\Delta(g_w,\mathbbm{1})\label{490}
\end{multline}

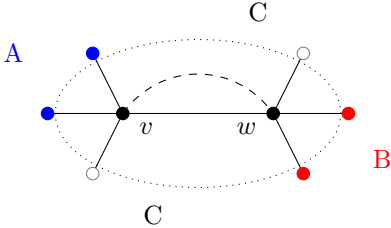
\begin{figure}[H]
    \centering
\begin{tikzpicture}
\tikzset{
a/.style={scale=.5, circle,minimum size=.1cm,fill=blue,draw=blue},
b/.style={scale=.5, circle,minimum size=.1cm,fill=red,draw=red},
c/.style={scale=.5, circle,minimum size=.1cm,fill=white!20,draw=gray,
                },
    v/.style={scale=.5, circle,minimum size=.1cm,fill=black,draw}}

\node[v] (0) at (1,0) {};
\node[v] (1) at (3,0) {};
\node[a] (2) at (0,0) {};
\node[a] (3) at (0.6,0.8) {};
\node[c] (4) at (0.6,-0.8) {};
\node[c] (5) at (3.4,0.8) {};
\node[b] (6) at (4,0) {};
\node[b] (7) at (3.4,-0.8) {};

\foreach \from/\to in {0/1, 2/0,3/0,4/0,5/1,6/1,7/1}
    \draw (\from) -- (\to);
 
\draw (1.1,0) node[anchor=north west]{$v$};
\draw (2.9,0) node[anchor=north east]{$w$};

\draw (2.8,1.1) node[above]{C} ;

\draw (1.4,-1.1) node[below]{C} ; 

\draw[blue] (-0.2,0.8)node[anchor=east]{A} ; 
\draw[red] (4.2,-0.6) node[anchor=west]{B} ;
\draw[dashed] (1,0) .. controls (1.5,0.7) and (2.5,0.7) .. (3,0);
 \draw[dotted] (2,0) ellipse (54 pt and 28 pt);
\end{tikzpicture}
\caption{Tree graph with two vertices and a pairwise correlation between the intertwiners.}
\label{2vertgraphcorr}
\end{figure}

In the following we {discuss the values of} $\mathcal{H}_k$ and $\mathcal{H}^c_k$ for $k=3$ and $k=4$.

\subsubsection{Order k=3 minimal Hamiltonians}

In order to determine the minimal value of $\mathcal{H}_3$, we study the energy cost of each configuration of generalised spins, for given graph and boundary conditions. Generally, we expect dominant configurations minimizing the action to correspond to a tripartition of the spin domains, with boundary condition percolating in the bulk up to a shared domain wall. 

In the tripartite case, however, interesting new equilibrium configurations are associated to the emergence of an extra spin domain, corresponding to set of vertices labelled by permutation elements that are \emph{geodesics} for the Cayley metric between $\mathbbm{1}$,$X$ and $X^{-1}$ (see \cref{permutation appendix}; a detailed discussion can be found in~\cite{Dong_21}). Being geodesics, the permutations in $T$ naturally minimizes the energy cost given by the Cayley distance whenever the $\mathbbm{1}$,$X$ and $X^{-1}$ elements meet, creating an energetically favourable bubble. Such domain is absent in the case of pure bipartite states previously studied in~\cite{Hayden:2016cfa}. 

For the simple graph under study, with $k=3$, we can have three geodesic configurations given by the three swap operators $S_{12}$, $S_{23}$ and $S_{13}$ (corresponding to the non-crossing partitions of $\mathbb{S}_3$, see \cref{permutation appendix}). We denote by $\tau$ the generic geodesic element. Using the results in \eqref{distancefromtau}, we then compute the value of the Hamiltonian of the configuration with swap operators in the bulk vertices $v$ and $w$. From \eqref{distancefromtau}, we get
\begin{eqnarray}
\Delta(\mathbbm{1},\tau)&=&1\nonumber\\
\Delta(X,\tau)&=&\Delta(X^{-1},\tau)=1\nonumber
\end{eqnarray}
These permutations define the new geodesic \emph{transition domain}, that we label by $T$.  Hence, we have that $g_v=g_w=\tau$  $\rightarrow$ $v,w\in\,T$ and
 \begin{equation}\label{2conf4nc}
\mathcal{H}_3=2\Delta(\tau,X)+2\Delta(\tau,X^{-1})+2\Delta(\tau,\mathbbm{1})=6
 \end{equation}
For the simple graph under study, the $T$ domain fills the whole bulk, preventing the $A,B,C$ domain walls to enter the graph.
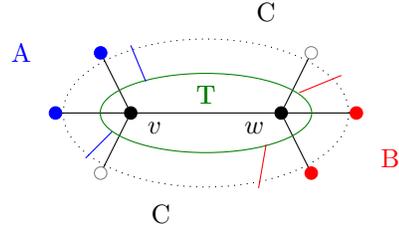
\begin{figure}[H]
    \centering
\begin{tikzpicture}
\tikzset{
a/.style={scale=.5, circle,minimum size=.1cm,fill=blue,draw=blue},
b/.style={scale=.5, circle,minimum size=.1cm,fill=red,draw=red},
c/.style={scale=.5, circle,minimum size=.1cm,fill=white!20,draw=gray,
                },
    v/.style={scale=.5, circle,minimum size=.1cm,fill=black,draw}}

\node[v] (0) at (1,0) {};
\node[v] (1) at (3,0) {};
\node[a] (2) at (0,0) {};
\node[a] (3) at (0.6,0.8) {};
\node[c] (4) at (0.6,-0.8) {};
\node[c] (5) at (3.4,0.8) {};
\node[b] (6) at (4,0) {};
\node[b] (7) at (3.4,-0.8) {};

\foreach \from/\to in {0/1, 2/0,3/0,4/0,5/1,6/1,7/1}
    \draw (\from) -- (\to);
 
\draw (1.1,0) node[anchor=north west]{$v$};
\draw (2.9,0) node[anchor=north east]{$w$};

\draw (2.8,1.1) node[above]{C} ; 
\draw (1.4,-1.1) node[below]{C} ; 
\draw[blue] (-0.2,0.8)node[anchor=east]{A} ; 
\draw[red] (4.2,-0.6) node[anchor=west]{B} ;

 \draw[dotted] (2,0) ellipse (54 pt and 28 pt) node[above,green!50!black]{T};
 \draw[green!50!black] (2,0) ellipse (40 pt and 15 pt) node[above,green!50!black]{T}; 
   \draw[blue] (1.2,.42) --  (1.,.9);
   \draw[blue] (.75,-0.25) --(0.4,-0.6); 

   \draw[red] (2.8,-0.42) --(2.7,-1.);
   \draw[red] (3.24,0.27) --(3.8,0.5);
\end{tikzpicture}
\caption{Transition domain on a tree graph with two vertices.}
\label{transition domain two vertices}
\end{figure}

For the same graph, we can repeat the analysis with the insertion of a bulk intertwiner correlations. Again, we report the numerical results in \cref{Hamiltonian appendix}. 

In this case, due to the edge-like form of the bulk correlations, the effect of the bulk is an increase of the value of the Hamiltonian for configurations with different permutations on correlated intertwiner pairs. Because of the spin homogeneity assumption, the pairwise intertwiner correlations contributes to the Hamiltonian as additional domain wall crossing edges (dashed line in \cref{2vertgraphcorr}) inside the graph. If the same permutation is inserted on the two adjacent vertices, their interaction gives no contribution. Both the edge and the bulk terms of the Hamiltonian in \eqref{488} in facts vanish.

Further, bulk correlations remove the degeneracy of the minimal Hamiltonian. For instance, we see that if the bulk entropy vanishes (i.e. there is no correlations between intertwiners) the minimal value of Hamiltonian ($6$ in the example above) is shared by the configurations in \eqref{2conf2nc} and \eqref{2conf4nc}: in this sense we have a \emph{degenerate minimum value} of $\mathcal{H}_3$. The degeneracy is removed once we insert the bulk correlations as the second configuration has an higher energy cost because of the additional edge, crossing the domain walls between $A$ and $B$. 
Similar behaviour was already discussed in \cite{Chirco:2021chk}.

Finally, when the transition domain $T$ extends over the two vertices and the domain wall is pushed out of the bulk region, the additional link does not cross any domain wall. The Hamiltonian is left unchanged and it has a \emph{unique} minimum.

\subsubsection{Order k=4 minimal Hamiltonians} 
All the considerations about $\mathcal{H}_3$ hold for the case of $\mathcal{H}_4$: \begin{itemize}
    \item the bulk correlations increase the value of the Hamiltonian for  configurations with vertices in different domains;
    \item the degeneracy of the minimal Hamiltonian is reduced by such correlations;
    \item the lowest energy configuration is characterized by a \emph{transition domain} filling the whole bulk: in this configuration, the domain walls are pushed out of the bulk. In the large spin limit, such configuration gives the dominant contribution of the Ising action.
\end{itemize}
Looking at our results for $k=3,4$, we see how the values of the Hamiltonian scale both with the number of links crossing the domain walls and with the order $k$. In fact, the same configuration assumes larger and larger values of $\mathcal{H}_k$ as the order of the replicas increases. 
The Hamiltonian of our generalised Ising model can be written in terms of the number of edges crossing each domain wall, respectively denoted as $\abs{\gamma_A}$,  $\abs{\gamma_B}$ and  $\abs{\gamma_C}$. If we look at the values of $\mathcal{H}_k$ for each configuration (see \cref{Hamiltonian appendix}), it is possible to verify that
(notice hereafter we focus on $k$ even only)
\begin{equation}
    \mathcal{H}_k= \qty(\frac{k}{2}-1)\qty( \abs{\gamma_A}+ \abs{ \gamma_B})+\frac{k}{2} \abs{\gamma_C}\,,  \quad k\;\, \text{even} \label{HamilArea}
\end{equation}
The partition function $\overline{Z_k}$ is dominated by the lowest energy configuration term
\begin{equation}
\overline{Z^{(k)}_1}=\sum_{\{g_v\}}e^{-\beta\mathcal{H}_k}\simeq e^{-\beta\mathcal{H}_k^{(\min)}} \, , 
\end{equation}
which is unique when we have bulk correlations. We write the Rényi log negativity as 
\begin{equation}
    \log \overline{N}_k=\log \overline{Z}^{(k)}_1-\log \overline{Z}^{(k)}_0
\end{equation}
Looking at $\mathcal{H}_0=\sum_E \Delta(g_v,g_w)+\sum_{\partial \gamma}\Delta(\mathbbm{1},g_v)$, we notice that its minimal value is 0, corresponding to the configuration with all vertices labeled by the identity permutation. Then
\begin{equation}
   \log \overline{N}_k=\log \overline{Z^{(k)}_1}=-\beta\mathcal{H}_k^{(\min)}  
\end{equation}
Eventually, the log negativity is given by the analytic continuation of the logarithm of the even momenta. We get
\begin{multline}
\log\overline{\mathcal{N}}=\lim_{k\to \frac{1}{2}}\log\overline{\mathcal{N}_{2k}}=-\beta\lim_{k\to \frac{1}{2}}\mathcal{H}_{2k}\\
=\beta\qty[\frac{1}{2}\qty(\abs{\gamma_A}+\abs{\gamma_B})-\frac{1}{2}\abs{\gamma_C}]\label{5115}
\end{multline}
The entanglement of the boundary random spin network mixed state $\rho_{AB}$ scales with the areas of the domain walls of the dual generalised Ising model, corresponding to minimal energy configurations. The measure of log negativity depends on the local correlation structure only in relation to the connectivity of the graph, which necessarily plays a role in defining the minimal energy configurations. On the other hand, the area scaling is directly affected by the bulk correlations between intertwiners.  A similar behaviour has been pointed out in recent works \cite{Colafranceschi:2020ern,Chirco:2021chk} regarding $2$-nd R\'enyi entropy for a pure bipartite random spin network. Note that since we are dealing with the trivial case of a homogeneous spin network with edgewise intertwiner correlations, the extra  contributions to the area law \eqref{HamilArea} are hidden in the increased values of the domains, while the correlations become effective additional connectivity.

The form of our results is consistent with the behavior obtained for random tensor network in \cite{Dong_21}, where the right-hand term of the \eqref{5115} is interpreted as the \emph{quantum mutual information} $I_{A:B}$ of the two subsystems. In our case, the domain of the third region is $\gamma_C=\qty(\gamma_A\cup\gamma_B)^c=\gamma_{AB}$ and we have the equality $\beta\abs{\gamma_C}=S_{AB}$. 
The effect of inserting quantum correlations among intertwiners is the removal of the degeneracies of $\mathcal{H}^{(\min)}$ and the appearance of the intermediate region (the \emph{transition region} T) filling the bulk. Since $\gamma_C$ is the complement of the two remaining boundary regions, the log negativity can be written as the mutual information between $A$ and $B$.
In this terms, the typical value of the log negativity is given by
\begin{equation}\label{lognegativity}
\log\overline{\mathcal{N}}_{AB}=\frac{1}{2}\qty[S_A+S_B-S_{AB}]\, .
\end{equation}

\subsection{Open tree-graph with 3 vertices}
\label{three vertices graph}
We now move to the slightly more articulate case of an open tree-graph with one extra vertex. This setting allows us to consider the effect of \emph{non-local} edge-like correlations in the bulk. 

Consider a graph $\gamma$ composed by three 4-valent vertices $x$, $y$ and $z$, with tripartite boundary conditions as pictured in the figure below.

\begin{figure}[H]
    \centering
\begin{tikzpicture}
\tikzset{
a/.style={scale=.5, circle,minimum size=.1cm,fill=blue,draw=blue},
b/.style={scale=.5, circle,minimum size=.1cm,fill=red,draw=red},
c/.style={scale=.5, circle,minimum size=.1cm,fill=white!20,draw=gray},
    v/.style={scale=.5, circle,minimum size=.1cm,fill=black,draw}}

\node[v] (0) at (1,0) {};
\node[v] (1) at (3,0) {};
\node[v] (8) at (5,0) {};
\node[a] (2) at (0,0) {};
\node[a] (3) at (0.6,0.8) {};
\node[c] (4) at (0.6,-0.8) {};
\node[c] (5) at (5.4,.8) {};
\node[c] (9) at (3.,-1.2) {};
\node[a] (10) at (3.,1.2) {};
\node[b] (6) at (6,0) {};
\node[b] (7) at (5.4,-0.8) {};

\foreach \from/\to in {0/1,1/8, 2/0,3/0,4/0,5/8,6/8,7/8,1/9, 1/10}
    \draw (\from) -- (\to);
  
\draw (1.1,0) node[anchor=north west]{$x$};
\draw (2.9,0) node[anchor=north east]{$y$};
\draw (4.9,0) node[anchor=north east]{$z$};

\draw (4.8,1.3) node[above]{C} ; 
\draw (1.4,-1.3) node[below]{C} ;
\draw[blue] (-0.2,0.8)node[anchor=east]{A} ; 
\draw[red] (5.6,-1.2) node[anchor=west]{B} ; 
\draw[dotted] (3,0) ellipse (90 pt and 40 pt); 

\end{tikzpicture}
\end{figure}

The uncorrelated Hamiltonian is
\begin{eqnarray}
\mathcal{H}_k&=&\Delta(g_x,g_y)+\Delta(g_y,g_z)+2\Delta(g_x,X)+\Delta(g_y,X)\nonumber \\ &+&\Delta(g_x,\mathbbm{1})++\Delta(g_y,\mathbbm{1})+\Delta(g_z,\mathbbm{1})+2\Delta(g_z,X^{-1})\nonumber \\
\end{eqnarray}
If we insert local correlations between adjacent vertices ($xy$ and $yz$ equivalently) the two Hamiltonians become
\begin{eqnarray}
    \mathcal{H}_k^{(xy)}&=&\pmb{2}\Delta(g_x,g_y)+\Delta(g_y,g_z)+2\Delta(g_x,X)+\Delta(g_y,X)\nonumber \\ &+&\Delta(g_x,\mathbbm{1})++\Delta(g_y,\mathbbm{1})+\Delta(g_z,\mathbbm{1})+2\Delta(g_z,X^{-1})\nonumber \\
\end{eqnarray}

\begin{figure}[H]
    \centering
\begin{tikzpicture}
\tikzset{
a/.style={scale=.5, circle,minimum size=.1cm,fill=blue,draw=blue},
b/.style={scale=.5, circle,minimum size=.1cm,fill=red,draw=red},
c/.style={scale=.5, circle,minimum size=.1cm,fill=white!20,draw=gray
                },
    v/.style={scale=.5, circle,minimum size=.1cm,fill=black,draw}}

\node[v] (0) at (1,0){} ;
\node[v] (1) at (3,0){} ;
\node[v] (8) at (5,0){} ;
\node[a] (2) at (0,0){} ;
\node[a] (3) at (0.6,0.8){} ;
\node[c] (4) at (0.6,-0.8){};
\node[c] (5) at (5.4,.8){} ;
\node[c] (9) at (3.,-1.2){};
\node[a] (10) at (3.,1.2){} ;
\node[b] (6) at (6,0){} ;
\node[b] (7) at (5.4,-0.8){} ;

\foreach \from/\to in {0/1,1/8, 2/0,3/0,4/0,5/8,6/8,7/8,1/9, 1/10}
    \draw (\from) -- (\to);
    \draw[dashed] (0) to[bend left] (1);

\draw (1.1,0) node[anchor=north west]{$x$};
\draw (2.9,0) node[anchor=north east]{$y$};
\draw (4.9,0) node[anchor=north east]{$z$};

\draw (4.8,1.3) node[above]{C} ; 
\draw (1.4,-1.3) node[below]{C} ; 
\draw[blue] (-0.2,0.8)node[anchor=east]{A} ; 
\draw[red] (5.6,-1.2) node[anchor=west]{B} ; 

\draw[dotted] (3,0) ellipse (90 pt and 40 pt); 
\end{tikzpicture}
\end{figure}

\begin{eqnarray}
    \mathcal{H}_k^{(yz)}&=&\Delta(g_x,g_y)+\pmb{2}\Delta(g_y,g_z)+2\Delta(g_x,X)+\Delta(g_y,X)\nonumber \\ &+&\Delta(g_x,\mathbbm{1})++\Delta(g_y,\mathbbm{1})+\Delta(g_z,\mathbbm{1})+2\Delta(g_z,X^{-1})\nonumber \\
\end{eqnarray}

\begin{figure}[H]
    \centering
\begin{tikzpicture}
\tikzset{
a/.style={scale=.5, circle,minimum size=.1cm,fill=blue,draw=blue},
b/.style={scale=.5, circle,minimum size=.1cm,fill=red,draw=red},
c/.style={scale=.5, circle,minimum size=.1cm,fill=white!20,draw=gray},
    v/.style={scale=.5, circle,minimum size=.1cm,fill=black,draw}}
 
\node[v] (0) at (1,0){} ;
\node[v] (1) at (3,0){} ;
\node[v] (8) at (5,0){} ;
\node[a] (2) at (0,0){} ;
\node[a] (3) at (0.6,0.8){} ;
\node[c] (4) at (0.6,-0.8){};
\node[c] (5) at (5.4,.8){} ;
\node[c] (9) at (3.,-1.2){};
\node[a] (10) at (3.,1.2){} ;
\node[b] (6) at (6,0){} ;
\node[b] (7) at (5.4,-0.8){} ;
 
\foreach \from/\to in {0/1,1/8, 2/0,3/0,4/0,5/8,6/8,7/8,1/9, 1/10}
    \draw (\from) -- (\to);
        \draw[dashed] (1) to[bend left] (8);

\draw (1.1,0) node[anchor=north west]{$x$};
\draw (2.9,0) node[anchor=north east]{$y$};
\draw (4.9,0) node[anchor=north east]{$z$};

\draw (4.8,1.3) node[above]{C} ; 
\draw (1.4,-1.3) node[below]{C} ; 
\draw[blue] (-0.2,0.8)node[anchor=east]{A} ; 
\draw[red] (5.6,-1.2) node[anchor=west]{B} ; 
\draw[dotted] (3,0) ellipse (90 pt and 40 pt); 
\end{tikzpicture}
\end{figure}
The neat result is an increase of the local connectivity of the graph, while the Hamiltonians do not get new terms. 
Differently, by inserting non-local correlation between the intertwiner in $x$ and $z$, the Hamiltonian is modified by the new term $\Delta(g_x,g_z)$. 
\begin{figure}[H]
    \centering
\begin{tikzpicture}
\tikzset{
a/.style={scale=.5, circle,minimum size=.1cm,fill=blue,draw=blue},
b/.style={scale=.5, circle,minimum size=.1cm,fill=red,draw=red},
c/.style={scale=.5, circle,minimum size=.1cm,fill=white!20,draw=gray},
    v/.style={scale=.5, circle,minimum size=.1cm,fill=black,draw}}

\node[v] (0) at (1,0){} ;
\node[v] (1) at (3,0){} ;
\node[v] (8) at (5,0){} ;
\node[a] (2) at (0,0){} ;
\node[a] (3) at (0.6,0.8){} ;
\node[c] (4) at (0.6,-0.8){};
\node[c] (5) at (5.4,.8){} ;
\node[c] (9) at (3.,-1.2){};
\node[a] (10) at (3.,1.2){} ;
\node[b] (6) at (6,0){} ;
\node[b] (7) at (5.4,-0.8){} ;

\foreach \from/\to in {0/1,1/8, 2/0,3/0,4/0,5/8,6/8,7/8,1/9, 1/10}
    \draw (\from) -- (\to);
        \draw[dashed] (0) to[bend left] (8);
 
\draw (1.1,0) node[anchor=north west]{$x$};
\draw (2.9,0) node[anchor=north east]{$y$};
\draw (4.9,0) node[anchor=north east]{$z$};

\draw (4.8,1.3) node[above]{C} ; 
\draw (1.4,-1.3) node[below]{C} ; 
\draw[blue] (-0.2,0.8)node[anchor=east]{A} ;
\draw[red] (5.6,-1.2) node[anchor=west]{B} ; 
\draw[dotted] (3,0) ellipse (90 pt and 40 pt); 
\end{tikzpicture}
\end{figure}
We get
\begin{multline}
\mathcal{H}_k^{(xz)}=\Delta(g_x,g_y)+\Delta(g_y,g_z)+\pmb{ \Delta(g_x,g_z)}+\\ +2\Delta(g_x,X)+\Delta(g_y,X)+ \Delta(g_x,\mathbbm{1})+\\ +\Delta(g_y,\mathbbm{1})+\Delta(g_z,\mathbbm{1})+2\Delta(g_z,X^{-1})
\end{multline}
Despite the minimal change made to the graph with respect to the previous case, we now find many more spin configurations of interest. In \cref{Hamiltonian appendix} we report the examples of the low energy configurations which are very close to the minimal value of the Hamiltonian and the degenerate configurations of the latter. Using the relations on distances in \eqref{593}, the value of the Hamiltonian is easily calculated. Once again, as an explicit example, let us look at the case of $\mathcal{H}_k$ for $k=3$ and $k=4$.

\subsubsection{Order k=3 minimal Hamiltonians} 

It easy to see that if we insert a local bulk correlation between adjacent vertices ($xy$ or $yz$), only some degeneracies are removed. In fact, consider the case with a bulk correlation between $x$ and $y$: the Hamiltonian has an additional term that increases its value only if the vertices $x$ and $y$ belong to different domains, \emph{i.e.} the distance between the permutations $g_x$ and $g_y$ is not vanishing.
The minimal energy configurations (reported in \eqref{3neg3ver1}, \eqref{3neg3ver2} and \eqref{3neg3ver3}) are left untouched, since the two spin variables in $x$ and $y$ are the same. The Hamiltonians of \eqref{3neg3ver4} and \eqref{3neg3ver5} must be discussed. 
As we pointed out, \eqref{3neg3ver4} actually corresponds to three different degenerate configurations with $\mathcal{H}_3=8$: the ones with the transition region filling only a couple of vertices of the bulk, e.g. $xy$, $yz$ and $xz$. The first of these configurations will have a vanishing contribution from the bulk link insertion, so one of these degeneracies will not be eliminated.
The same result holds for \eqref{3neg3ver5}, since we have the three cases with the transition region only filling one vertex in the bulk; it is now clear that the configuration with $z\in\,T$ and $x,y\in\,A$ will still have $\mathcal{H}_3=8$.\par
Consider now the other possibile of local correlation, \emph{i.e.} the couple $yz$. The Hamiltonians of the configurations \eqref{3neg3ver1} and \eqref{3neg3ver3} remain the same. The Hamiltonian \eqref{3neg3ver2} becomes
\begin{equation}
    \mathcal{H}_3=8+\Delta(X,X^{-1})=10
\end{equation}
Similarly to the previous case, the degenerate configurations with pair of vertices in the transition regions are \emph{partially} removed: the Hamiltonians of the configurations with $x$ and $y$ or $x$ and $z$ in the $T$ region increase their value by one, while the other is unchanged.\\
The same result (degeneracy partially removed) is obtained for the configurations with the transition region filling only one vertex. 

We insert now a \emph{non-local correlation} in the bulk, that is an edge between $x$ and $z$ and study the effect of such correlation on the different configurations:  

\begin{enumerate}
    \item $g_x=g_y=g_z=\mathbbm{1}$ $\rightarrow$ $x,y,z\in\,C$
    \begin{equation}
        \mathcal{H}_3= 2\Delta(g_x,X)+\Delta(g_y,X)+2\Delta(g_z,X^{-1})=10
    \end{equation}

   \begin{figure}[H]
    \centering
	\includegraphics[width=0.45\textwidth]{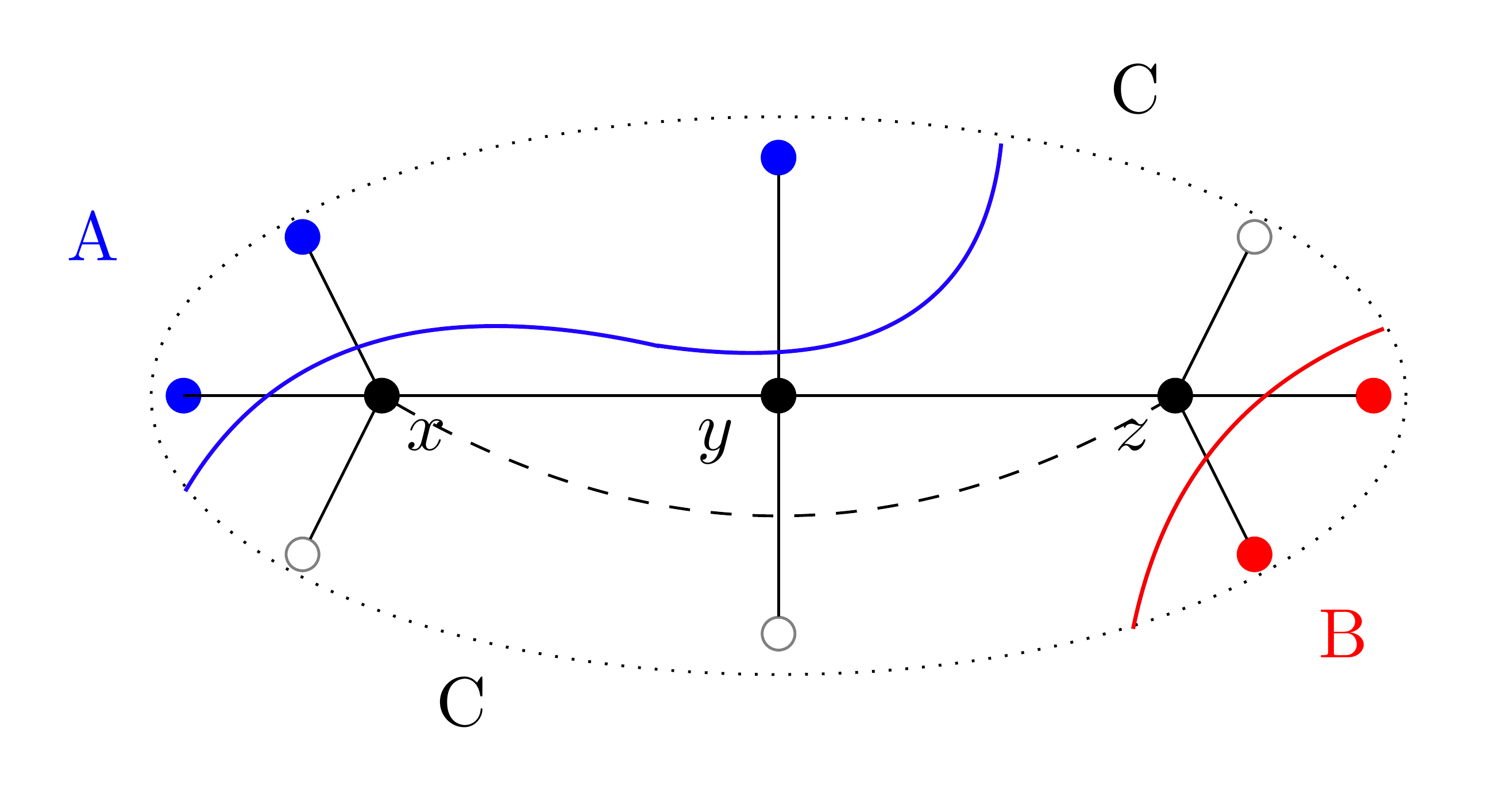}
\end{figure} 

    \item $g_x=g_y=X$, $g_z=X^{-1}$  $\rightarrow$ $x,y\in\,A$, $z\in\,B$
  \begin{equation}
      \mathcal{H}_3=\Delta(g_y,g_z)+\Delta(g_x,g_z)+\Delta(\mathbbm{1},X)+\Delta(\mathbbm{1},X)+\Delta(\mathbbm{1},X^{-1})=10
  \end{equation}

\begin{figure}[H]
    \centering
	\includegraphics[width=0.45\textwidth]{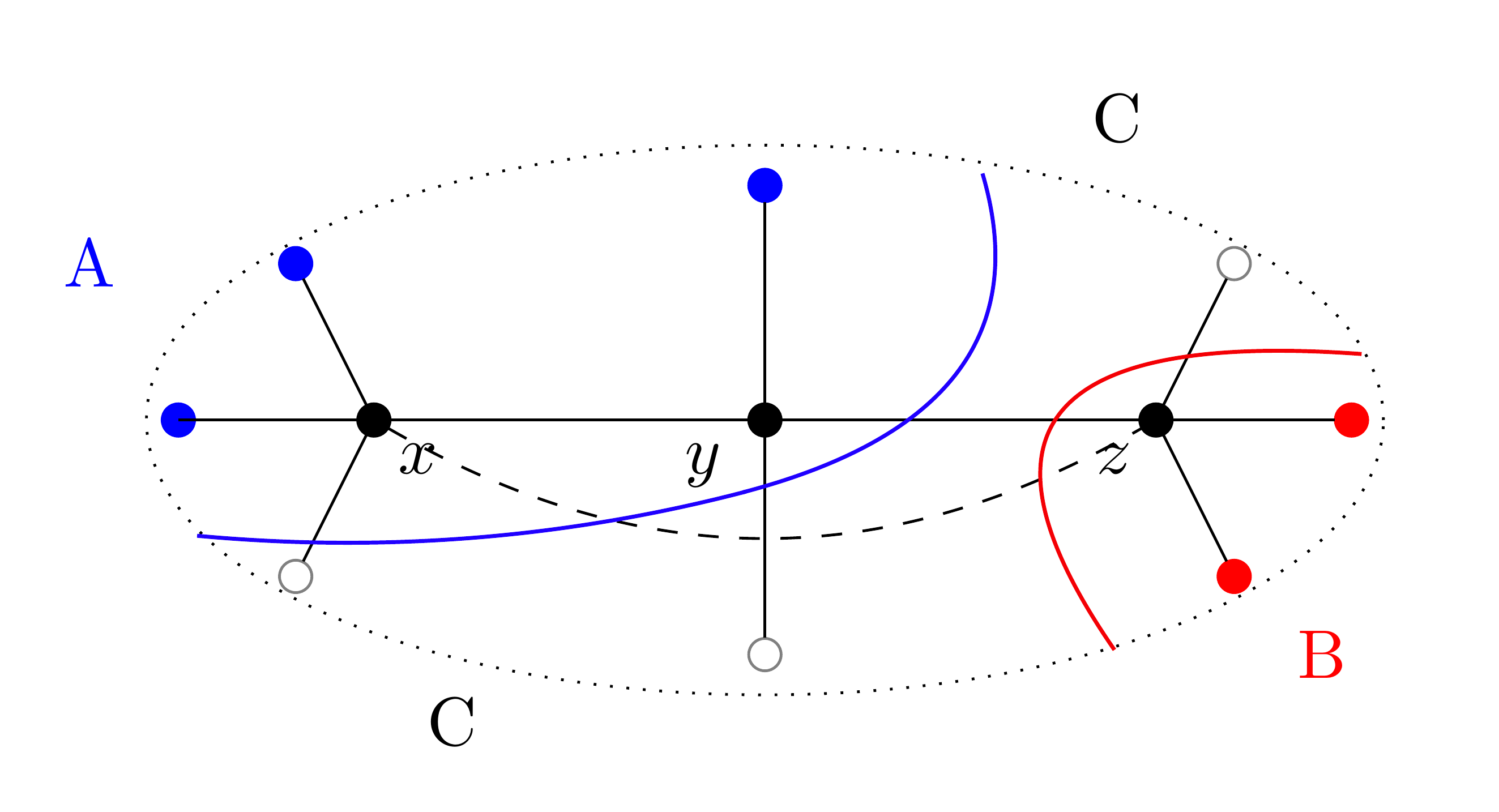}
\end{figure} 

\item $g_x=g_y=g_z=\tau$ $\rightarrow$ $x,y,z\in\,T$
 \begin{eqnarray}
      \mathcal{H}_3&=&2\Delta(\tau,X)+\Delta(\tau,X)+3\Delta(\mathbbm{1},\tau)+\nonumber \\ &+&2\Delta(\tau,X^{-1})=8 \label{3neg3ver3c}
  \end{eqnarray}
\begin{figure}[H]
    \centering
	\includegraphics[width=0.45\textwidth]{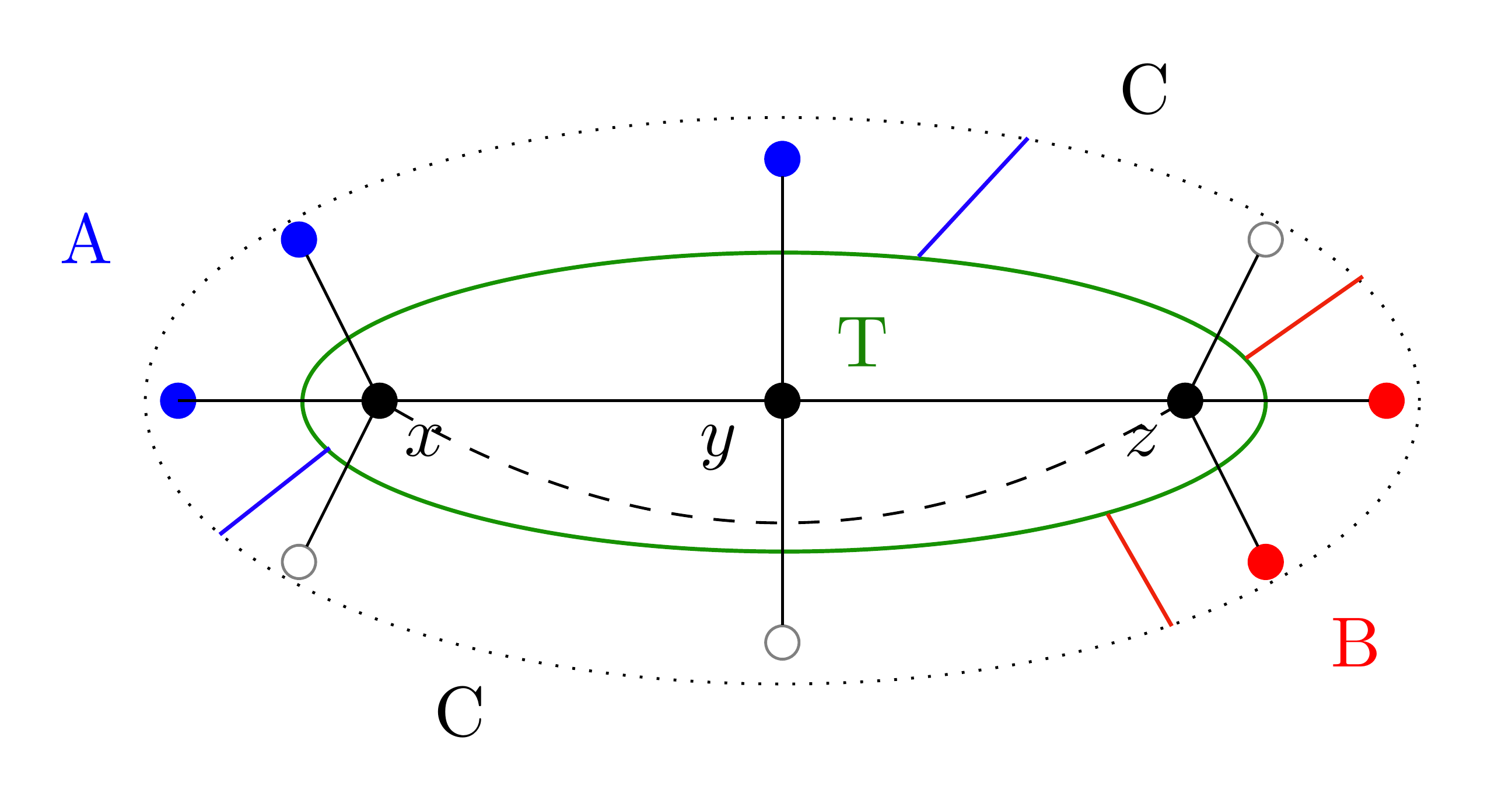}
\end{figure} 
The configuration with the transition region filling the whole bulk is unchanged.

\item \label{3neg3ver4c} $g_x=X$, $g_y=g_z=\tau$ $\rightarrow$ $x\in\,A$, $y,z\in\,T$
 \begin{eqnarray}
      \mathcal{H}_3&=&2\Delta(\tau,X)+\Delta(\mathbbm{1},X)+\nonumber \\ &+&2\Delta(\mathbbm{1},\tau)+2\Delta(\tau,X)+\Delta(X,\tau)=9
\end{eqnarray}
\begin{figure}[H]
    \centering
	\includegraphics[width=0.45\textwidth]{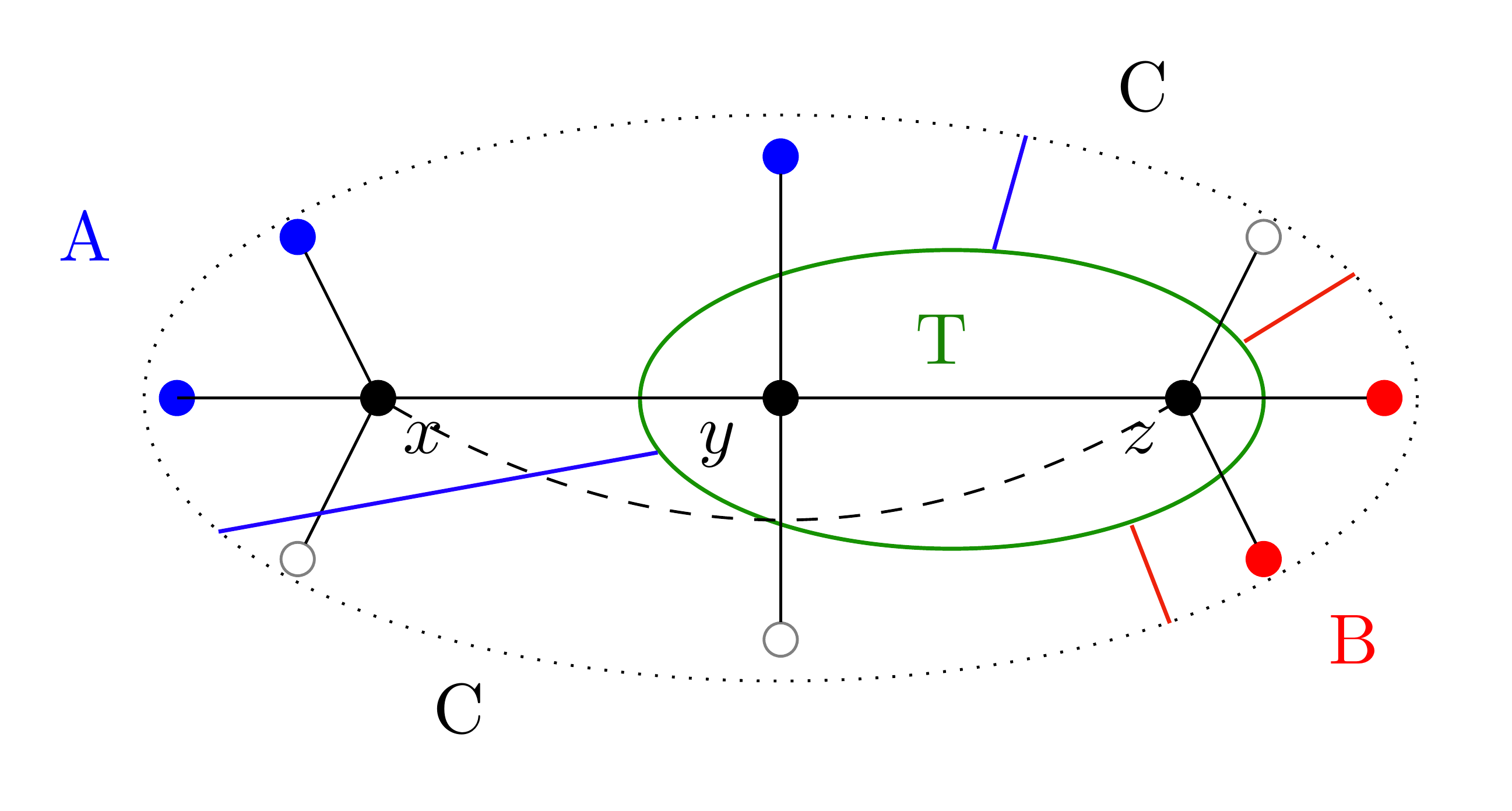}
\end{figure} 

\item\label{3neg3ver5c} $g_x=g_y=X$, $g_z=\tau$ $\rightarrow$ $x,y\in\,A$, $z\in\,T$

 \begin{eqnarray}
      \mathcal{H}_3&=&\Delta(\tau,X)+\Delta(\tau,X^{-1})+\Delta(X,\tau) 
      +\Delta(\mathbbm{1},X)+\nonumber\\&+&\Delta(\tau,X)+\Delta(X^{-1},\mathbbm{1})+\Delta(X,\tau)=9
\end{eqnarray}
\begin{figure}[H]
    \centering
	\includegraphics[width=0.45\textwidth]{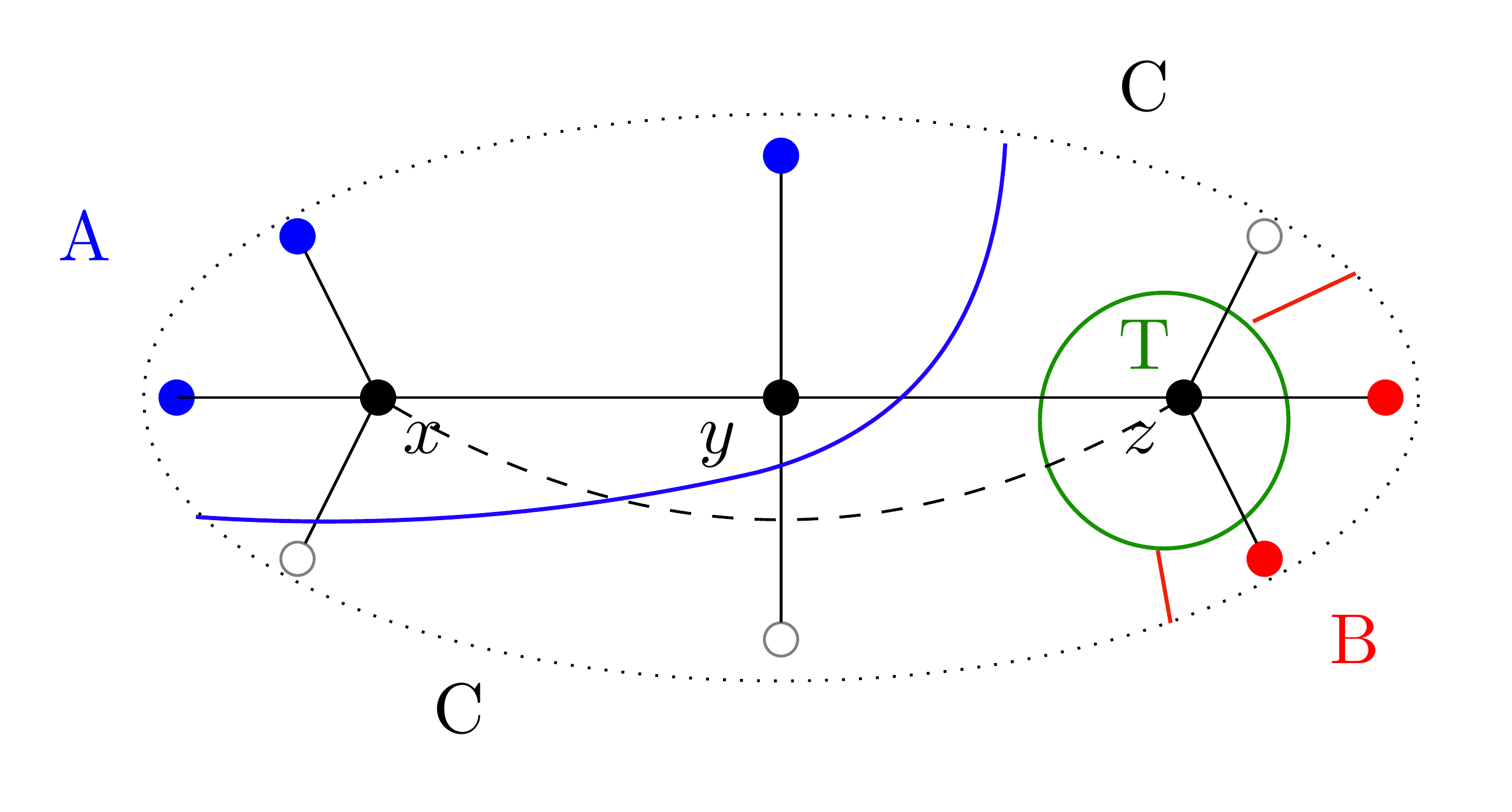}
\end{figure} 

\end{enumerate}
The lowest energy configuration is given in \eqref{3neg3ver3c}. Once again the dominant term is given by the configuration with the transition region filling the whole bulk and domain walls pushed out of the latter. 

The same arguments hold for the fourth order Hamiltonian and give exactly the same results: local correlations partially remove degeneracies of the lowest energy configuration while non-local correlations tend to favour equilibrium configurations where the boundary spins domains are prevented to enter the correlated bulk. As a consequence, there is one non-degenerate configuration with minimal surface area $\abs{\gamma_T}$, which correspond to the whole spin network boundary surface. In the above results, we see that non-locality completely removes all the degeneracies between minimal configurations. This is due to the simplicity of the considered graphs. In more complex graphs, non-locality might not have the same effect: by looking at such non-local correlations as additional links (beside their different nature), the connectivity of the three-vertices graph we analysed is such that the additional link between vertices $x$ and $z$ spreads throughout all the graph, intersecting the domain walls of both A and B. If we consider a graph with more vertices, or if we choose a different tripartition of the boundary, even non-local correlations might not be strong enough to push the domain walls out of the bulk. In these cases, not all the degeneracy would be eliminated and the minimal energy configurations could still be found via direct computation of Hamiltonians, with the foresight of taking into account the number of degenerate contributions in the statistical model.

In this sense, non-local correlations must be further investigated in graphs with an arbitrary number of vertices in order to provide a deeper understanding of their role in the statistical model and their physical interpretation.  

\section{Discussion}\label{discussion}

In this paper, we study the entanglement structure of a class of spin network states describing a quantum patch of $3d$ space with boundary. We consider a tripartition of such boundary, and focus on the \emph{mixed} state description of two patches $A$ and $B$, generically disconnected and immersed in an environment $C$. We investigate the entanglement between $A$ and $B$ via a measure of logarithmic negativity and its Rényi generalization. Negativity is a measure of entanglement well defined for both pure and mixed states, hence suitable for generalising most recent results on spin network entanglement entropy from a bipartite to a simple \textit{tripartite} system. The introduction of the notion of negativity from the recent literature on multipartite random tensor networks (see e.g. \cite{Dong_21}) to the quantum gravity setting is a first element of novelty of the work. 

We provide an explicit calculation of the negativity by restricting to a class of extremely simple quantum spin network states. First of all, we limit our analysis to fixed area and fixed graphs spin networks, corresponding at most to a truncation of a quantum $3d$ geometry state, hence disregarding all possible quantum correlations involving sums over the spins, as well as the effect of any graph superposition. Further, we use the generalized notion of \emph{random} spin network and bulk to boundary mapping to engineer a very transparent structure of bulk correlations for our states. In facts, we construct the spin networks in two ideal steps. First, we build what we call a random \emph{entanglement background} state, defined by a collection of single \emph{random} vertex tensors -- dual to randomised states of quantum tetrahedra with extra boundary spin information -- glued together via maximal entangled edge-like pairs of spins. By taking the rank of the vertex tensors regular (four) and the value of the spin homogeneous throughout the graph, the resulting spin network is a gauge symmetric analogue of a projected entangled pairs tensor network state (s-PEP TN) with random nodes. Secondly, we plug non random bulk correlations on such background via a projection, which realises a bulk to boundary mapping procedure. This eventually turns our random spin networks into the analogue of an error-correcting code, incorporating bulk degrees of freedom in the boundary and adding the desired bulk entanglement. 

Such two levels of construction correspond to two different entanglement structures of the spin network state we want to make apparent. The first layer of local entanglement between spin states located on the network’s edges involves non gauge invariant degrees of freedom. Despite playing a key role in the definition of the spin network graph connectivity, such entanglement cannot be measured in a proper gauge invariant setting involving quantum geometry operators. In this sense, most recent literature considers this layer of entanglement unphysical and irrelevant for the entanglement/geometry correspondence. Here, we shall understand such entanglement layer by considering the edge spins as higher energy degrees of freedom, coexisting with the single vertex quanta description of GFT and disappearing once the gauge invariance has been established and the spin network graph has formed at some lower energy scale. In this sense, the ultra-local entanglement layer do contribute to any measure of entanglement in a spin network as long as it is responsible of the connectivity of the spin network graph. 
The second layer of entanglement consists of the non-local, gauge-invariant quantum correlations between intertwiners, intended as genuine correlations of excitations of quantum geometry in the bulk. To keep the bulk entanglement structure of our states under control, we define the bulk correlations in terms of pairs of maximally entangled intertwiner states. This choice eventually allows us to avoid departures from the holographic scaling of the entanglement negativity, while making the effect of a non-locally correlated bulk apparent.    

Concretely, we first compute the typical $k$-th Rényi negativity for a generic open random spin networks. Thanks to the random character of the states, the computation of the averaged $k$-th R\'enyi negativity is mapped to the evaluation of the partition functions of a \emph{generalized Ising model}, defined by a two-body interaction model between permutation elements, which act as \emph{generalized spins} attached to the spin network vertices. The tripartite boundary conditions are set by the permutations fields $X$, $X^{-1}$ and $\mathbbm{1}$ attached to virtual pinning vertices comprising the $A,B,C$ boundary subregions. For $k=3$, we can write the generalized Ising model for triples of spins, given by swaps operators at each vertex, by which we parameterize the symmetric group $\mathbb{S}_3$ (in \cref{spinn}). The use of spin variables for the model allows for a direct comparison of the similar model for the bipartite boundary entanglement entropy studied in~\cite{Colafranceschi:2021acz}. For the generic $k$, we rederive the approach of~\cite{Dong_21} which effectively describes the two-body interaction via the Cayley distance on the permutation group. This action prefers neighbouring generalised spins to be parallel. This means that in the large spin dimension limit, namely the strong coupling or low temperature regime, the dominant configurations that minimize the action contain large domains separated by domain walls, which in turn give the energy cost of the configuration.
Differently from \cite{Dong_21}, the models we define contain new terms deriving from the bulk correlations between intertwiners degrees of freedom. Such correlations give new relevant contributions to the analysis of the minimal surfaces. Also, our model is characterised by \emph{a priori} different interactions strengths, given by the logarithm of the dimension of the link, semi-link or intertwiner space according to the term we consider. Nevertheless, under the homogeneity assumption, we end up dealing with a single (inverse) temperature $\log d$.

For a restricted class of spin network states with support on two-vertices and three-vertices tree-graphs, we compute the logarithmic negativity as the analytic continuation of the $k$-th R\'enyi negativity, for even $k=4$. We find that the logarithmic negativity is nontrivial whenever the entanglement wedge of $A$ and $B$ are connected in the bulk. Differently from the bipartite case, where the two entanglement wedges always share a unique extremal surface in the bulk, in the tripartite case, the appearance of a transition domain depends on the degree of non-locality of the bulk correlations and generically on the effect of the environment $C$. Whenever such environment is not too big, for any (even) $k$, the negativity is nontrivial and it displays a holographic scaling given by the formula,
\begin{equation}
\log\overline{\mathcal{N}}=\beta\qty[\frac{1}{2}\qty(\abs{\gamma_A}+\abs{\gamma_B})-\frac{1}{2}\abs{\gamma_{C}}]
\end{equation}
This is a natural generalisation of the RT formula for the entanglement entropy of a pure bipartite state $\rho_{AB}$. 

In particular, we see that the bulk intertwiner entanglement generally increases the negativity of the boundary by favouring the formation of transitory regions in the bulk, while at the same time tends to break up the holographic behaviour. In our setting, we do not see the deviation from the area scaling, just due to the special form of bulk correlations considered, defined by Bell-like pairs of intertwiner states. In this case the effect of the bulk is limited to an increase of the value of the minimal domain wall surfaces in the bulk.

Further, in the cases of two and three-vertices tree-graphs, we find that the dominant configurations always correspond to the appearance of transition domains $T$, which englobes the whole bulk, hence preventing the $A$ and $B$ domain walls to enter the network. For such configurations, we identify the domain $C$ as the complementary boundary region to the union of $A$ and $B$, thereby identifying typical log negativity as half of the quantum mutual information between the two subsystems $A$ and $B$. In particular, such configurations are unique, while in general the expression of the negativity is corrected by a degeneracy factor.  

Despite the very simple spin network states considered, the obtained results open a series of new instances which deserve further investigation. First of all, the richer entanglement wedge structure which appears in the case of a tripartite state. By using the dual simplicial description of our spin networks, we have a first geometric interpretation of the domain walls emerging in the statistical model as extremal surfaces, with areas measured by the value of the spins of the crossing edges. However, we see that maximally entangled intertwiner correlations in the bulk can mimic the same result, while now relating the value of the extremal surface areas to some effective dimension of the correlated intertwiner spaces. The latter does not have a clear geometric understanding yet. The very simplicial (dual) interpretation of the \emph{entanglement wedges} involved in the definition of the mutual information is open. Nevertheless, we expect the quantum \emph{volume} of such wedges to play a role from an information theoretic viewpoint, for instance in relation to a possible measure of \emph{quantum complexity} (see e.g.~\cite{Susskind_16, Ben-Ami:2016qex}) for quantum geometry described via spin network states. 

Already a straightforward generalisation to more generic graphs, with a higher number of vertices and some degree of inhomogeneity in the spins, would allow to investigate the richer entanglement phase diagram appearing in the tripartite setting, as a function of the ratios of the dimensions of the three subregions $A$, $B$ and $C$, along the lines of  \cite{Shapourian_21}. For a bipartite random open spin network system, the entanglement entropy follows the Page curve \cite{Chirco:2021chk}. In the tripartite case, the Page curve is modified due to the presence of a third subsystem $C$. Typically, one finds that the Page curve admits a \emph{plateau} in the intermediate regime where the log  negativity depends only on the size
of the system but not on how the system is partitioned~\cite{Shapourian_21}. In our setting, the result in \eqref{lognegativity} can be interpreted as a confirmation of such a plateau regime for the spin network framework. Having larger graphs with bigger environment $C$ degrees of freedom would further allow to study the vanishing of the entanglement of the two $A,B$ patches due to the large dimensionality of the region $C$. Moreover, a different factorisation of the spin network Hilbert space, for instance with bulk curvature degrees of freedom playing the role of the $C$ environment, would allow to describe the degree of thermalization of the boundary system (the AB reduced state) as a function of the dimension of the bulk environment.  
In this sense, the tripartite setting seems to provide a natural minimal framework to further investigate the very notion of \emph{quantum black hole} in non-perturbative quantum gravity from an information theoretic viewpoint (see e.g.~\cite{Livine:2005mw, PhysRevD.81.104032, Pranzetti_2013, Anza:2017dkd,PhysRevD.97.066017} for a diverse set of ideas in this sense).

\begin{acknowledgments}
The authors are grateful to the members of the  Quantum Space and Quantum Information Group of the Department of Physics E.Pancini at the University of Naples Federico II, for the useful discussions on the preliminary results of the work.
\end{acknowledgments}
\appendix

\section{Permutation approach} \label{permutation appendix}
The permutation group plays a crucial role in the computation of the $k$-th R\'enyi negativity via statistical mapping. We briefly recall some useful definitions and tools regarding this group that have been used in the derivation of our results.

\subsection{Permutation Group}
The permutation group of order $n$, $S_n$, is a finite group of cardinality $n!$. Each element $\sigma \in S_n$ is defined to be a bijection of a given set M to itself, i.e. 
\begin{equation}
\sigma(i)=j\,\,\,\, \forall\,\, i,j\in M\,.
\end{equation}

There are different notations to represent a permutation. We use two of them: the first one is {\itshape Cauchy's two lines notation}, which is very practical in the calculation of the composition of two group element; the second one is the cyclic form, which turns out to be very useful in the calculation of the cycles and of the Cayley metric that we define later. In particular, {\itshape Cauchy's two lines notation} consists in writing two rows. Say $\sigma\in S_n$ is a permutation acting on a given set of $n$ element $\{x_1,\ldots,x_n\}$. In the first row we list all the element of the set. In the second row we write the image under the permutation below each number, that is
\begin{equation}
\sigma=\qty(\begin{array}{cccc}
          x_1 & x_2 & \ldots & x_n \\
          \sigma(x_1) & \sigma(x_2) & \ldots & \sigma(x_n)
    \end{array})\,.
\end{equation}
For instance a particular permutation of the set $\{1,2,3,4\}$ can be written as
\begin{equation}
    \sigma =\begin{pmatrix}
1 & 2 & 3 & 4\\
2 & 4 & 3 & 1
\end{pmatrix}\,.
\end{equation}
In the {\itshape cyclic notation}, we write in round brackets the chains of numbers such that the second one is the image of the first one, the third element is the image of the second on and so on. For example the previous permutation is written as $(124)(3)$ where $(3)$ is a trivial cycle because it is left unchanged by the permutation.

\subsection{Geodesics on permutation group}
Permutation group in equipped with a natural metric that plays a key role in the generalised Ising model described in \cref{setting}. In the computation of the negativity, the minimization of the action of the statistical model requires to study geodesics on permutation group. In particular, we focus on the class of permutations that are geodesics between the identity, the cyclic permutation and its inverse \cite{Dong_21}. 

Consider a permutation $g\in \mathbb{S}_k$, we can define the {\it length} of $g$ as the minimum number of \emph{swaps} to get g starting from the identity.
For example in $S_3$ the permutations $(12)(3)$ and $(123)$ have respectively length 1 and 2.
Similarly we can define the number of disjoint {\it cycles} of a permutation as $\chi(g)=\#(cycles)$. The permutation $(12)(3)$ and $(123)$ have respectively 2 and 1 cycles. By this example it is easy to see that
\begin{equation}
    l(g)+\chi(g)=k\,.
\end{equation}
We can thus introduce a natural metric, given by
\begin{equation}
    \Delta(g,h)=l(g^{-1}h)=k-\chi(g^{-1}h)\,.
\end{equation}
As mentioned before, we will be interested in studying the distance between 3 particular permutations of $\mathbb{S}_k$:
\begin{equation}
    \begin{split}
        &\mathbbm{1}=(1)(2)\ldots(k)\,, \\
        &X=(12\ldots k)\,,\\
        &X^{-1}=(k\ldots 21)\,.
    \end{split}
\end{equation}
Since both lenghts and cycles are easy to calculate for these permutations, we can immediately obtain the distances between them:
\begin{align}
    \Delta(\mathbbm{1},X)&=k-1\,,\\
    \Delta(\mathbbm{1},X^{-1})&=k-1\,,\\
    \Delta(X,X^{-1})&=\biggl\{\begin{array}{cc}
        k-1 & k\,\,\,odd \\
         k-2 & k\,\,\, even
    \end{array}\,.\label{438}
\end{align}
A set of permutation $(g_1,g_2,\ldots g_n)$ is a {\bfseries geodesic} on the permutation group if
\begin{equation}
    \Delta(g_1,g_2)+\Delta(g_2,g_3)+\ldots +\Delta(g_{n-1},g_n)=\Delta(g_1,g_n)\,.
\end{equation}
The set of permutations that are on a geodesic between $\mathbbm{1}$ and $X$ , i.e. $\Delta(\mathbbm{1},g)+\Delta(g,X)=\Delta(\mathbbm{1},X)=k-1$, is known to be in bijection with the set of \textit{Non-Crossing Partitions} of the set $[k]=\{1,2,\ldots k\}$. A $NCP$ is a set of non-empty pairwise disjoint subsets called ``blocks", such that no two blocks cross each other: consider the following permutations in $\mathbb{S}_5$ 
\begin{equation}
    g=(1)(25)(34)\quad , \quad h=(1)(24)(35)\,.
\end{equation}
These permutations can be diagrammatically drawn as
\begin{figure}[h]
    \centering
 \begin{tikzpicture}
\draw (0,0) node[anchor=south]{1} -- (1,0) node[anchor=south]{2} -- (2,0) node[anchor=south]{3} -- (3,0) node[anchor=south]{4} -- (4,0) node[anchor=south]{5}; \draw (0,0) .. controls (-0.4,-0.8) .. (0,-1) .. controls (0.4,-0.8)..  (0,0); 
\draw (2,0) .. controls (2.2,-0.5) and (2.8,-0.5) ..(3,0);
\draw (1,0) .. controls (2,-1) and (3,-1).. (4,0);
\end{tikzpicture}
\end{figure}
\begin{figure}[h]
    \centering
 \begin{tikzpicture}
\draw (0,0) node[anchor=south]{1} -- (1,0) node[anchor=south]{2} -- (2,0) node[anchor=south]{3} -- (3,0) node[anchor=south]{4} -- (4,0) node[anchor=south]{5}; \draw (0,0) .. controls (-0.4,-0.8) .. (0,-1) .. controls (0.4,-0.8)..  (0,0); 
\draw (2,0) .. controls (2.7,-0.5) and (3.3,-0.5) ..(4,0);
\draw (1,0) .. controls (1.5,-1) and (2.5,-1).. (3,0);
\end{tikzpicture}
\end{figure}
We can see that only $(1)(25)(34)$ is a non crossing pairing.
The number of NCP of $\mathbb{S}_k$ is given by the \textit{Catalan number}:
\begin{equation}
    C_k=\abs{NC(k)}=\frac{1}{k+1}\binom{2k}{k}\,.
\end{equation}
We will use this information to obtain the set of permutations that are simultaneously geodesics for the three distances \eqref{438}. Namely we are looking for the permutations such that:

\begin{equation}
    \left\{\begin{array}{ll}
    \Delta(\mathbbm{1},\tau)+\Delta(\tau,X)=k-1  \\  \Delta(\mathbbm{1},\tau)+\Delta(\tau,X^{-1})=k-1  \\ \Delta(X,\tau)+\Delta(\tau,X^{-1})=\biggl\{\begin{array}{ll}
            k-1 \,\,\,\,\, k\,\,odd \\
             k-2 \,\,\,\,\, k\,\,even
        \end{array}
    \end{array}\right.\,.
\end{equation}
 
 These conditions are equivalent to
\begin{equation}
    \begin{split}
       & \Delta(\mathbbm{1},\tau)=\left\lfloor\frac{k}{2}\right\rfloor\,,\\
       & \Delta(X,\tau)=\Delta(X^{-1},\tau)=\left\lceil\frac{k}{2}\right\rceil -1\,.
       \label{distancefromtau}
    \end{split}
\end{equation}
where $\left\lfloor\frac{k}{2}\right\rfloor$ and $\left\lceil\frac{k}{2}\right\rceil$ represent respectively the floor and ceiling functions. By solving these conditions, it is possible to prove that $\tau$ is on the geodesic only if it is a permutation corresponding to a non-crossing partition of the set $[k]$ containing only blocks of length 2 plus a single block of length 1 if $k$ is odd. We call this set \textit{Non Crossing Pairings}
$NC_2(k)$ and its cardinality is denoted by $a_k$. For even $k$ there exists a bijection $NC_2(k)\longleftrightarrow NC(\frac{k}{2})$, and for odd $k$ we have $a_k=ka_{k-1}$. Thus we can calculate cardinality in the two different cases

\begin{equation}
    a_k=\left\{ \begin{array}{ll}
       k\, C_{\frac{k-1}{2}} &\,\,\,\, odd\\
       C_{\frac{k}{2}}&\,\,\,\, even
    \end{array}\right.\,,
\end{equation}
whose limit for $k\to 1$ is given by
\begin{equation}
    \lim_{k\to 1}a_k=\left\{\begin{array}{ll}
        1\,\,\, odd\\
        \frac{8}{3\pi}\,\,\, even
    \end{array}\right.\,. \label{545}
\end{equation}
This detailed discussion of permutations properties provides a quick recap of all the mathematics we need to develop a statistical model for R\'enyi negativity. Indeed, there exists an alternative way to describe this class of states. Following the previous works on Random Spin Network \cite{Chirco:2021chk,Colafranceschi:2021acz}, we can consider a reference state $\ket{0}\in\mathcal{H}$ and consider all the states that can be obtained acting with a unitary operator. Negativity exhibits a natural internal symmetry related to the permutation group. So it is natural to consider the unitary operator acting on the reference state as a unitary representation of the group $\mathbb{S}_n$. Since we are interested in ensemble averaging for induced mixed states, we can consider a preliminary example to see how permutations arise in the calculation \cite{38}. \newline
Consider a bipartite system
\begin{equation}
    \mathcal{H}=\mathcal{H}_A\otimes\mathcal{H}_B\,,
\end{equation}
described by a density matrix $\rho$. Tracing over $B$ we have an induced mixed state. If we calculate the average over $\alpha$ copies of $\rho_A$, since trace and averaging are commuting operations, we get
\begin{equation}
\overline{\rho_A^{\otimes\alpha}}=\Tr_B\bigg[\overline{\ket{\psi}\bra{\psi}^{\otimes\alpha}}\bigg]\,.
\end{equation}
Group averaging is given by the sum over all the possible permutations acting on $\alpha$ copies of the system, so:
\begin{equation}
    \overline{\rho_A^{\otimes\alpha}}=\frac{\sum_{\tau\in S_{\alpha}}g_{\tau_{A}}\Tr\qty[g_{\tau_B}]}{\sum_{\tau\in S_{\alpha}}\Tr\qty[g_{\tau}]}\label{448}
\end{equation}
where $\tau_A$ and $\tau_B$ are permutation acting only on the subsystems $A$ and $B$. The trace of a permutation is easy to calculate since it is equal to the dimension of the Hilbert space ($d_A$ or $d_B$) to the number of the cycles $\chi(\tau)$. So the denominator becomes:
\begin{equation}
    \sum_{\tau}\Tr\qty[g_{\tau}]=\sum_{\tau}\qty(d_Ad_B)^{\chi(\tau)}
\end{equation}
This quantity can be summed exactly: the number of permutation of $\alpha$ element with $k$ cycles is given by the well known Stirling number of first kind $\qty[\begin{array}{c}
     n  \\
      k
\end{array}]$. Since we are interested in the regime of large dimensions, only the permutation that maximizes $\chi(\tau)$ will contribute at leading order. This permutation is the Identity, i.e. the only permutation of $S_{\alpha}$ with $\alpha$ cycles. Thus the denominator can be approximated to $\qty(d_Ad_B)^{\alpha}$.

\section{Mapping R\'enyi k-th Negativity to generalized Ising model}\label{mapping}

In this appendix we show the detailed calculations for the partition function $\overline{Z^{(k)}_1}$ in \eqref{z1 with permutations}. We will derive the explicit form of the actions with a particular bulk state $\ket{\zeta}$ which however is easily generalized to a generic state. We will write the state $\ket{\zeta}$ as
\begin{equation}
    \ket{\zeta} = \ket{\zeta_{\Omega}}\otimes\qty(\bigotimes_{v\in \bar{\Omega}}\ket{\zeta_v})
\end{equation}
where $\Omega \subset \dot{\gamma}$ and $\bar{\Omega} \cup \Omega \equiv \dot{\gamma}$. The intertwiner degrees of freedom in $\bar{\Omega}$ will not enter in the trace computation. To see this observe that

\begin{equation}
    \ket{\zeta_v} = \sum_{\iota} \zeta^{(v)}_\iota \ket{\{j\}_v,\iota}
\end{equation}
gives
\begin{widetext}
\begin{multline}
\ket{f_v(\zeta)}=\braket{\zeta_v}{f_v} = \sum_{\iota_1} \sum_{\{m\},\iota_2} f^{\{j\}}_{\{m\}\, \iota_2} (\tau^{(v)}_{\iota_1})^* \, \bigotimes_{i}\ket{j_i, m_i}\braket{\iota_1,\{j\}_v}{\{j\}_v,\iota_2} = \sum_{\{m\}} f(\tau)^{\{j\}}_{\{m\}}\, \bigotimes_{i}\ket{j_i, m_i}\, ,
\end{multline}
\end{widetext}
where $f(\zeta)^{\{j\}}_{\{m\}} = \sum_\iota f^{\{j\}}_{\{m\}\, \iota} (\zeta^{(v)}_{\iota})^*$. Now $\ket{f_v(\zeta)}$ are random states and with the randomization we get
\begin{equation}
\mathbb{E}\Big[\big(\ket{f_v(\zeta)}\bra{f_v(\zeta)}\big)^{\otimes k}\Big] = \frac{\qty(D_{\partial v} -1)!}{\qty(D_{\partial v}+k -1)!}\sum_{g_v \in S_k} P_{\partial v}(g_v)\, ,
\end{equation}
where
\begin{equation}
    P_{\partial v}(g_v) =  \bigotimes_{e^i_{vw} \in L} P_{v,i}(g_v) \otimes \bigotimes_{e^i_{v\bar{v}} \in \partial \gamma} P_{v,i}(g_v)
\end{equation}
and $D_{\partial v} = \prod_{i} d_{j^v_i} = \frac{D_v}{D_{\{j\}_v}}$.
Therefore we see that the bulk intertwiners that get contracted with a state $\ket{\zeta_v} \, \in \, \bar{\Omega}$ do not contribute to the trace.

Keeping this in mind, the partition function $\overline{Z^{(k)}_1}$ can be written as
\begin{widetext}
\begin{equation}\label{z1 with permutations appendix}
\overline{Z^{(k)}_1} = \mathcal{C} \Tr\Bigg\{\rho_{\zeta_\Omega}^{\otimes k}\otimes\rho_E^{\otimes k}\, \qty(\bigotimes_{v\in \Omega}\overline{\qty(\ket{f_v}\bra{f_v})^{\otimes k}}\otimes \bigotimes_{v\in \bar{\Omega}} \overline{\qty(\ket{f_v(\zeta)}\bra{f_v(\zeta)})^{\otimes k}}) \,  P_A\qty(X)\otimes P_B\qty(X^{-1})\otimes P_C\qty(\mathbbm{1})\Bigg\} \end{equation}
\end{widetext}
where $\mathcal{C} = \qty(\prod_{v\in \Omega}\frac{\qty(D_v-1)!}{\qty(D_v+k-1)!})\qty(\prod_{v\in \bar{\Omega}}\frac{\qty(D_{\partial v}-1)!}{\qty(D_{\partial v}+k-1)!})$.
This trace factorizes over a) internal edges, b) boundary spins and c) bulk intertwiners degrees of freedom. At each vertex, the sum over the permutation operators factorises according to the product structure of the single vertex Hilbert space. Thus, the computation of the trace defining the averaged partition function can be decomposed in three contributions, following \cite{Dong_21}, as follows
\begin{widetext}
\begin{itemize}
\item[a)]{{\bf Edges contribution}
\begin{eqnarray}
\Tr_E\qty[\rho_E^{\otimes k}\otimes\bigotimes_v\sum_{g_v\in S_k}P_v(g_v)]&=& \sum_{\{g_v\}}\Tr\qty[\qty(\ket{e^i_{vw}}\bra{e^i_{vw}})^{\otimes k}\otimes\qty(P_{v,i}(g_v)\otimes P_{w,i}(g_v))]= \nonumber\\ &=&\sum_{\{g_v\}}\prod_{e^i_{vw}\in E} d_{j^i_{vw}}^{-k+\chi(g_v^{-1}g_w)}=\sum_{\{g_v\}}\prod_{e^i_{vw}\in E} d_{j^i_{vw}}^{-\Delta(g_v,g_w)}\, ,
\end{eqnarray}
}
where the term $d_{j^i_{vw}}^{-k}$ comes from the normalization of the link states $\ket{e^i_{vw}}$.
\item[b)]{{\bf Boundary contributions}\\
We can rewrite the constant $\mathcal{C}$ as $\mathcal{C} = \qty[\prod_{v\in \Omega}D_v^{-k} \qty(1+O(D_v^{-1}))]\qty[\prod_{v\in\bar{\Omega}}D_{\partial v}^{-k} \qty(1+O(D_{\partial v}^{-1}))]$. Taking the term $d_{j^i_v}^{-k}$ out of $D_v^{-k}$ and $D_{\partial v}$ for the boundary contributions, for $\partial A$ we get 
\begin{multline}
\qty(\prod_{v\in A} d_{j^i_v}^{-k})\Tr_{\partial A}\qty[\bigotimes_{e^{i}_{v\bar{v}}\in \partial A}\qty(\sum_{g_v\in S_k} P_v(g_v))\otimes P_A(X)]= \qty(\prod_{v\in A} d_{j^i_v}^{-k})\prod_{e^i_{v\bar{v}}\in\partial A}\sum_{g_v \in S_k} \Tr_{\partial A}\qty[P_{v,i}(g_v)\otimes P_A(X)] =\\=\sum_{\{g_v\}}\prod_{e^i_{v\bar{v}}\in\partial A} d_{j^i_v}^{-k+\chi\qty(g_v^{-1}X)}=\sum_{\{g_v\}} \prod_{\partial A}d_{j^i_v}^{-\Delta(g_v,X)}\, .
\end{multline}
Similarly, for $\partial B$ and $\partial C$, we have
\begin{equation}
    \qty(\prod_{v\in B} d_{j^i_v}^{-k})\Tr_{\partial B}\qty[\bigotimes_{e^i_{v\bar{v}}\in \partial B}\qty(\sum_{g_v\in S_k}P_v(g_v))\otimes P_B(X^{-1})]=\sum_{\{g_v\}} \prod_{\partial B}d_{j^i_v}^{-\Delta(g_v,X^{-1})}
\end{equation}
\begin{equation}
    \qty(\prod_{v\in C} d_{j^i_v}^{-k})\Tr_{\partial C}\qty[\bigotimes_{e^{i}_{v\bar{v}}\in \partial C}\qty(\sum_{g_v\in S_k}P_v(g_v))\otimes P_C(\mathbbm{1})]=\sum_{\{g_v\}} \prod_{\partial C}d_{j^i_v}^{-\Delta(g_v,\mathbbm{1})}\, .
\end{equation}
}
\item[c)]{{\bf Bulk contribution}\\
The contribution of the bulk to the partition function is
\begin{equation}
    \sum_{\{g_v\}}\Tr_{\Omega} \Bigg\{\ketbra{\zeta_\Omega}{\zeta_\Omega}^{\otimes k}\qty(\bigotimes_{v\in \Omega} P_{v,0}(g_v))\Bigg\}\, ,
\end{equation}
which depends on the form chosen for the bulk state $\ket{\zeta_\Omega}$.
}
\end{itemize}
Putting these terms all together, we can write $\overline{Z_1^{(k)}}$ as
\begin{multline}
\overline{Z^{(k)}_1}=\mathcal{C}' \sum_{\{g_v\}}\Bigg\{\qty(\prod_{e^i_{vw} \in E}d_{j^i_{vw}}^{-\Delta(g_v,g_w)}) \qty(\prod_{e^i_{v\bar{v}} \in \partial A} d_{j_i^v}^{-\Delta(g_v,X)})\qty(\prod_{e^i_{v\bar{v}} \in \partial B} d_{j_i^v}^{-\Delta(g_v,X^{-1})})\qty(\prod_{e^i_{v\bar{v}} \in \partial C} d_{j_i^v}^{\Delta(g_v,\mathbbm{1})})\cdot\\ \cdot\Tr_{\Omega} \Bigg[\ketbra{\zeta_\Omega}{\zeta_\Omega}^{\otimes k}\qty(\bigotimes_{v\in \Omega} P_{v,0}(g_v))\Bigg]\Biggr\}\, ,
\label{partition function order k}
\end{multline}
where $\mathcal{C}'$ is another constant whose value is not relevant since it will be simplified. Indeed, $\overline{Z^{(k)}_0}$ will have the same form of \eqref{partition function order k} but with $X$ and $X^{-1}$ replaced by $\mathbbm{1}$.
\end{widetext}
Finally, we can write these partition functions as
\begin{equation}
     \overline{Z^{(k)}_{1/0}} = \sum_{\{g_v\}} e^{-A^{(k)}_{1/0}\big[\{g_v\}\big]}\, ,
\end{equation}\\ \\
where\\
\begin{multline}
    A^{(k)}_1\big[\{g_v\}\big] = \sum_{e^i_{vw} \in E}\Delta(g_v,g_w)\log d_{j^i_{vw}} + \\
    +\sum_{e^i_{v\bar{v}} \in A} \Delta(g_v, X)\log d_{j^i_v} +  \sum_{e^i_{v\bar{v}} \in B} \Delta(g_v, X^{-1})\log d_{j^i_v} +\\+ \sum_{e^i_{v\bar{v}} \in C} \Delta(g_v, \mathbbm{1})\log d_{j^i_v} + A(\zeta_\Omega) + \xi 
   \end{multline}
   and 
   \begin{multline}
   A^{(k)}_0\big[\{g_v\}\big] = \sum_{e^i_{vw} \in E}\Delta(g_v,g_w)\log d_{j^i_{vw}} + \\   + \sum_{e^i_{v\bar{v}} \in \partial \gamma} \Delta(g_v, \mathbbm{1})\log d_{j^i_v} +  A(\zeta_\Omega) + \xi \, ,
\label{481}
\end{multline}
$\xi$ being a constant term and

\begin{equation}\label{bulk contribution appendix}
    A(\zeta_\Omega) = -\log \qty[\Tr_{\Omega} \Bigg\{\ketbra{\zeta_\Omega}{\zeta_\Omega}^{\otimes k}\qty(\bigotimes_{v\in \Omega} P_{v,0}(g_v))\Bigg\}]
\end{equation}
is the bulk state contribution.

\section{Statistical modelling of $\mathcal{N}_3$ using spins}\label{spinn}
\subsection{Setup}
We here show an alternative way of dealing with the computation problem for the third order negativity which is more similar to what some of the authors have previously done \cite{Colafranceschi:2021acz}. To do so, we will parameterize the symmetric group $\mathbb{S}_3$ using swaps operator, namely

\begin{equation}
    \mathbb{S}_3 = \Big\{\mathbbm{1}, S_{12}, S_{13}, S_{23}, S_{12}S_{13}, S_{12}S_{23}\Big\}\, .
\end{equation}
Here, the combination $S_{12}S_{13}$ is the cyclic permutation $X$ while $S_{12}S_{23}\equiv X^{-1}$. The boundary of the graph will always be divided in three regions: $C$ will be traced to obtain a mixed state and we will compute the correlations between $A$ and $B$ (over the latter we perform the partial transpose).\\
We want to compute the third negativity momentum of $\rho_{AB}\qty(\zeta) = \Tr_{C}\{\rho_{\partial \gamma}(\zeta)\}$, namely

\begin{equation}
    m_3 = \Tr_{A,B} \bigg\{\mathbb{E}\Big[\big(\rho_{AB}(\zeta)^{T_B}\big)^3\Big] \bigg\}\, .
\end{equation}
The bulk state $\zeta$ is written in the same way as before

\begin{equation}
    \ket{\zeta} = \ket{\zeta_{\Omega}}\otimes\qty(\bigotimes_{v\in \bar{\Omega}}\ket{\zeta_v})\, ,
\end{equation}
thus the intertwiner degrees of freedom that are contracted with a product state will not give a contribution as before. Dividing the vertices into the two regions $\Omega$ and $\bar{\Omega}$ we have
from Schur's lemma

\begin{eqnarray}
\mathbb{E}\Big[(\ket{f_v}\bra{f_v})^{\otimes3}\Big] &=& \frac{1}{D_v (D_v +1)(D_v +2)} \sum_{\pi \in \mathbb{S}_3} S_\pi^v\,, \nonumber \\ \\
\mathbb{E}\Big[(\ket{f_v}\bra{f_v})^{\otimes3}\Big] &=& \frac{1}{D_{\partial v} (D_{\partial v} +1)(D_{\partial v} +2)} \sum_{\pi \in \mathbb{S}_3} S_\pi^{\partial v}\,.\nonumber
\end{eqnarray}
For $v\in \Omega$ and $v \in \bar{\Omega}$ respectively. In the above $D_v = \prod_{i} d_{j^v_i} D_{\vec{j}_v}$ and $D_{\partial v} = \prod_{i} d_{j^v_i}$ are the dimensions of the vertices Hilbert spaces in $\Omega$ and $\bar{\Omega}$ respectively. $S_\pi^v$ ($S_\pi^{\partial v}$) is a swap operator acting on the tensor product of three copies of the Hilbert space of the vertex $v\in \Omega$ ($v\in \bar{\Omega}$).\\
The third momentum can therefore be written as
\begin{equation}
    m_3 = \frac{Z_1}{Z_0}\,,
\end{equation}
with
\begin{multline}
    Z_1 = \mathcal{C} \Tr \bigg\{\rho_{\zeta_\Omega}^{\otimes 3}\otimes\rho_E^{\otimes3} \Big(\bigotimes_{v\in \Omega} \sum_{\pi \in \mathbb{S}_3} S_{\pi}^v\Big)\cdot \\ \cdot \Big(\bigotimes_{v\in \bar{\Omega}} \sum_{\pi \in \mathbb{S}_3} S_{\pi}^{\partial v}\Big) S_{12}^{A}S_{13}^{A}S_{12}^{B}S_{23}^{B}\bigg\}
    \label{Partition function 1 appendix}
\end{multline}
and $Z_0=\Tr \Big\{\mathbb{E}\big[\rho^{\otimes3}\big]\Big\}$. In the above, $\rho_\zeta$ gives the contribution from the bulk insertions and, for bulk states of the form \eqref{483}, it will give the same contribution of the $\rho_E$ term, just computed, in general, on a different set of vertices, $\rho_{\zeta_\Omega} = \bigotimes_{\langle vw\rangle} \ket{e^\iota_{vw}}\bra{e^\iota_{vw}}$.\\
Clearly, for the swap operators on the vertices we have a factorization over the degrees of freedom

\begin{equation}
    S^v_{\pi}= S^{v,0}_{\pi}\otimes\bigotimes_{e^i_{vw} \in E} S^{v,i}_{\pi} \otimes \bigotimes_{e^i_{v\bar{v}} \in \partial \gamma} S^{v,i}_{\pi}
\end{equation}
and

\begin{equation}
    S^{\partial v}_{\pi}=\bigotimes_{e^i_{vw} \in E} S^{v,i}_{\pi} \otimes \bigotimes_{e^i_{v\bar{v}} \in \partial \gamma} S^{v,i}_{\pi}\,.
\end{equation}

\subsection{Explicit computations}
We now assign a spin variable to each of the three swap operators, namely $\sigma_1$ to $S_{12}$, $\sigma_2$ to $S_{13}$ and $\sigma_3$ to $S_{23}$, so that we can rewrite the trace in \eqref{Partition function 1 appendix} as
\begin{widetext}
\begin{eqnarray}
    Z_1 &=& \mathcal{C}\,\sum_{\{\vec{\sigma}'\} } \Tr \Bigg\{\rho_{\zeta_\Omega}^{\otimes 3}\otimes\rho_E^{\otimes3} \left(\bigotimes_{v\in\Omega : \sigma_{1}^v = -1}  S_{12}^v\right)\left(\bigotimes_{v\in\Omega : \sigma_{2}^v = -1}  S_{13}^v\right) 
    \left(\bigotimes_{v\in\Omega : \sigma_{3}^v = -1}  S_{23}^v\right)\cdot \nonumber \\&\cdot&\left(\bigotimes_{v\in\bar{\Omega} : \sigma_{1}^v = -1}  S_{12}^v\right)\left(\bigotimes_{v\in\bar{\Omega} : \sigma_{2}^v = -1}  S_{13}^v\right) 
    \left(\bigotimes_{v\in\bar{\Omega} : \sigma_{3}^v = -1}  S_{23}^v\right) S_{12}^{A}S_{13}^{A}S_{12}^{B}S_{23}^{B}\Bigg\}\, ,
    \label{Partition function 1 after replica trick}
\end{eqnarray}
where
\begin{equation}\label{prefactor dimension}
  \mathcal{C}=\qty[\prod_{v\in \Omega} D_v^{-3}\frac{1}{\qty(1+D_{v}^{-1})\qty(1+2D_{v}^{-1})}]\qty[\prod_{v\in \Omega} D_{\partial v}^{-3}\frac{1}{\qty(1+D_{\partial v}^{-1})\qty(1+2D_{\partial v}^{-1})}]
\end{equation}
\end{widetext}
and $\{\vec{\sigma}'\} \equiv \{\vec{\sigma}_1,\vec{\sigma}_2,\vec{\sigma}_3\}$ stands for all the configurations of the three spins attached to each vertex, without $(1,-1,-1)$ and $(-1,-1,-1)$ corresponding to $S_{13}S_{23} \equiv S_{12}S_{13}$ and $S_{12}S_{13}S_{23} \equiv S_{13}$ that are not to be considered in \eqref{Partition function 1 appendix}. \\
This trace can be factorized with respect to the different degrees of freedom, therefore
\begin{widetext}
\begin{eqnarray}
m_3 &=& \mathcal{C}\,\sum_{\{\vec{\sigma}\}'}\Bigg\{\Tr_E \Bigg[\rho_E^{\otimes3}\left(\bigotimes_{e^i_{vw} \in E : \sigma_{1}^v = -1}  S_{12}^{v,i}\right)\left(\bigotimes_{e^i_{vw} \in E : \sigma_{2}^v = -1}  S_{13}^{v,i}\right) \cdot \left(\bigotimes_{e^i_{vw} \in E : \sigma_{3}^v = -1}  S_{23}^{v,i}\right)\Bigg]\nonumber \\ &\cdot&  \Tr_{\Omega} \Bigg[\rho_{\zeta_\Omega}^{\otimes 3}\left(\bigotimes_{\langle vw\rangle : \sigma_{1}^v = -1}  S_{12}^{v,0}\right)\left(\bigotimes_{\langle vw\rangle : \sigma_{2}^v = -1}  S_{13}^{v,0}\right) \cdot 
\left(\bigotimes_{\langle vw\rangle : \sigma_{3}^v = -1}  S_{23}^{v,0}\right)\Bigg] \cdot \\ &\cdot& \Tr_{\partial \gamma} \Bigg[S_{12}^{A}S_{13}^{A}S_{12}^{B}S_{23}^{B} \otimes \left(\bigotimes_{e^i \in \partial \gamma : \sigma_{1}^v = -1}  S_{12}^{v,i}\right)\left(\bigotimes_{e^i \in \partial \gamma : \sigma_{2}^v = -1}  S_{13}^{v,i}\right) \cdot \left(\bigotimes_{e^i \in \partial \gamma : \sigma_{3}^v = -1}  S_{23}^{v,i}\right)\Bigg]\Bigg\}\nonumber \, .
\end{eqnarray}
Moreover, the three traces above factorize over all the degrees of freedom, therefore
\begin{equation}
\Tr_E = \prod_{e^i_{vw} \in E}\Tr_{e^i_{vw}}\;\,,\;\,
\Tr_{\Omega} = \prod_{\langle vw \rangle}\Tr_{e^\iota_{vw}}\;\,,\;\, 
\Tr_{\partial \gamma} = \prod_{e^i_{v\bar{v}} \in \partial \gamma}\Tr_{e^i_{v\bar{v}}}\,.
\end{equation}
\end{widetext}

\subsubsection{Bulk links and intertwiners}
For the trace over $E$ we have to compute

\begin{eqnarray}
&\prod_{e^i_{vw} \in E} \Tr \Bigg\{\big(\ket{e^i_{vw}}\bra{e^i_{vw}}\big)^{\otimes3} \cdot \\ 
&\prod_{ij=12,13,23} \bigg[\frac{1}{2}\big(1+\sigma_{ij}^v\big)\mathbbm{1}+\frac{1}{2}\big(1-\sigma_{ij}^v\big)S_{ij}^{v,i}\bigg] \otimes\nonumber \\ &\otimes \prod_{hk=12,13,23} \bigg[\frac{1}{2}\big(1+\sigma_{hk}^w\big)\mathbbm{1}+\frac{1}{2}\big(1-\sigma_{hk}^w\big)S_{hk}^{w,i}\bigg]\Bigg\}  \nonumber\, .
\end{eqnarray}
By computing this trace over all the possible configurations $\{\vec{\sigma}\}'$ we get that there are three possible results, namely $d_{j_{vw}^i}^n$ with $n=1,2,3$.\\
To model this trace in terms of a spin action we write the result as $d_{j_{vw}^i}^{K(\vec{\sigma}^v,\,\vec{\sigma}^w)}$ with $K(\vec{\sigma}^v,\vec{\sigma}^w)$ being the most general interaction term between the six spins involved in the trace. This will be made up by a constant term, linear terms in all the spins, pairwise interactions and so on up to six order interaction terms. If we enforce the symmetry between the exchange of the two vertices, we have $64$ parameters in this unknown function that can be determined by enforcing that this function takes on the values ($1,2,3$) we computed before on the configurations $\{\vec{\sigma}\}'$ and that it is $0$ on all the other configurations ($64$ equations in total). Doing so we get

\begin{widetext}
\begin{eqnarray}
    K(\vec{\sigma}^v,\vec{\sigma}^w) &=& \frac{1}{32} \Big\{-63+\sigma^v_1 \sigma^w_1(5+3\sigma^w_2)+ \sigma^w_3\big[\sigma^v_1\sigma^w_1(3+\sigma^w_2 - 11\sigma^w_2)\big]+ 11(\sigma^w_2+\sigma^w_3) + \sigma^v_3\big[11+\sigma^w_2 + 9\sigma^w_3 +\nonumber\\ &-&\sigma^w_2\sigma^w_3 + \sigma^v_1\sigma^w_1(3+5\sigma^w_3 -3\sigma^w_2-\sigma^w_2\sigma^w_3)\big] + \sigma^v_2\big[11+9\sigma^w_2+\sigma^w_3-\sigma^w_2\sigma^w_3 + \sigma^v_1 \sigma^w_1(3+5\sigma^w_2 - 3\sigma^w_3 +\nonumber \\&-& \sigma^w_2\sigma^w_3)\big] -\sigma^v_3\big(11+\sigma^w_2+\sigma^w_3 -9\sigma^w_2\sigma^w_3 +\sigma^v_1\sigma^w_1(\sigma^w_2 -\ \sigma^w_3 +3\sigma^w_2\sigma^w_3-1)\big)\big]\Big\}\, ,
\end{eqnarray}
\end{widetext}
therefore the trace over the bulk links gives

\begin{equation}
     \prod_{e^i_{vw} \in L} d_{j^i_{vw}}^{K(\vec{\sigma}^v,\vec{\sigma}^w)}\, .
\end{equation}

A similar result holds for the bulk contribution from the intertwiners. Indeed the trace that should be computed is the same, the only difference being the set over which the links are considered. Thus for the intertwiners we have

\begin{equation}
     \prod_{\langle vw\rangle \in \Omega} D_{\vec{j}_{vw}}^{K(\vec{\sigma}^v,\vec{\sigma}^w)}\, .
\end{equation}

\subsubsection{Boundary edges}
The other trace is given by

\begin{multline}
    \prod_{e^i_{v\bar{v}} \in \partial \gamma}\Tr \Bigg\{S^{A,i}_{12}S^{A,i}_{13}S^{B,i}_{12}S^{B,i}_{23} \quad \otimes \\ \otimes  \left(\bigotimes_{e^i \in \partial \gamma : \sigma_{1}^v = -1}  S_{12}^{v,i}\right)\left(\bigotimes_{e^i \in \partial \gamma : \sigma_{2}^v = -1}  S_{13}^{v,i}\right)\\ \quad \left(\bigotimes_{e^i \in \partial \gamma : \sigma_{2}^v = -1}  S_{23}^{v,i}\right)\Bigg\}\, ,
\end{multline}
where the operators $S^{A,i}$ and $S^{b,i}$ act on the Hilbert space of the virtual vertex $\bar{v}$.
We can model this trace in terms of spin actions by introducing a pinning field $\vec{\mu}$ which model the presence or absence of the the additional swap operators coming from the replica trick. In particular, we have
\begin{equation}
    \vec{\mu}^v =\left\{
\begin{array}{ll}
 (1,1,1) \;\, , \;\, v \in C\\(-1,-1,1) \;\, , \;\, v \in A \\ (-1,1-,1) \;\, , \;\, v \in B
\end{array} \right.\, .
\label{Configurations mu}
\end{equation}

To compute the action for the boundary edges, we follow the same steps as before. Indeed, we can imagine that the boundary edges link the boundary vertices to virtual vertices on which the pinning fields are assigned, propagating on the link. Therefore the conceptual framework is the same of the bulk links calculation in which the interaction occurs between two vertices. Thus, we compute the trace over all possible configurations of the spins $\vec{\sigma}$ and the pinning fields $\vec{\mu}$, obtaining as results $d_{j_i^v}^n$ with $n=1,2,3$; then, we write the most general interaction term $G(\vec{\sigma},\vec{\mu})$ and we enforce that this function vanish on the spin configurations $\vec{\sigma} = (1,-1,-1),(-1,-1,-1)$ and on the pinning fields configurations that are not in \eqref{Configurations mu} and that it gives the results $1,2,3$ that we obtain on the other configurations. Subtracting the constant $3$ coming from \eqref{prefactor dimension} we get
\begin{widetext}
\begin{eqnarray}
    G(\vec{\sigma}^v,\vec{\mu}^v) &=& \frac{1}{64}\bigg\{-159+11 (\mu^v_3+\sigma^v_2 +\sigma^v_3) + \sigma^v_1 + 3 \mu^v_3\sigma^v_1 + \mu^v_3\sigma^v_2-\sigma^v_1\sigma^v_2+ 5 \mu^v_3\sigma^v_1\sigma^v_2 \nonumber \\ &-& \sigma^v_3\Big[\sigma^v_1 + 11 \sigma^v_2 + 3 \sigma^v_1 \sigma^v_2 + \mu^v_3(-9+\sigma^v_2+3\sigma^v_1 + \sigma^v_1 \sigma^v_2)\Big]+ \mu^v_2 \Big[11+9\sigma^v_2 + \nonumber \\ &+&\sigma^v_3 -\sigma^v_2\sigma^v_3+\sigma^v_1(3+5\sigma^v_3-3\sigma^v_2 -\sigma^v_2 \sigma^v_3)+\mu^v_3(-11 -\sigma^v_2-\sigma^v_3+ \nonumber \\ &+& 9\sigma^v_2\sigma^v_3 + \sigma^v_1(5+3\sigma^v_2 +3\sigma^v_3+\sigma^v_2\sigma^v_3)\Big]-\mu^v_1\Big[-11 + 5\sigma^v_1 - \sigma^v_2+3\sigma^v_1\sigma^v_2+ \nonumber \\ &+& \sigma^v_3(-1+9\sigma^v_2+3\sigma^v_1+\sigma^v_1\sigma^v_2)+\mu^v_3(11+9\sigma^v_2+\sigma^v_3-\sigma^v_2\sigma^v_3+\sigma^v_1(3+ \nonumber \\ &+& 5\sigma^v_3-3\sigma^v_2-\sigma^v_2\sigma^v_3))+\mu^v_2(11+3\sigma^v_1 +\sigma^v_2+5\sigma^v_1\sigma^v_2-\sigma^v_3(-9+\sigma^v_2+ \nonumber \\ &+& 3\sigma^v_1+\sigma^v_1\sigma^v_2))+\mu^v_3(11(3+\sigma^v_2+\sigma^v_3-\sigma^v_2\sigma^v_3))-\sigma^v_1(-1+\sigma^v_2) + \nonumber \\ &+& \sigma^v_3+3\sigma^v_2\sigma^v_3\Big]\bigg\}\, .
\end{eqnarray}
To simplify this cumbersome formula, we write it as the sum over the three regions $A$, $B$ and $C$ and in which our boundary graph is divided. In particular we have $\vec{\mu}=(-1,-1,1)$, $\vec{\mu} = (-1,1,-1)$ and $\vec{\mu} = (1,1,1)$ respectively for the three regions. Doing so we obtain
\begin{equation}
     G\qty(\vec{\sigma}^v,\vec{\mu}^v) = G_A\qty(\vec{\sigma}^v) + G_B\qty(\vec{\sigma}^v)+G_C\qty(\vec{\sigma}^v)\, ,
\end{equation}
where
\begin{eqnarray}
    G_A(\vec{\sigma}^v) &=& \frac{1}{8}\Big\{-13+\sigma^v_2+5\sigma^v_3\qty(1-\sigma^v_2)+\sigma^v_1\big[\sigma^v_2-1-\sigma^v_3\qty(3+\sigma^v_2)\big] \Big\}\, , \\
     G_B(\vec{\sigma}^v) &=& \frac{1}{8}\Big\{-13+\sigma^v_3 +5\sigma^v_2\qty(1-\sigma^v_3) +\sigma^v_1\big[\sigma^v_3-1-\sigma^v_2(3+\sigma^v_3)\big]\Big\}\,, \\
       G_C(\vec{\sigma}^v) &=& \frac{1}{8}\Big\{-13+\sigma^v_1(3+\sigma^v_2+\sigma^v_3) + 5\sigma^v_2+\sigma^v_3 \big[5-\sigma^v_2(1+\sigma^v_1)\big]\Big\}\, .
\end{eqnarray}
\end{widetext}
Therefore the trace over the boundary half-links gives
\begin{equation}
     \prod_{e^i_{v\bar{v}} \in \partial \gamma} d_{j_i^v}^{G(\vec{\sigma}^v,\vec{\mu}^v)}\, ,
\end{equation}
with $G(\vec{\sigma}^v,\vec{\mu}) =  G_A(\vec{\sigma}^v)+ G_B(\vec{\sigma}^v)+ G_C(\vec{\sigma}^v)$\,.
\subsection{The action}

We now write the result as

\begin{equation}
    \log m_3 = \log Z_1 - \log Z_0 = F_0 -F_1\,,
\end{equation}
where $F_i$ are the free energies that we write as

\begin{equation}
    F_i = - \log Z_i = - \log \left[\sum_{\{\vec{\sigma}\}} e^{-A_i(\vec{\sigma})}\right]\,,
\end{equation}
with $A_i(\vec{\sigma})$ given by $-$ the logarithm of the trace computed in the previous subsection. In particular we have

\begin{multline}\label{Ising spin action k3}
    -A_1(\vec{\sigma}) =  \sum_{\langle vw\rangle \in \Omega} K(\vec{\sigma}^v,\vec{\sigma}^w) \log D_{\vec{j}_{vw}} +\\ + \sum_{v \in \partial \gamma} G(\vec{\sigma}^v, \vec{\mu}^v) \log d_{j^i_v} + \\ + \sum_{e^i_{vw} \in E} K(\vec{\sigma}^v,\vec{\sigma}^w) \log d_{j^i_{vw}} -\xi 
\end{multline}
and $A_0(\vec{\sigma})$ has the same functional form but with all the pinning fields set to $1$, $\vec{\nu} = (1,1,1)$. $\xi$ is just a constant whose value is not relevant, since in the large spin limit the dominant configuration for $Z_1$ ($Z_0$) is the one that minimizes the action $A_1$ ($A_0$), thus we have $\log m_3 = A_1-A_0$. In computing this difference, $\xi$ cancels out. \par
In the case of homogeneous spin network, all the above dimensions are the same, $D_{\vec{j}_{vw}} = d_{j^i_{vw}}=d_{j^i_v} = d = 2j+1$. Therefore we can rewrite the action as

\begin{equation}
    A_1\qty(\vec{\sigma}) = \beta\, \mathcal{H}\qty(\vec{\sigma},\vec{\mu})\,,
\end{equation}
with $\beta \vcentcolon= \log d$. Since we are in the large $d$ limit, namely the low temperature regime, the dominant will be the one that minimizes the Hamiltonian $\mathcal{H}$, while $A_0 = 0$. We show this for the configurations that minimize the Hamiltonian for the two vertices and the three vertices cases analyzed in the main text.

For the two vertices case we have that the configuration that minimizes the Hamiltonian is the one where the two vertices are coloured with the same non crossing pairing. For the case $k=3$ the non crossing pairings are given by the swap operators $S_{13}$, $S_{12}$ and $S_{23}$. These correspond to the spin configurations
\begin{eqnarray}
    g_v=g_w=S_{12}&:& \quad \vec{\sigma_v}=\vec{\sigma_w}=(-1,1,1) \nonumber\,,\\
    g_v=g_w=S_{13}&:& \quad \vec{\sigma_v}=\vec{\sigma_w}=(1,-1,1)\,,\\
    g_v=g_w=S_{23}&:& \quad \vec{\sigma_v}=\vec{\sigma_w}=(1,1,-1)\nonumber\,.
\end{eqnarray}
With this in mind, it is possible to easily compute \eqref{Ising spin action k3} to obtain the value of the Hamiltonian in these three configurations. These give the same value of $\mathcal{H}=6$ both with and without the intertwiner link.

For the three vertices case the minimal configuration still corresponds to the one where the three vertices are all coloured with the same non crossing pairing (one of the swap operators) and the value of the Hamiltonian is easily obtained from \eqref{Ising spin action k3} to be $\mathcal{H}=8$ both with and without the non local intertwiner link.

\section{Computation of k-th order Hamiltonian}
\label{Hamiltonian appendix}
The Hamiltonian associated to the Ising model can be written as 
\begin{eqnarray}
    \mathcal{H}_k&=&\sum_{e^i_{vw}\in L}\Delta(g_v,g_w)+\sum_{\partial A}\Delta(g_v,X)+\\ &+&\sum_{\partial B}\Delta(g_v,X^{-1})
    +\sum_{\partial C}\Delta(g_v,\mathbbm{1})\,,
\end{eqnarray}
for each order $k$. 
Inserting bulk correlations, $\mathcal{H}_k$ exhibits a corrective term
\begin{multline}
     \mathcal{H}_k^c=\sum_{e^i_{vw}\in L}\Delta(g_v,g_w)+\sum_{\partial A}\Delta(g_v,X)+\\+\sum_{\partial B}\Delta(g_v,X^{-1})+\sum_{\partial C}\Delta(g_v,\mathbbm{1})+\sum_{\langle vw\rangle\in\Omega} \Delta(g_v,g_w)\,.
\end{multline}
We can consider the generic boundary tripartition examined in \cref{two vertices graph} and \cref{three vertices graph} for the case of a two and three vertices graph, and compute the value of the $k-th$ Hamiltonian for different configurations of the domain walls associated to the Ising model.  

\subsection{Hamiltonian of the two vertices graph}

\begin{figure}[H]
    \centering
\begin{tikzpicture}
\tikzset{
a/.style={scale=.5, circle,minimum size=.1cm,fill=blue,draw=blue},
b/.style={scale=.5, circle,minimum size=.1cm,fill=red,draw=red},
c/.style={scale=.5, circle,minimum size=.1cm,fill=white!20,draw=gray},
    v/.style={scale=.5, circle,minimum size=.1cm,fill=black,draw}}

\node[v] (0) at (1,0){} ;
\node[v] (1) at (3,0){} ;
\node[a] (2) at (0,0){} ;
\node[a] (3) at (0.6,0.8){} ;
\node[c] (4) at (0.6,-0.8){};
\node[c] (5) at (3.4,0.8){} ;
\node[b] (6) at (4,0){} ;
\node[b] (7) at (3.4,-0.8){} ;

\foreach \from/\to in {0/1, 2/0,3/0,4/0,5/1,6/1,7/1}
    \draw (\from) -- (\to);
  
\draw (1.1,0) node[anchor=north west]{$v$};
\draw (2.9,0) node[anchor=north east]{$w$};

\draw (2.8,1.1) node[above]{C} ; 
\draw (1.4,-1.1) node[below]{C} ; 
\draw[blue] (-0.2,0.8)node[anchor=east]{A} ; 
\draw[red] (4.2,-0.6) node[anchor=west]{B} ;
 \draw[dotted] (2,0) ellipse (54 pt and 28 pt) ;
\end{tikzpicture}
\end{figure}

\subsubsection*{Order k=2 Hamiltonians} This case has no particular interest, because of the trivial features of $S_2$, such as $X=X^{-1}$ and NCP given by only one permutation, whose domain is actually the same of the second configuration. More interesting results can be found investigating higher orders of negativity, since $X\not=X^{-1}$ if $k\not=2$. In this case the value of the minimal Hamiltonian is equal to the number of links that cross the domain wall. For higher orders such equality is lost, but a proportionality still holds, thus allowing us to find a formula that directly relates the value of the Hamiltonian to the number of links crossing the domain wall.
\subsubsection*{Order k=3 Hamiltonians}
The permutation group $S_3$ has $3!=6$ elements
\begin{multline}
    S_3=\{\mathbbm{1}=(1)(2)(3), S_{12}=(12)(3), S_{13}=(13)(2),\\ S_{23}=(1)(23),X=(123),X^{-1}=(321)\}\,.
\end{multline}
Cyclic and anticyclic permutations are now different. Using the relations \eqref{438} we can write some preliminary calculations on the distances:
\begin{align}
    \Delta(\mathbbm{1},X)&=k-1=2\,, \\
    \Delta(\mathbbm{1},X^{-1})&=k-1=2\,, \\
    \Delta(X,X^{-1})&=k-1=2\quad(\text{odd }  k)\,.
    \label{593}
\end{align}
We can now compute the Hamiltonian for some configurations emerging from the six possible ``spins" that can be attached to vertices.
To lighten the notation, we will only write the non-vanishing terms of the Hamiltonian.

\begin{enumerate}
 \item       $g_v=g_w=\mathbbm{1}$,  $\rightarrow$ $v,w\in\,C$,
 \begin{equation}
     \mathcal{H}_3=2\Delta(\mathbbm{1},X)+2\Delta(\mathbbm{1},X^{-1})=8\,.
 \end{equation}

\begin{figure}[H]
    \centering
	\includegraphics[width=0.35\textwidth]{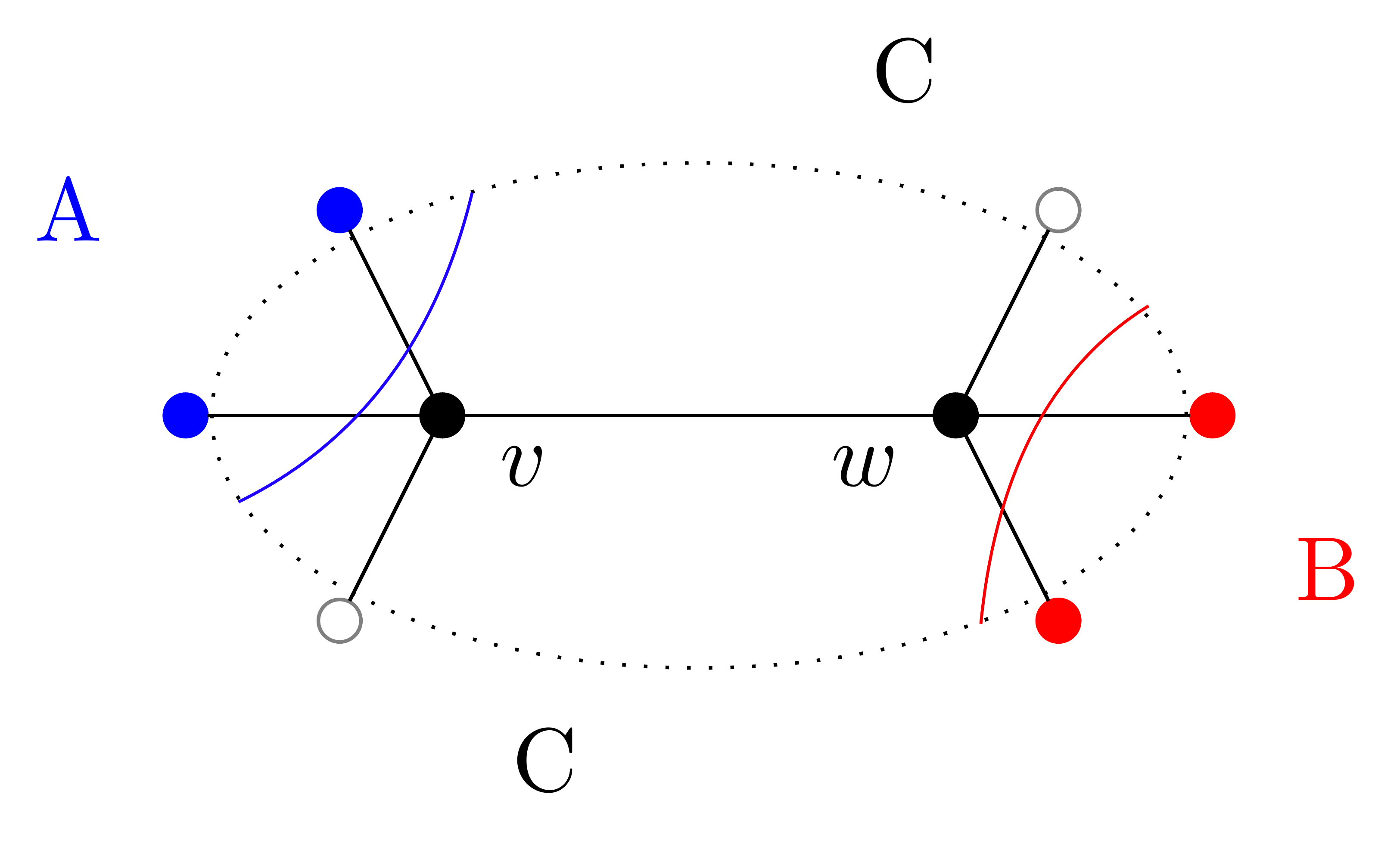}
\end{figure} 
\item      $g_v=X$,$g_w=X^{-1}$ $\rightarrow$ $v\in\,A$, $w\in\,B$,
 \begin{equation}\label{2conf2nc}
     \mathcal{H}_3=\Delta(X,X^{-1})+\Delta(X,\mathbbm{1})+\Delta(X^{-1},\mathbbm{1})=6\,.
 \end{equation}

  \begin{figure}[H]
    \centering
	\includegraphics[width=0.35\textwidth]{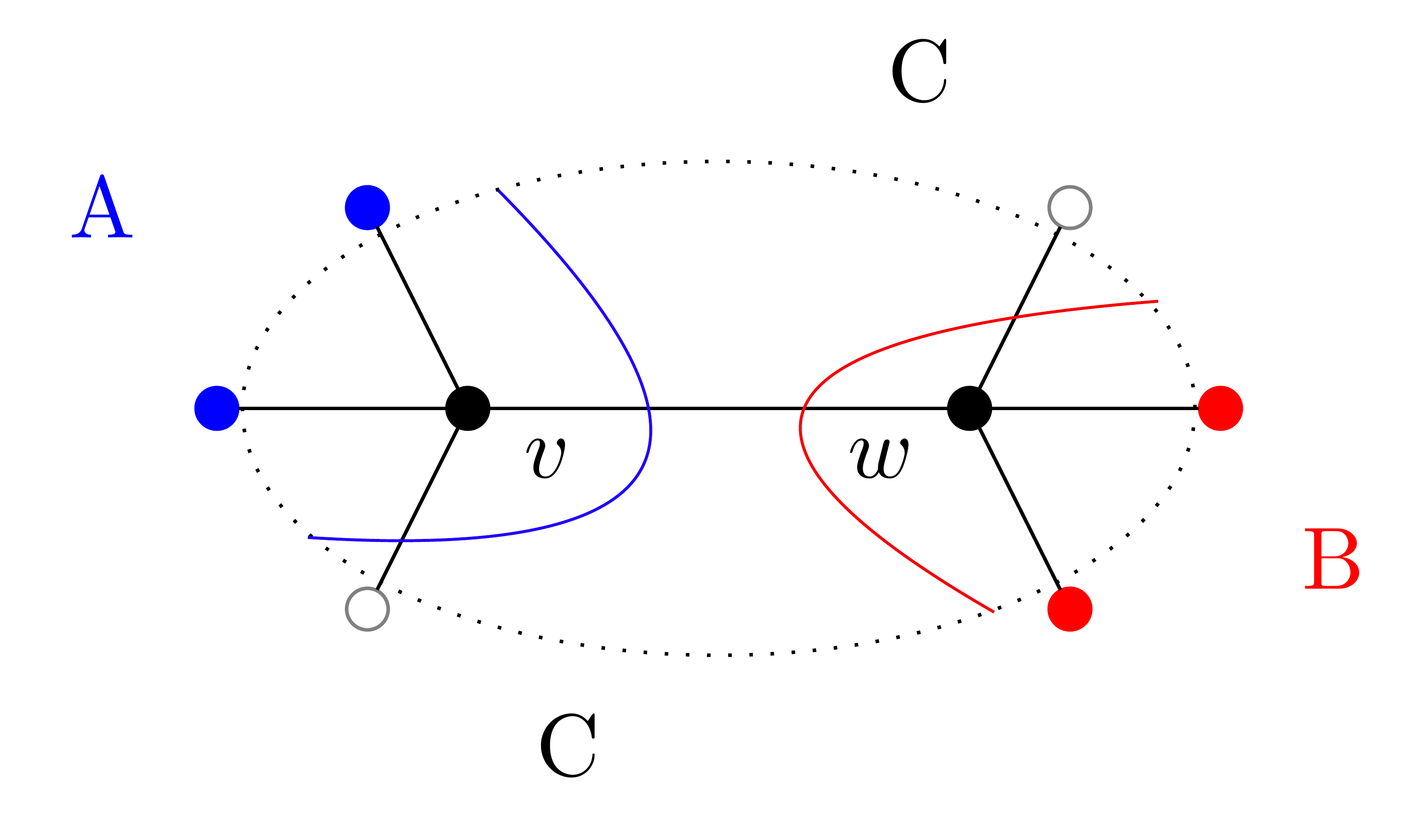}
\end{figure} 

\item       $g_v=X$,$g_w=\mathbbm{1}$  $\rightarrow$ $v\in\,A$, $w\in\,C$,
 \begin{equation}
     \mathcal{H}_3=\Delta(X,\mathbbm{1})+2\Delta(\mathbbm{1},X^{-1})+\Delta(X,\mathbbm{1})=8\,.
 \end{equation}

\begin{figure}[H]
    \centering
	\includegraphics[width=0.35\textwidth]{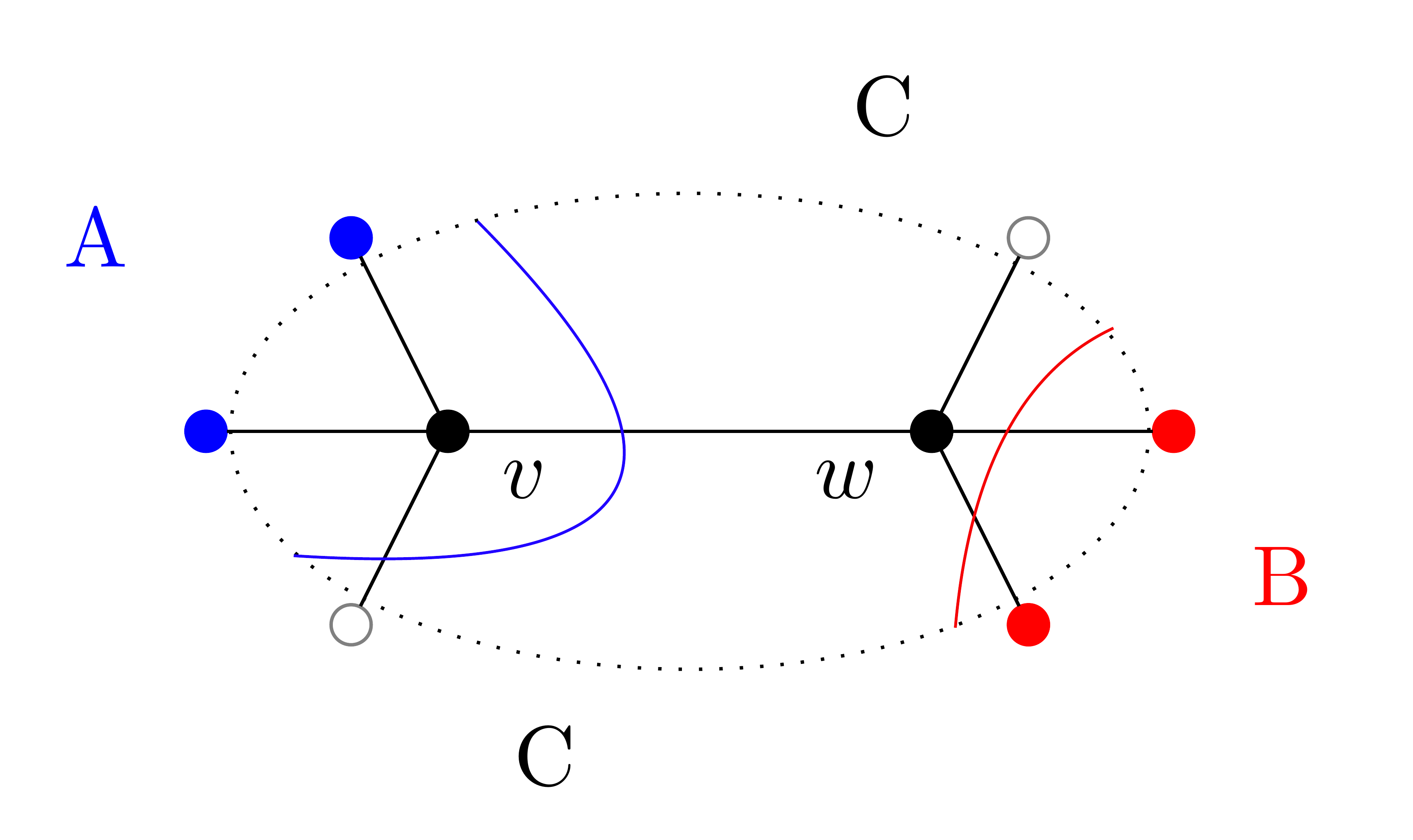}
\end{figure} 

\end{enumerate}
We can now consider correlations between intertwiners 
\begin{multline}
\mathcal{H}_k^c=2\Delta(g_v,g_w)+2\Delta(g_v,X)+\\ +2\Delta(g_w,X^{-1})+\Delta(g_v,\mathbbm{1})+\Delta(g_w,\mathbbm{1})\label{490a}
\end{multline} 
and investigate on how the values of the previous Hamiltonians are modified.
\begin{enumerate}
 \item       $g_v=g_w=\mathbbm{1}$,  $\rightarrow$ $v,w\in\,C$,
 \begin{equation}
     \mathcal{H}_3^c=2\Delta(\mathbbm{1},X)+2\Delta(\mathbbm{1},X^{-1})=8\,.
 \end{equation}

\begin{figure}[H]
    \centering
	\includegraphics[width=0.35\textwidth]{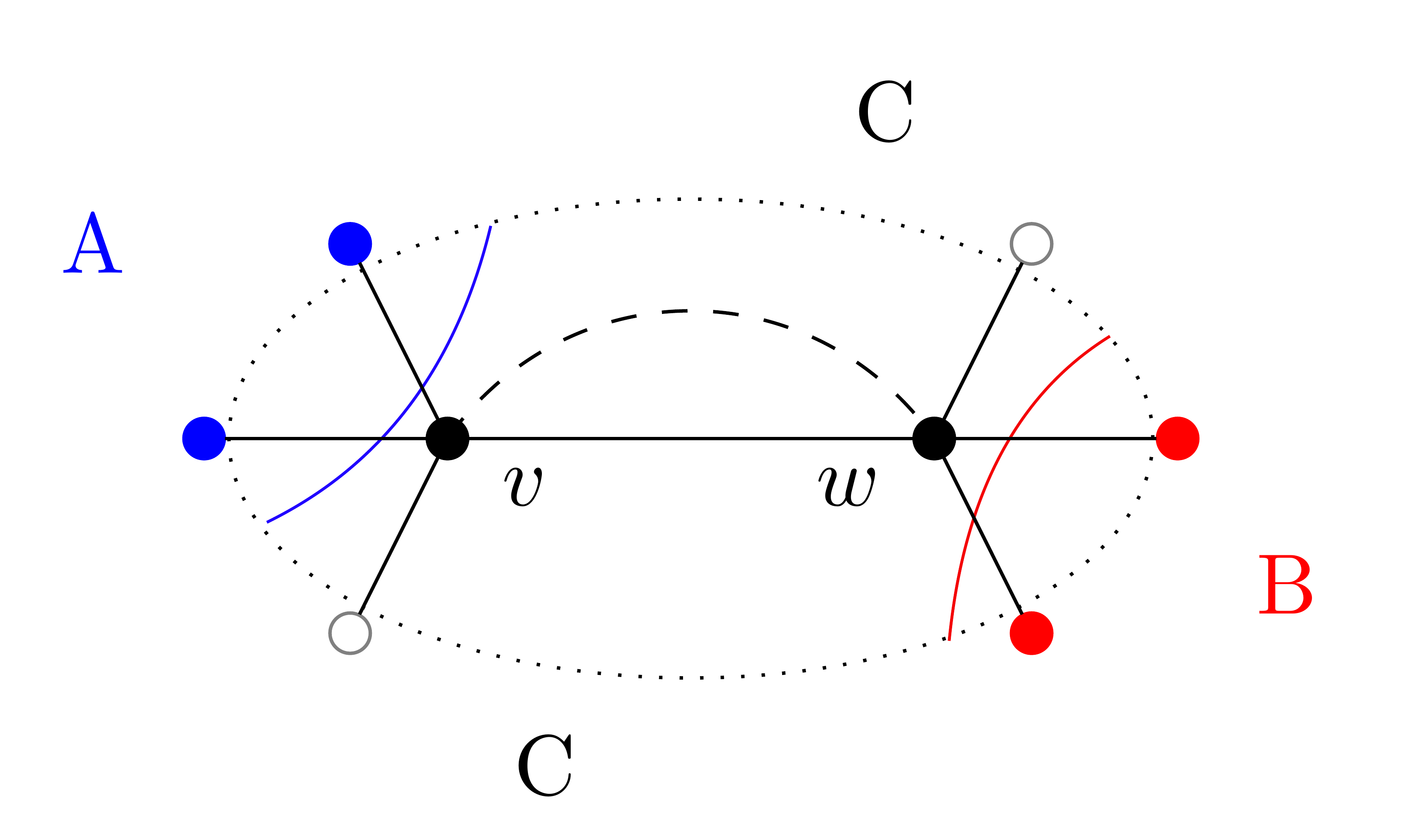}
\end{figure} 

\item       $g_v=X$,$g_w=X^{-1}$ $\rightarrow$ $v\in\,A$, $w\in\,B$,
 \begin{equation}
     \mathcal{H}_3^c=2\Delta(X,X^{-1})+\Delta(X,\mathbbm{1})+\Delta(X^{-1},\mathbbm{1})=8\,.
 \end{equation}
  
 \begin{figure}[H]
    \centering
	\includegraphics[width=0.35\textwidth]{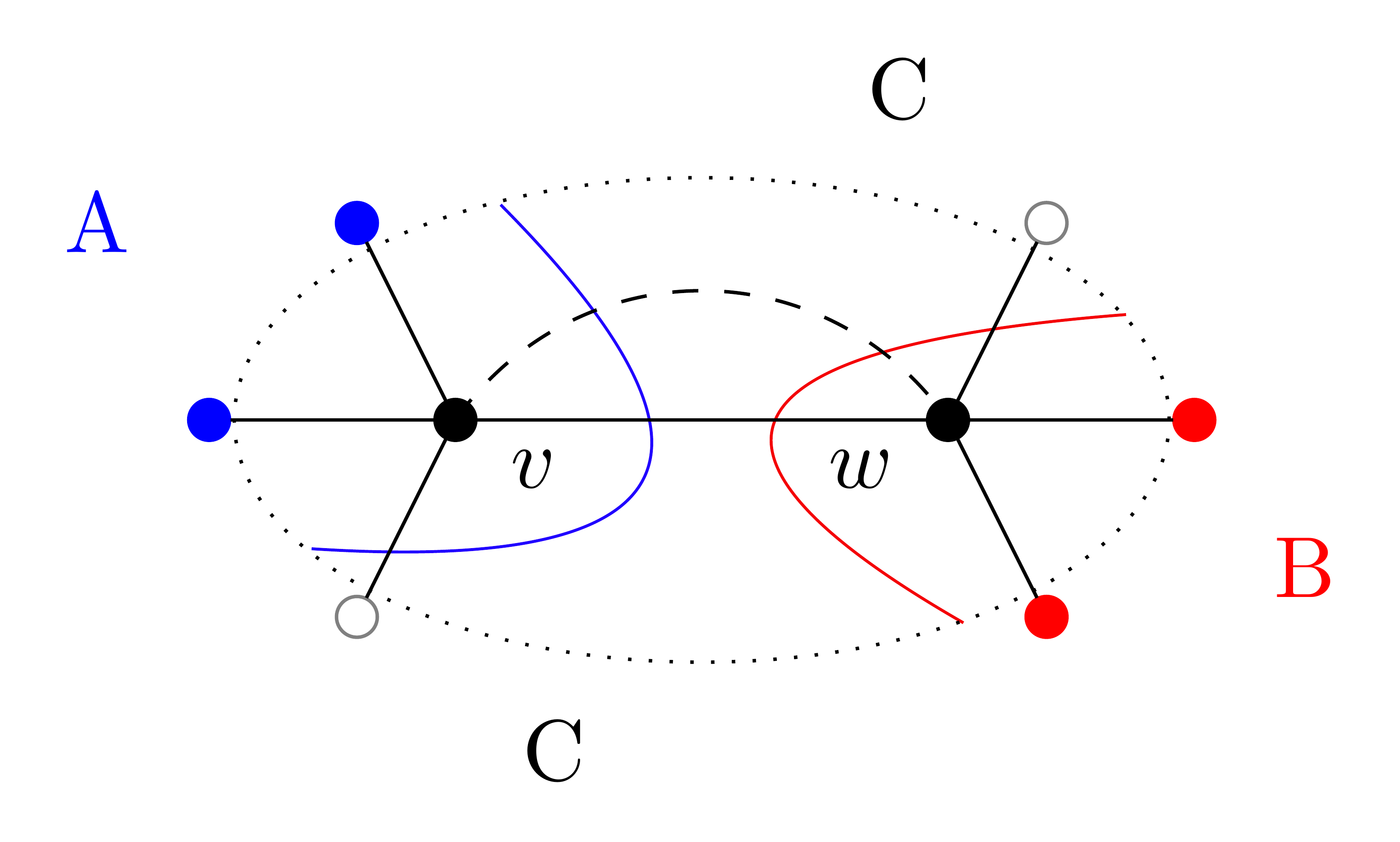}
\end{figure}  

\item       $g_v=X$,$g_w=\mathbbm{1}$  $\rightarrow$ $v\in\,A$, $w\in\,C$,
 \begin{equation}
     \mathcal{H}_3^c=2\Delta(X,\mathbbm{1})+2\Delta(\mathbbm{1},X^{-1})+\Delta(X,\mathbbm{1})=10\,.
 \end{equation}
\begin{figure}[H]
    \centering
	\includegraphics[width=0.35\textwidth]{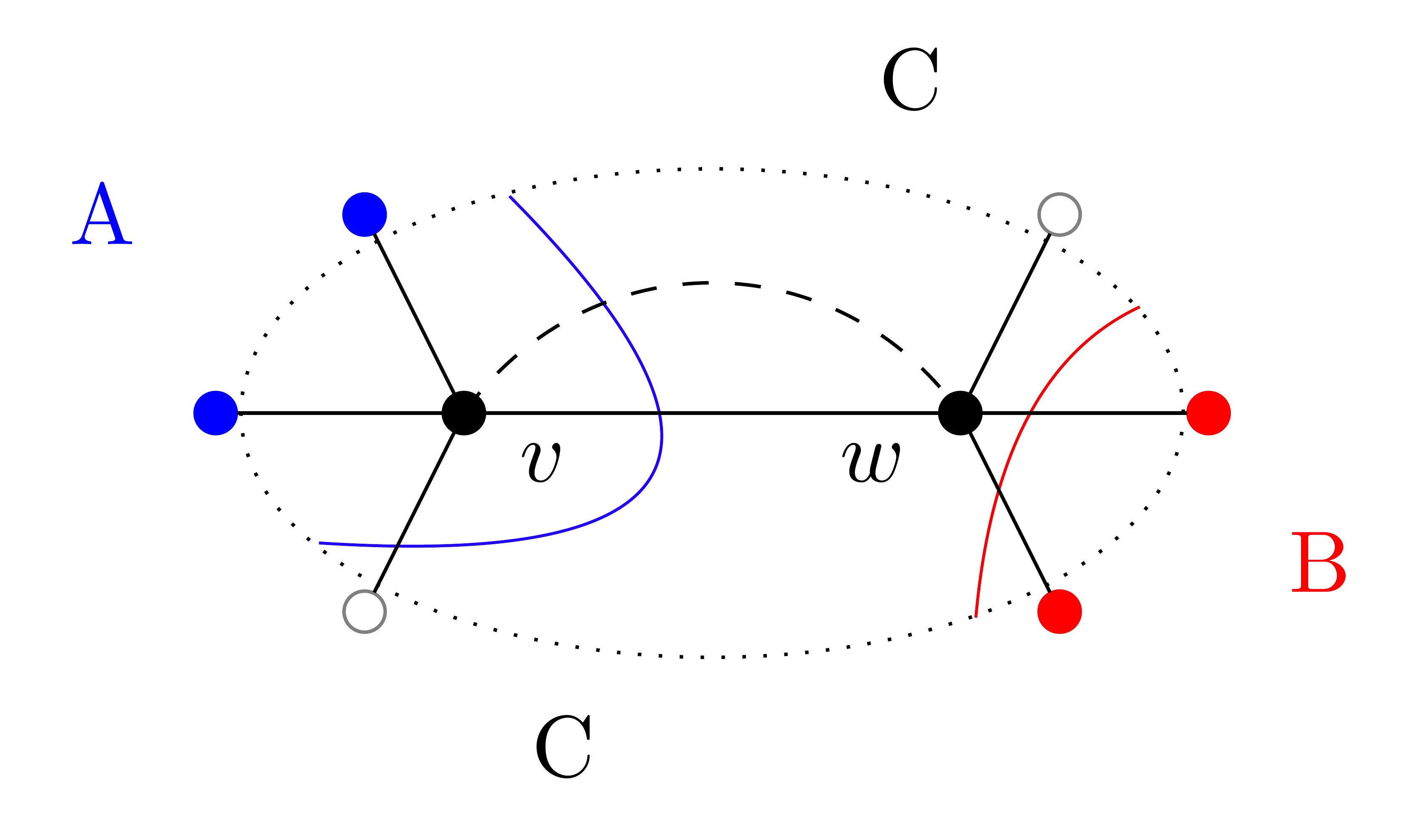}
\end{figure} 

\item $g_v=g_w=\tau$  $\rightarrow$ $v,w\in\,T$,
 \begin{equation}
     \mathcal{H}_3^c=2\Delta(\tau,X)+2\Delta(\tau,X^{-1})+2\Delta(\tau,\mathbbm{1})=6\,.
 \end{equation}
 \begin{figure}[H]
    \centering
	\includegraphics[width=0.35\textwidth]{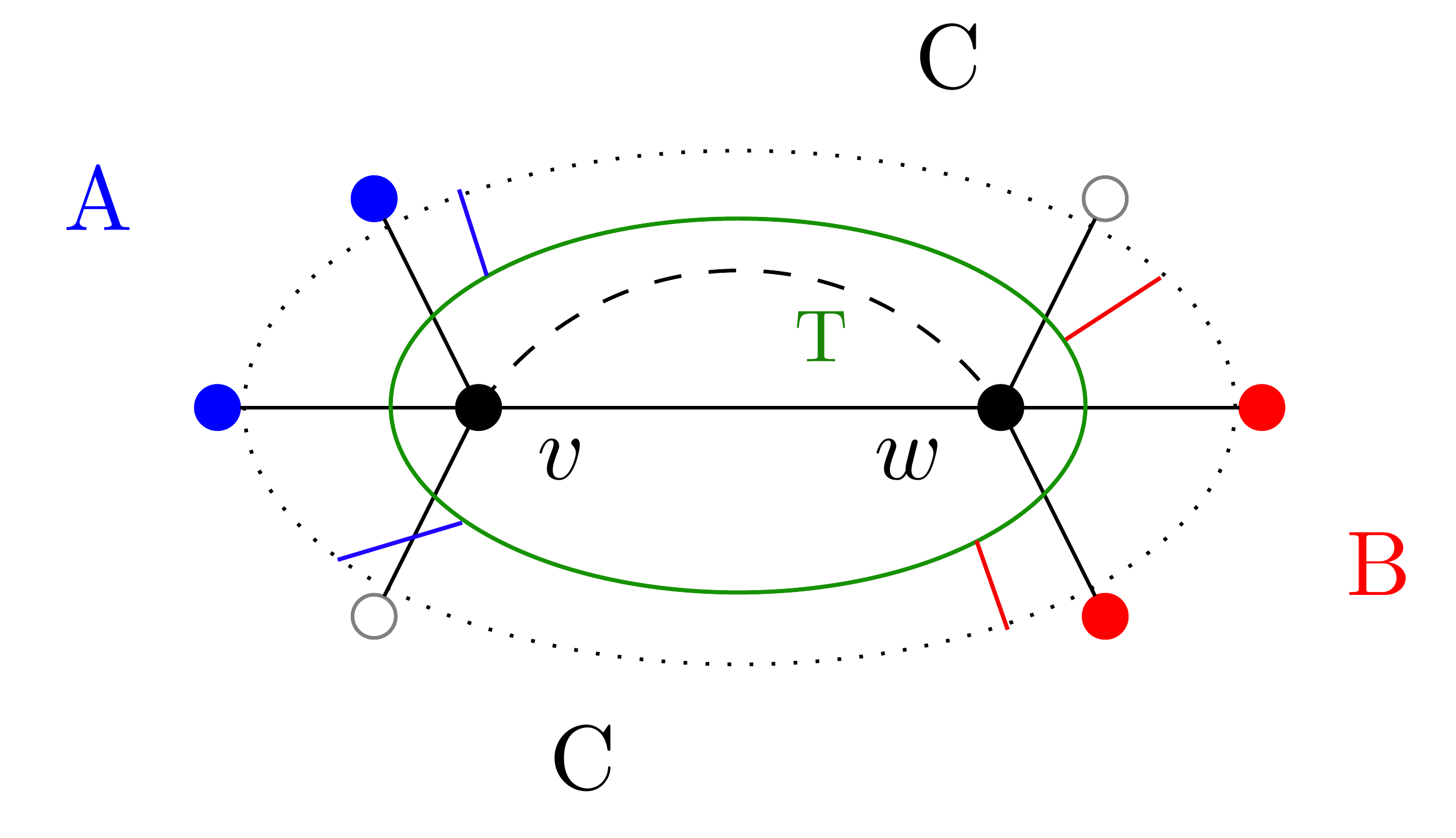}
\end{figure} 
\end{enumerate}

\subsubsection*{Order k=4 Hamiltonians} The permutation group $\mathbb{S}_4$ has $4!=24$ elements. We can focus on the set of permutations we are going to use in the calculation of $\mathcal{H}_4$:
\begin{eqnarray}
      X&=&(1234)\,\,\,\,\,X^{-1}=(4321)\,\,\,\,\,\mathbbm{1}=(1)(2)(3)(4)\,, \nonumber \\
      \tau&=&(12)(34)\;\;,\;\;(14)(23)\,, \nonumber \\
      \Delta(\mathbbm{1},X)&=&\Delta(\mathbbm{1},X^{-1})=4-1=3\,,\nonumber \\ \Delta(X,X^{-1})&=&4-2=2\,, \nonumber \\
      \Delta(\mathbbm{1},\tau)&=&\bigg\lfloor\frac{4}{2}\bigg\rfloor=2\,, \nonumber \\
      \Delta(\tau,X)&=&\Delta(\tau,X^{-1})=\bigg\lceil\frac{4}{2}\bigg\rceil-1=1\,.\nonumber
\end{eqnarray}
We can study in detail the cases of the two degenerate minimal configurations
\begin{enumerate}
 \item \label{2conf1nc}      $g_v=X$,$g_w=X^{-1}$ $\rightarrow$ $v\in\,A$, $w\in\,B$,
 \begin{equation}
     \mathcal{H}_4=\Delta(X,X^{-1})+\Delta(X,\mathbbm{1})+\Delta(X^{-1},\mathbbm{1})=8\,.
 \end{equation}
\begin{figure}[H]
    \centering
	\includegraphics[width=0.35\textwidth]{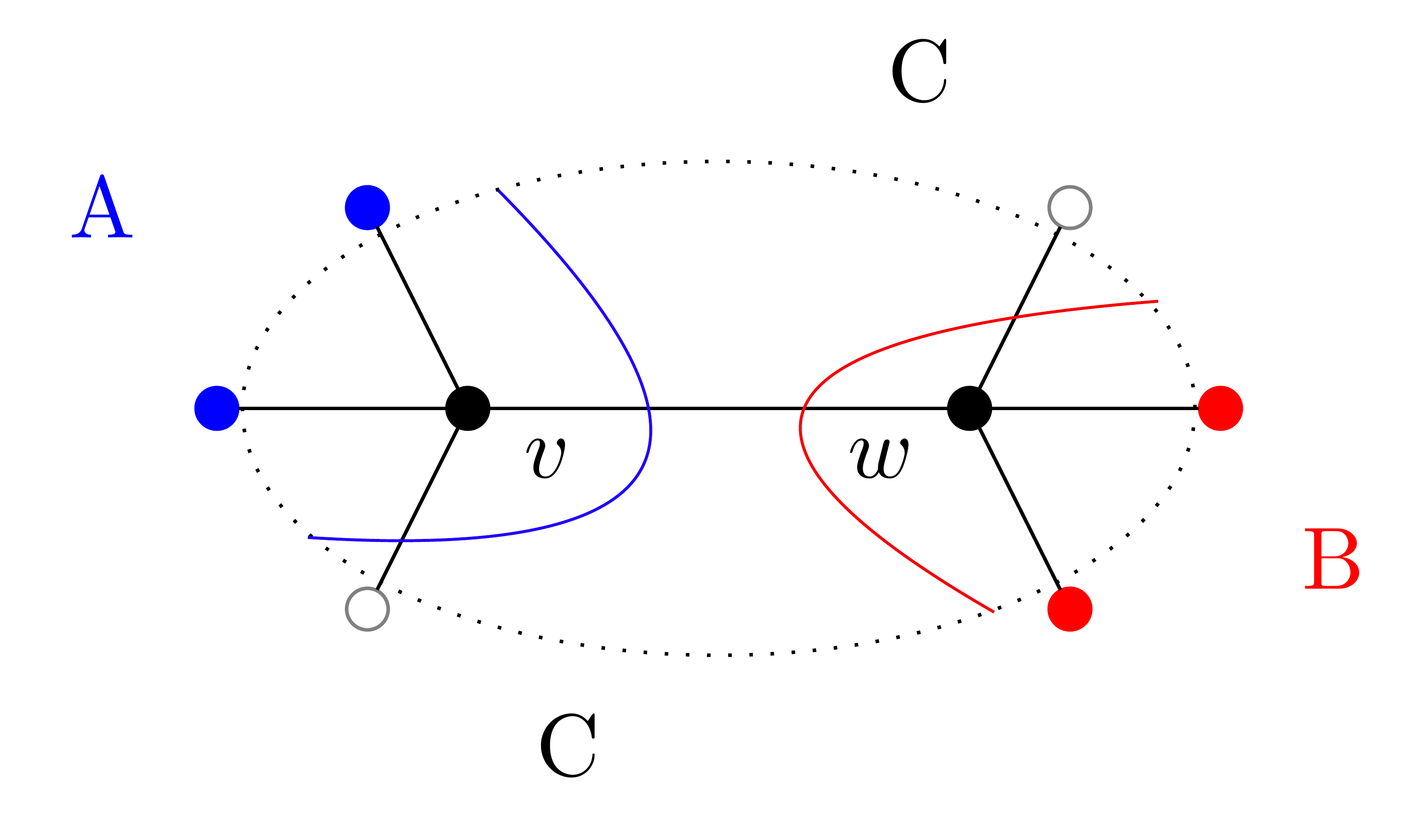}
\end{figure} 

\item\label{2conf2nc2} The set of  Non Crossing Pairings is made up by permutations with two blocks of length two. In $S_4$ there are two NCP, given by $S_{12}S_{34}$ and $S_{14}S_{23}$. Denoting by $\tau$ such permutations we have\\
\\
$g_v=g_w=\tau$  $\rightarrow$ $v,w\in\,T$,
 \begin{equation}
     \mathcal{H}_4=2\Delta(\tau,X)+2\Delta(\tau,X^{-1})+2\Delta(\tau,\mathbbm{1})=8\,.
 \end{equation}
\begin{figure}[H]
    \centering
	\includegraphics[width=0.35\textwidth]{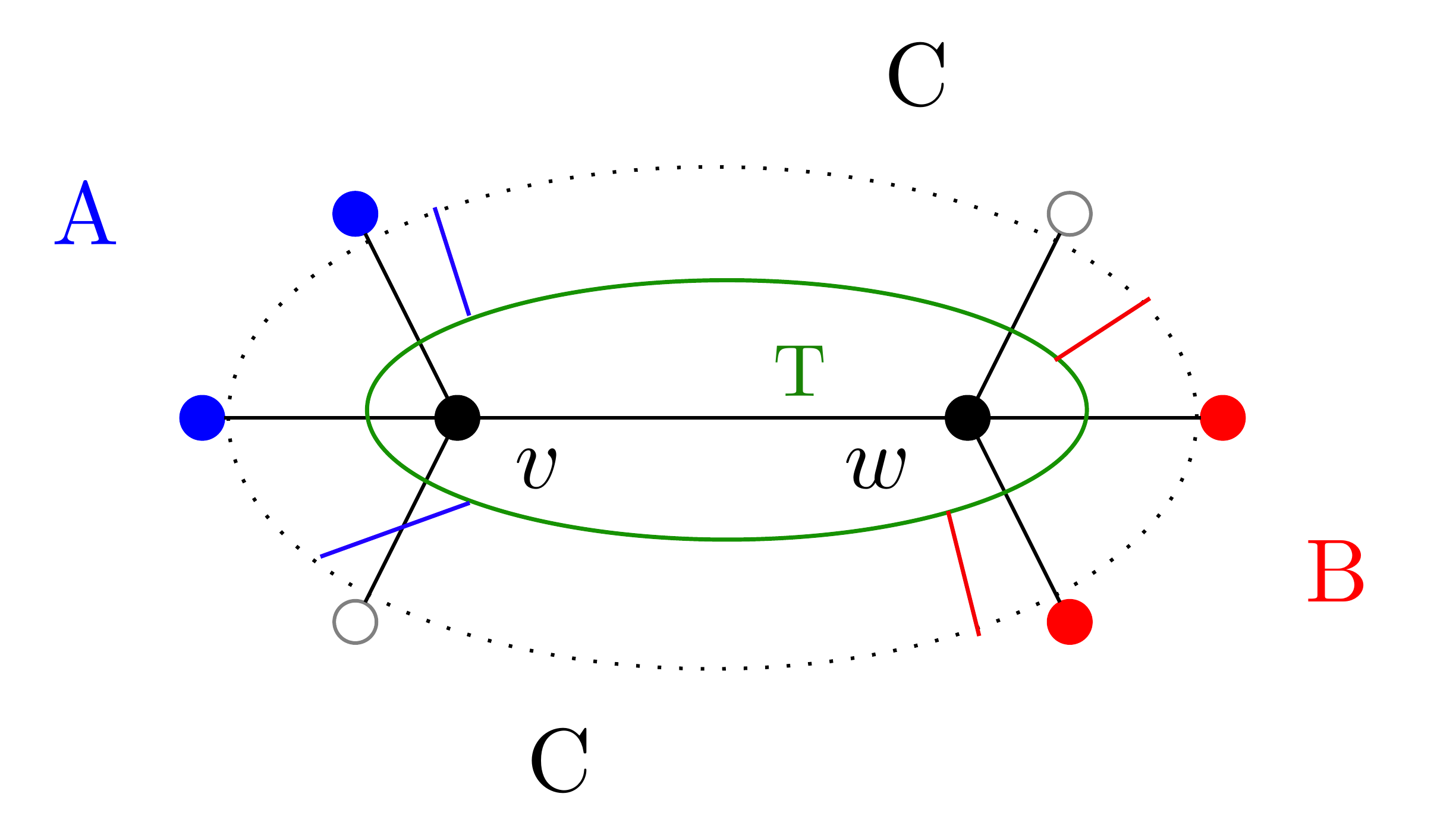}
\end{figure} 

\end{enumerate}

Considering now bulk correlation
\begin{enumerate}
 \item \label{2conf1nc2}      $g_v=X$,$g_w=X^{-1}$ $\rightarrow$ $v\in\,A$, $w\in\,B$,
 \begin{equation}
     \mathcal{H}_4=2\Delta(X,X^{-1})+\Delta(X,\mathbbm{1})+\Delta(X^{-1},\mathbbm{1})=10\,.
 \end{equation}
\begin{figure}[H]
    \centering
	\includegraphics[width=0.35\textwidth]{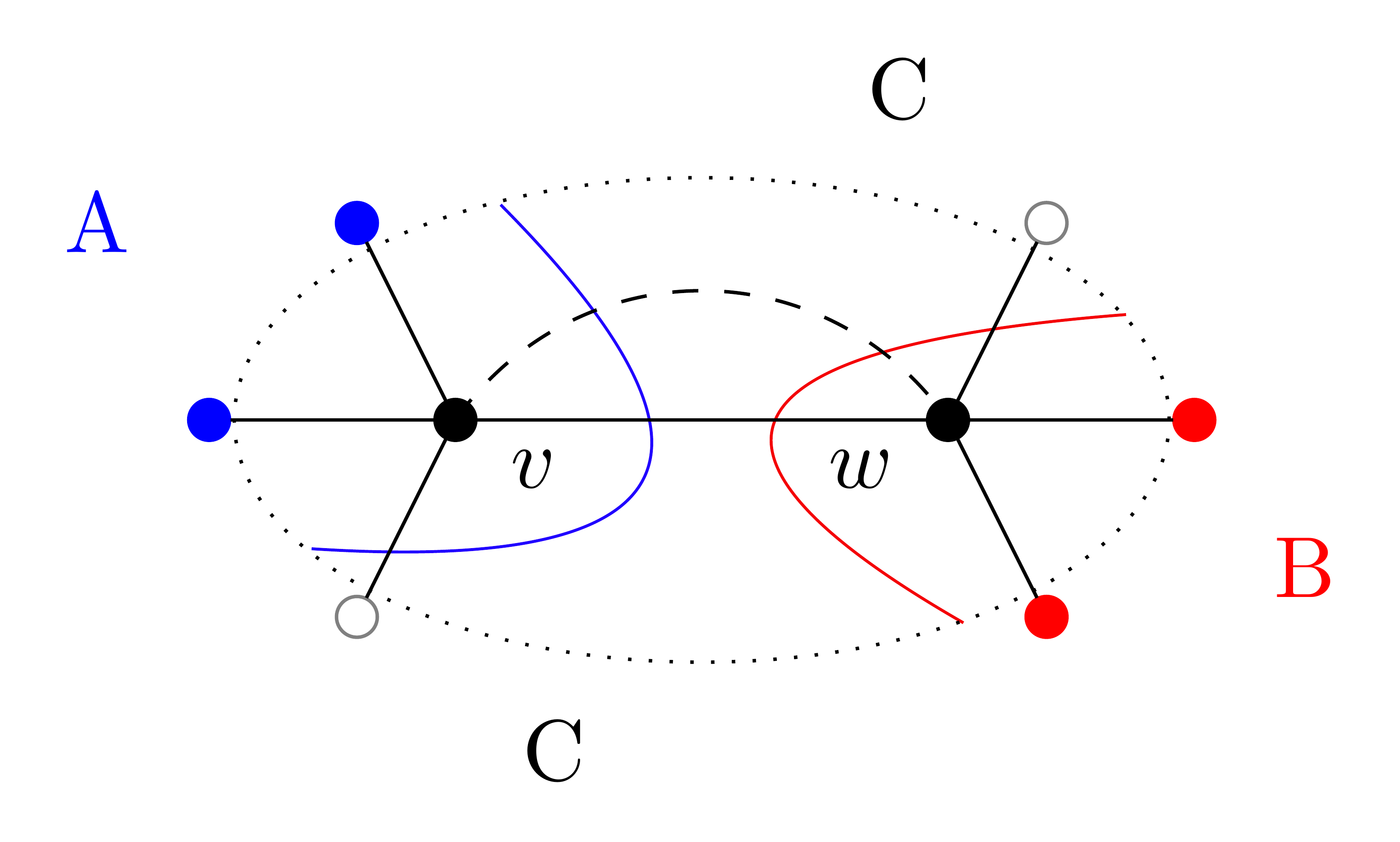}
\end{figure} 

\item\label{2conf2nc3} $g_v=g_w=\tau$  $\rightarrow$ $v,w\in\,T$,
 \begin{equation}
\mathcal{H}_4=2\Delta(\tau,X)+2\Delta(\tau,X^{-1})+2\Delta(\tau,\mathbbm{1})=8\,.
 \end{equation}
\begin{figure}[H]
    \centering
	\includegraphics[width=0.35\textwidth]{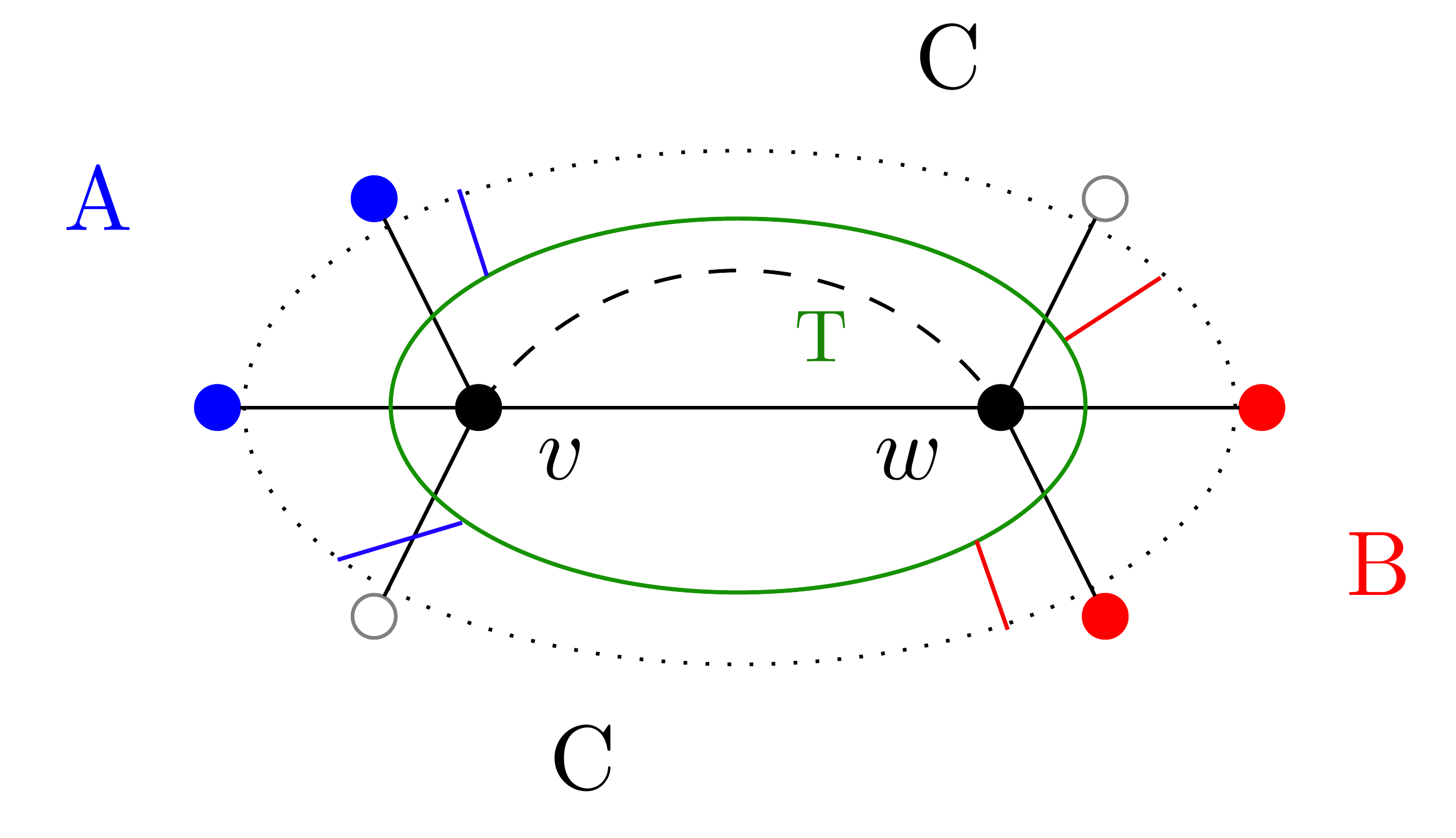}
\end{figure} 
\end{enumerate}

\subsection{Hamiltonian of the three vertices graph}
\subsubsection{Order k=3 Hamiltonians}
\begin{enumerate}
\item
  $g_x=g_y=g_z=\mathbbm{1}$,  $\rightarrow$ $x,y,z\in\,C$,
  \begin{equation}\label{3neg3ver1}
      \mathcal{H}_3=2\Delta(\mathbbm{1},X)+\Delta(\mathbbm{1},X)+2\Delta(\mathbbm{1},X^{-1})=10\,.
  \end{equation}
\begin{figure}[H]
    \centering
	\includegraphics[width=0.45\textwidth]{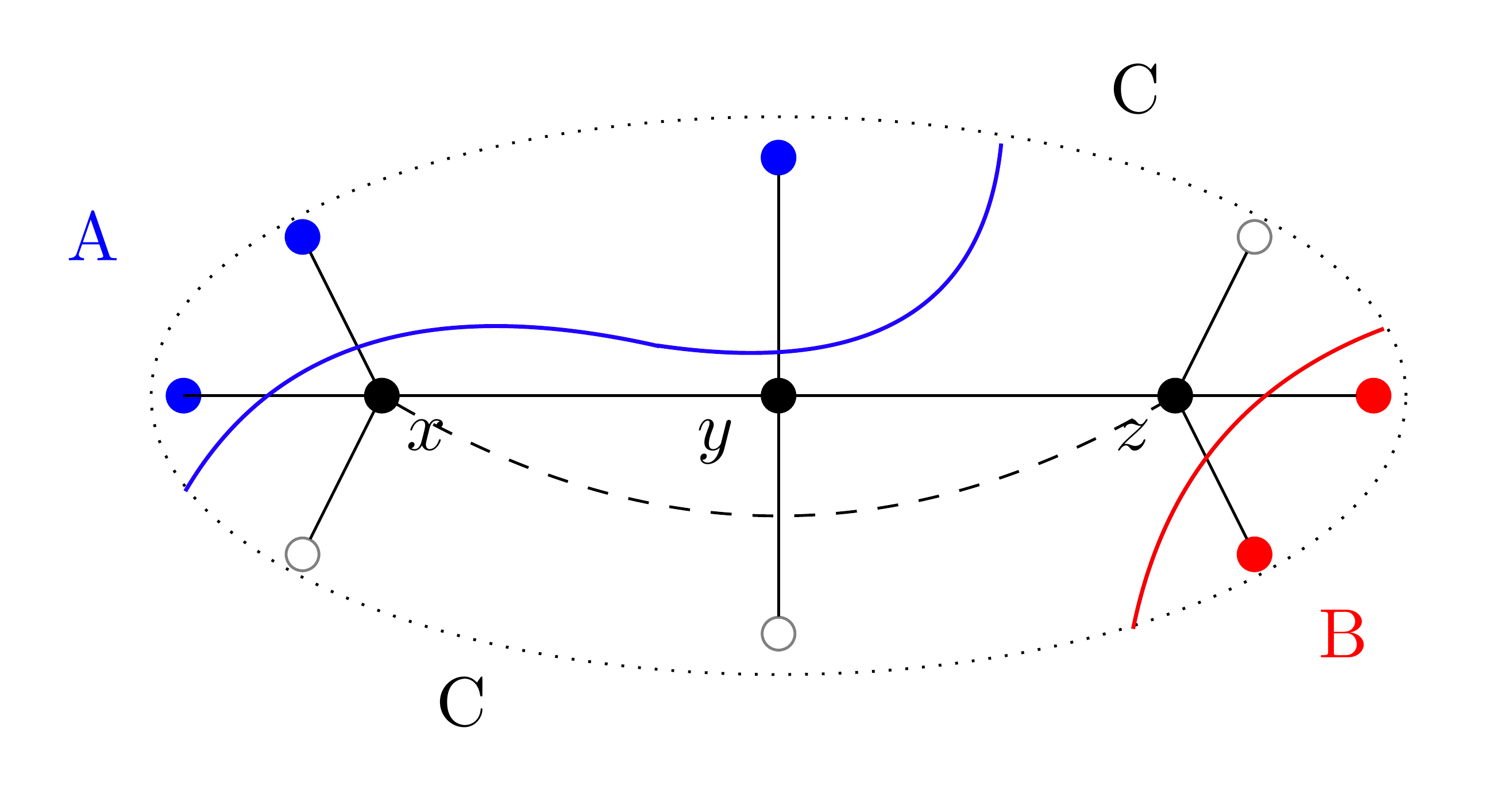}
\end{figure} 

\item 
$g_x=g_y=X$, $g_z=X^{-1}$  $\rightarrow$ $x,y\in\,A$, $z\in\,B$,
  \begin{equation}\label{3neg3ver2}
      \mathcal{H}_3=\Delta(g_y,g_z)+\Delta(\mathbbm{1},X)+\Delta(\mathbbm{1},X)+\Delta(\mathbbm{1},X^{-1})=8\,.
  \end{equation}
\begin{figure}[H]
    \centering
	\includegraphics[width=0.45\textwidth]{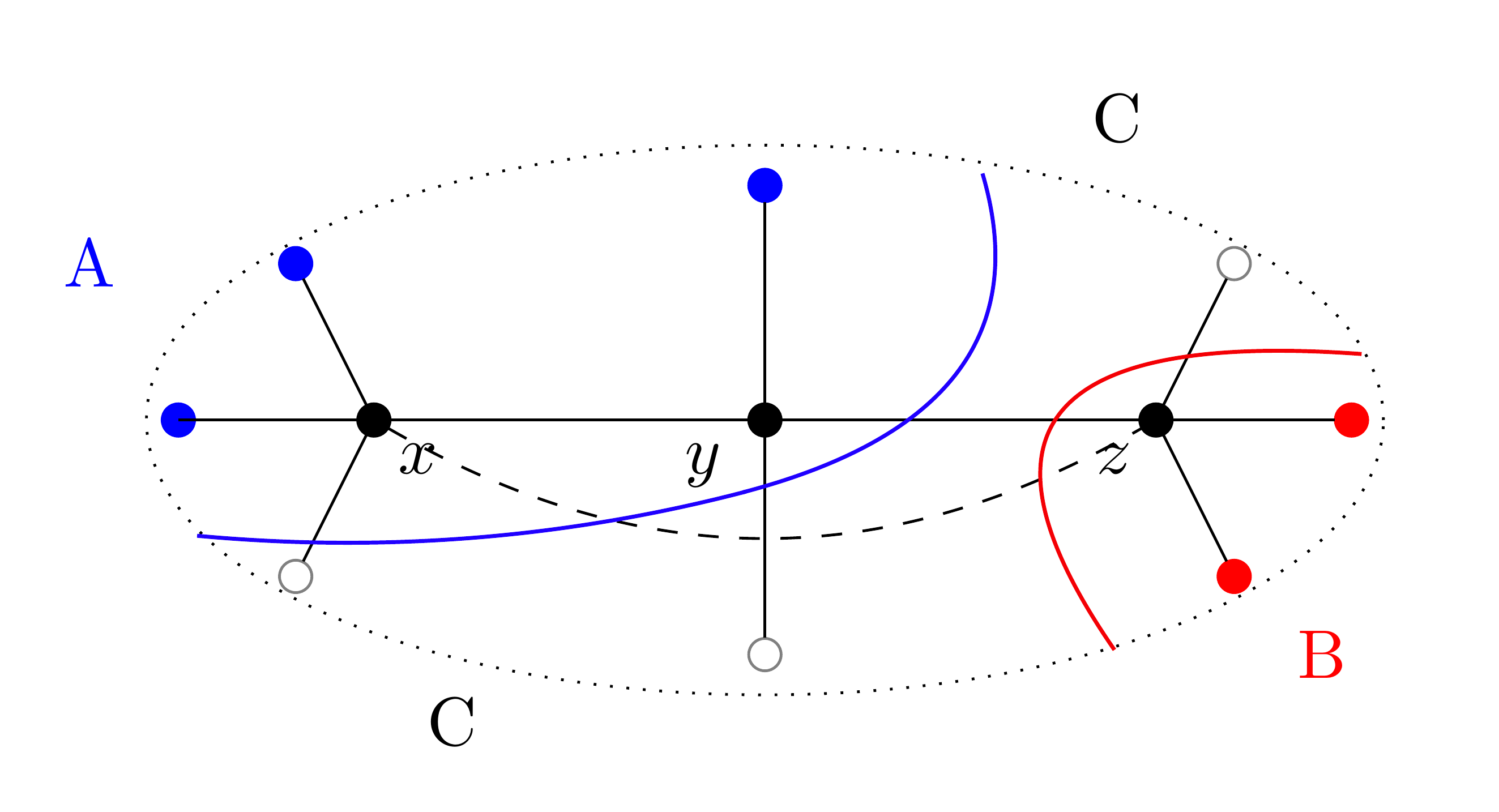}
\end{figure} 
\item $g_x=g_y=g_z=\tau$ $\rightarrow$ $x,y,z\in\,T$,
 \begin{equation}\label{3neg3ver3}
      \mathcal{H}_3=2\Delta(\tau,X)+\Delta(\tau,X)+3\Delta(\mathbbm{1},\tau)+2\Delta(\tau,X^{-1})=8\,.
  \end{equation}
\begin{figure}[H]
    \centering
	\includegraphics[width=0.45\textwidth]{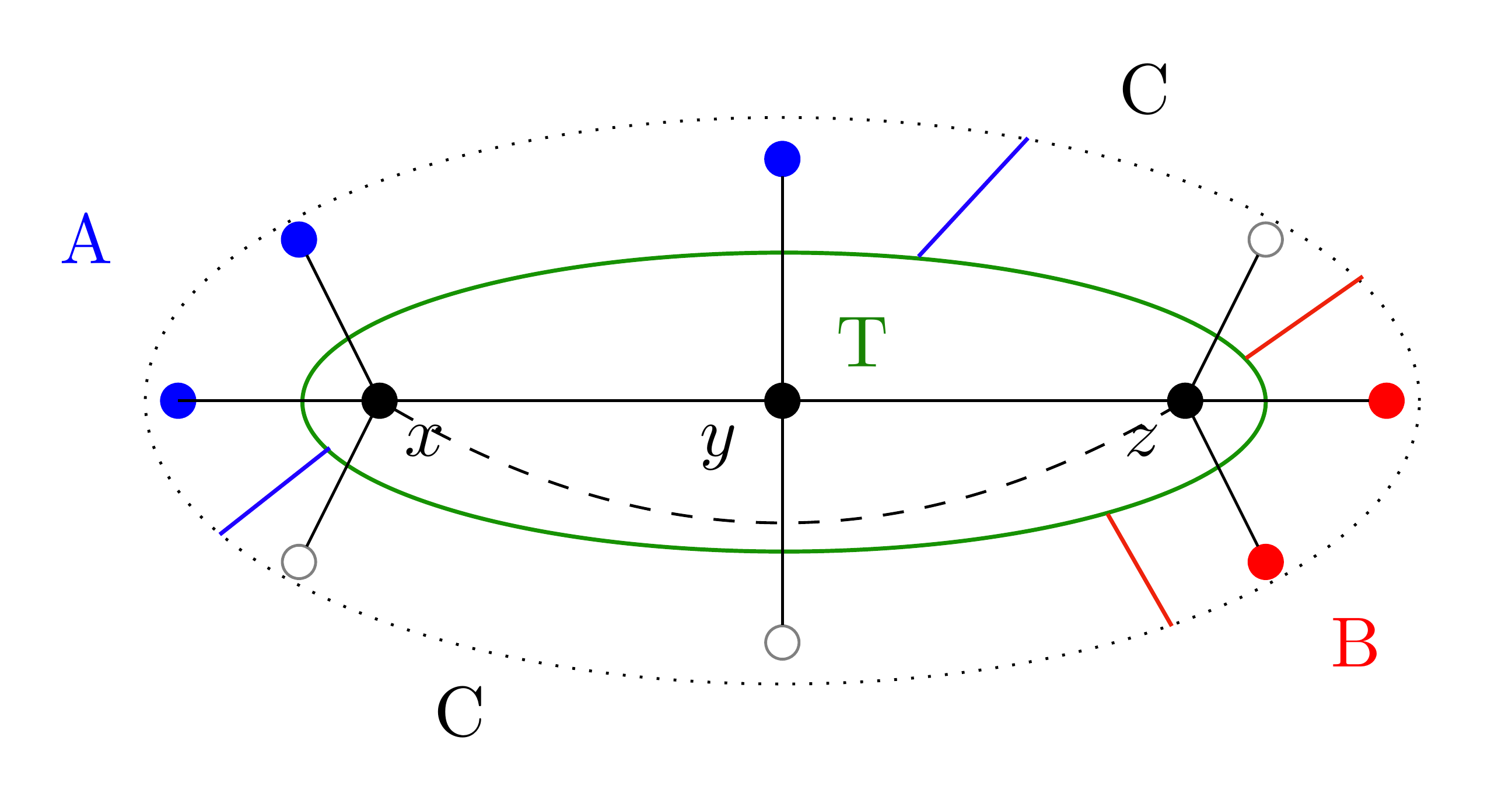}
\end{figure} 
\item  $g_x=X$, $g_y=g_z=\tau$ $\rightarrow$ $x\in\,A$, $y,z\in\,T$,
 \begin{equation}\label{3neg3ver4}
      \mathcal{H}_3=2\Delta(\tau,X)+\Delta(\mathbbm{1},X)+2\Delta(\mathbbm{1},\tau)+2\Delta(\tau,X)=8\,.
  \end{equation}
\begin{figure}[H]
    \centering
	\includegraphics[width=0.45\textwidth]{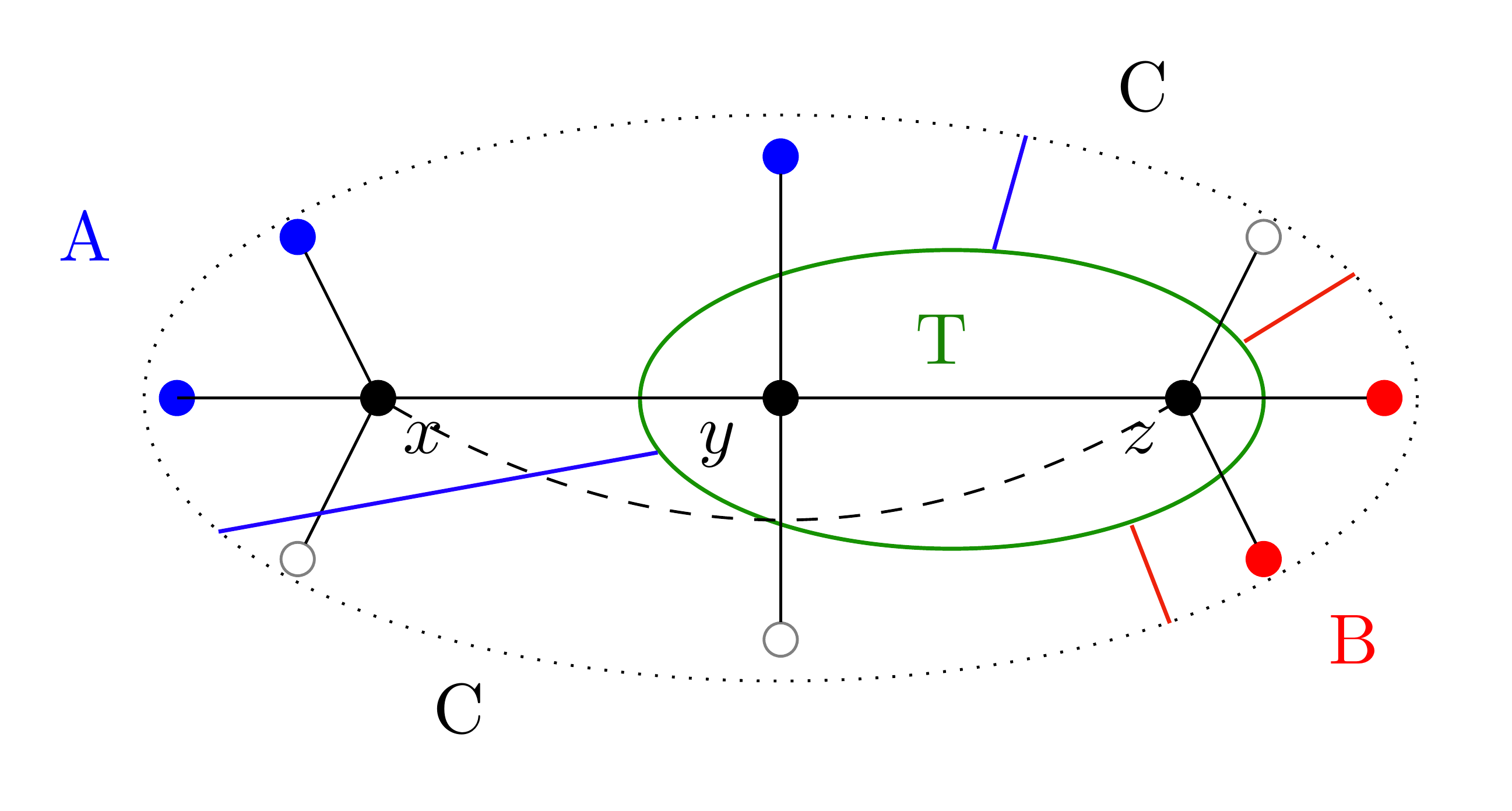}
\end{figure} 
The configuration with $x$ and $y$ in T gives the same value of $\mathcal{H}_3$.\\
\item $g_x=g_y=X$, $g_z=\tau$ $\rightarrow$ $x,y\in\,A$, $z\in\,T$,
\begin{multline}\label{3neg3ver5}
      \mathcal{H}_3=\Delta(\tau,X)+\Delta(\tau,X^{-1})+\Delta(X,\tau)+\\ +\Delta(\mathbbm{1},X)+\Delta(\tau,X)+\Delta(X^{-1},\mathbbm{1})=8\,.
  \end{multline}
\begin{figure}[H]
    \centering
	\includegraphics[width=0.45\textwidth]{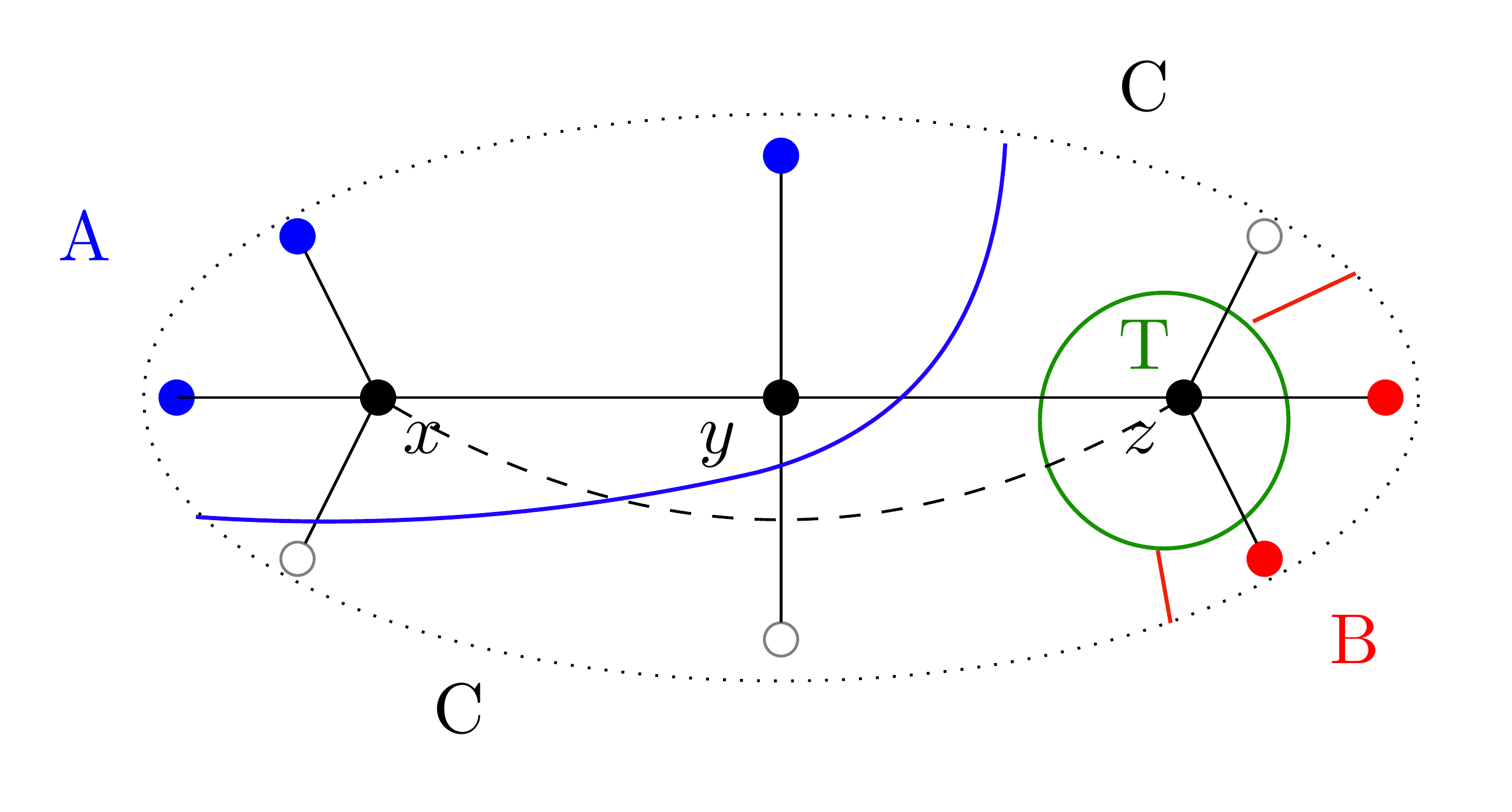}
\end{figure} 
\noindent
The configurations with $x$ or $y$ in T gives the same value of $\mathcal{H}_3$.
\end{enumerate}

\bibliographystyle{apsrev4-1}
\bibliography{bibliography_neg}

\end{document}